\documentclass{article}

\usepackage{arxiv}
\usepackage[utf8]{inputenc} 
\usepackage[T1]{fontenc}    
\usepackage{hyperref}       
\usepackage{url}            
\usepackage{booktabs}       
\usepackage{amsfonts}       
\usepackage{nicefrac}       
\usepackage{microtype}      
\usepackage{graphicx}       
\usepackage{subcaption}     
\usepackage{multirow}
\newcommand\numOfImages{780}
\usepackage{float}
\title{COVID-CXNet: Detecting COVID-19 in Frontal Chest X-ray Images using Deep Learning}

\author{
  Arman Haghanifar\\
  Div. of Biomedical Engineering\\
  University of Saskatchewan\\
  \texttt{arman.haghanifar@usask.ca}\\
   \And
  Mahdiyar Molahasani Majdabadi \\
  Department of Electrical \& Computer Engineering\\
  University of Saskatchewan\\
  \texttt{m.molahasani@usask.ca}\\
   \And
  Younhee Choi \\
  International Road Dynamics, Canada\\
  \texttt{younhee.choi@irdinc.com}\\
   \And
  S. Deivalakshmi \\
  National Institute of Technology, Trichy, India\\
  \texttt{deiva@nitt.edu}\\
   \And
  Seokbum Ko \\
  Department of Electrical \& Computer Engineering\\
  University of Saskatchewan\\
  \texttt{seokbum.ko@usask.ca}\\
}

\begin{document}
\maketitle

\begin{abstract}
One of the primary clinical observations for screening the novel coronavirus is capturing a chest x-ray image. In most patients, a chest x-ray contains abnormalities, such as consolidation, resulting from COVID-19 viral pneumonia. In this study, research is conducted on efficiently detecting imaging features of this type of pneumonia using deep convolutional neural networks in a large dataset. It is demonstrated that simple models, alongside the majority of pretrained networks in the literature, focus on irrelevant features for decision-making. In this paper, numerous chest x-ray images from various sources are collected, and the largest publicly accessible dataset is prepared. Finally, using the transfer learning paradigm, the well-known CheXNet model is utilized to develop COVID-CXNet. This powerful model is capable of detecting the novel coronavirus pneumonia based on relevant and meaningful features with precise localization. COVID-CXNet is a step towards a fully automated and robust COVID-19 detection system.
\end{abstract}


\section{Introduction}
Being declared as a pandemic, novel coronavirus is now a major emergency worldwide. The virus is transmitted person-to-person by respiratory droplets or close contact with a contaminated surface \cite{world2020coronavirus}. The most common symptoms are fever, cough, and dyspnea, which may appear 2-14 days after exposure to the virus. The standard diagnosis method, highly specific but with inconstant sensitivity \cite{kanne2020essentials}, is based on reverse transcription polymerase chain reaction (RT-PCR) \cite{rio2014reverse}. The RT-PCR test has certain shortcomings, such as availability and time-consumption. It needs special test-kits, which may not be widely available in some regions \cite{bai2020performance}, and the results are generally available within hours to days \cite{brueck2020coronavirus}. A diagnostic guideline proposed by Zhongnan Hospital of Wuhan suggests that the disease could be assessed by detecting clinical symptoms as well as radiological findings of pneumonia \cite{jin2020rapid}. Furthermore, Ai \textit{et al.} show that chest computed tomography (CT) scans have high sensitivity for COVID-19 diagnosis and can be considered as the primary diagnostic tools in epicenters \cite{ai2020correlation}.


Chest x-rays (CXRs) and CT scans have been used for COVID-19 screening and disease progression evaluation in hospital admitted cases \cite{rubin2020role}. Despite offering superior sensitivity to thoracic abnormality detection \cite{cellina2020false}, using CT has several challenges. CT scanners are non-portable and require sanitizing of the equipment and imaging room between patients. Besides, their radiation dose is considerably higher than x-rays \cite{furlow2010radiation}. On the contrary, portable x-ray units are widely available and can be easily accessed in most primary hospitals. Moreover, x-ray imaging can be operated in more isolated rooms with less staff exposure to the virus. In many cases, the patient's clinical situation does not allow a CT scan; hence, CXRs are a better choice for the initial assessment.

Since radiologists visit many patients every day and the diagnosis process takes significant time, errors may increase notably. As a result, there might be many more false negatives that will cost a lot to the patient and the medical staff. Therefore, automated computer-aided diagnostic (CAD) tools are of utmost importance. Automated deep learning-based CAD tools have previously shown promising results in pulmonary disease detection \cite{lakhani2017deep}. 



In this study, firstly, we collect a dataset of CXRs from COVID-19 patients from multiple publicly accessible sources. Our collected dataset is the largest source of COVID-19 CXRs, containing \numOfImages{} images. We then investigate the possibility of disease detection by an individual Convolutional Neural Network (CNN) model. On the next step, performance of prominent pretrained CNN models for fine-tuning on the dataset is investigated. Afterwards, the CheXNet pretrained model on the same type of medical images is introduced, and its efficiency is discussed. Finally, we develop our model based on the CheXNet and design a lung segmentation module to improve the model localization of lung abnormalities. Class activation map (CAM) is our main visualization leverage to compare our models. Main contributions can be summarized as:

\begin{itemize}
    \item Collecting the largest public dataset of COVID-19 CXR images from different sources
    \item Developing a robust detection model by training on a large dataset of COVID-19 pneumonia CXRs
    \item Precisely evaluating model performance by visualizing the results using CAMs
\end{itemize}

\section{Related Works} \label{related_works}
Identifying COVID-19 pneumonia using different types of medical images is a fast-growing topic of interest. ML-based methods, along with manual feature extraction algorithms, are used in a few articles to diagnose the disease \cite{barstugan2020coronavirus, hassanien2020automatic, dey2020social, al2020ai, gomes2020ikonos}. However, most studies are utilizing DL-based techniques. Researchers have tried to tackle the problem using CT images, reaching high scoring metrics and precise abnormality localization \cite{shan+2020lung, gozes2020rapid}. Contrarily, even though many studies have claimed to achieve excellent classification accuracy scores using CXRs, such as \cite{pereira2020covid} and \cite{khan2020coronet}, none of them have reported visualization results. Because pneumonia diagnosis is more challenging in CXRs and the available COVID-19 pneumonia CXR datasets are small, we investigate those studies with visual interpretability used as their metric.

Zhang \textit{et al.} used a dataset including 100 CXRs from COVID-19 cases and developed a ResNet-based model with pretrained weights from ImageNet as the backbone \cite{zhang2020covid}. Their best model achieved an f-score of $\approx$ 0.72 in classifying COVID-19 pneumonia from CAP. Li \textit{et al.} applied their multi-player model called COVID-MobileXpert on a dataset of 537 images equally divided into normal, CAP, and COVID-19 pneumonia samples \cite{li2020covidmobilexpert}. Their main goal was to achieve acceptable accuracy using lightweight networks, such as SqueezeNet, for pneumonia detection on mobile devices capturing noisy snapshots. Rajaraman \textit{et al.} collected a more expanded dataset containing 313 COVID-19 pneumonia CXRs from two different sources \cite{rajaraman2020iteratively}. Lung segmentation was then applied using a U-Net-based model. Finally, an ensemble of different fine-tuned models was implemented and pruned iteratively to reduce parameters. Their best single pruned architecture was Inception-V3, and their best ensemble model was by weighted averaging strategy. They have achieved f-scores of 0.9841 and 0.99 detecting COVID-19 pneumonia from CAP and normal CXRs, respectively. However, their final generated visualization maps are not precisely discussed, and their model suffers some implementation drawbacks due to the significant number of parameters.

In a more advanced effort, COVID-Net was introduced by Wang and Wong \cite{wang2020covid}. It was trained on COVIDx, a dataset with 358 CXR images from 266 COVID-19 patient cases. Their architecture was first trained on ImageNet and then achieved a best f-score of 0.9480 in three-class classification. Their model visualization is not properly presented nevertheless. A most recent similar research study was CovidAID conducted by Mangal \textit{et al.} \cite{mangal2020covidaid}. CovidAID is a DenseNet model built upon CheXNet weights. They compared their results with COVID-Net on the same test set. Their findings suggest that CovidAID surpassed COVID-Net with a notable margin, 0.9230 f1 score, compared with 0.3591. CovidAID image visualization shows more precise performance compared to previous studies. Consequently, developed models suffer a lack of robustness, mainly related to the insufficient number of images.

\section{Data and preprocessing}
\begin{figure}[H]
    \begin{subfigure}[b]{0.3\linewidth}
        \centering
        \includegraphics[width=0.7\linewidth]{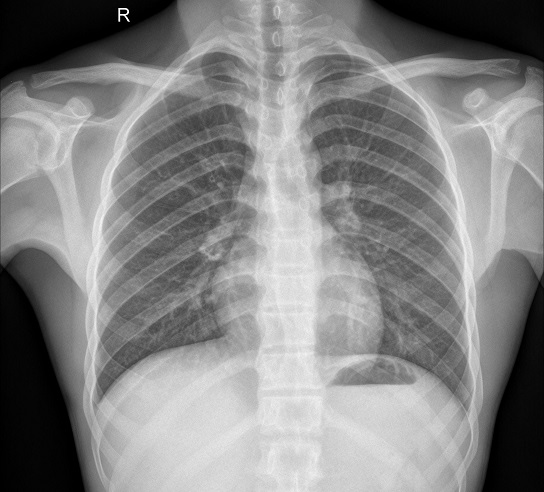}

        \label{fig:test1}
    \end{subfigure}
    \hfill
    \begin{subfigure}[b]{0.3\linewidth}
        \centering
        \includegraphics[width=0.7\linewidth]{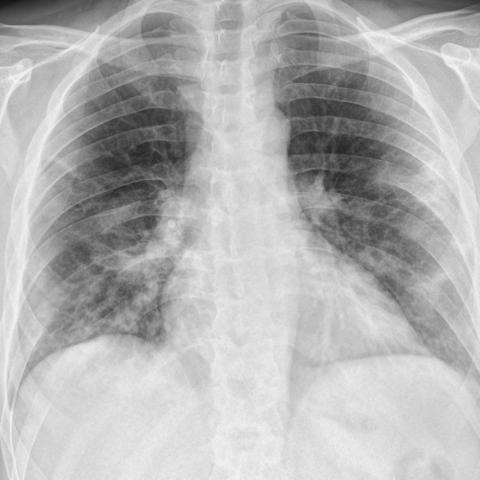}

        \label{fig:test2}
    \end{subfigure}
    \hfill
    \begin{subfigure}[b]{0.3\linewidth}
        \centering
        \includegraphics[width=0.7\linewidth]{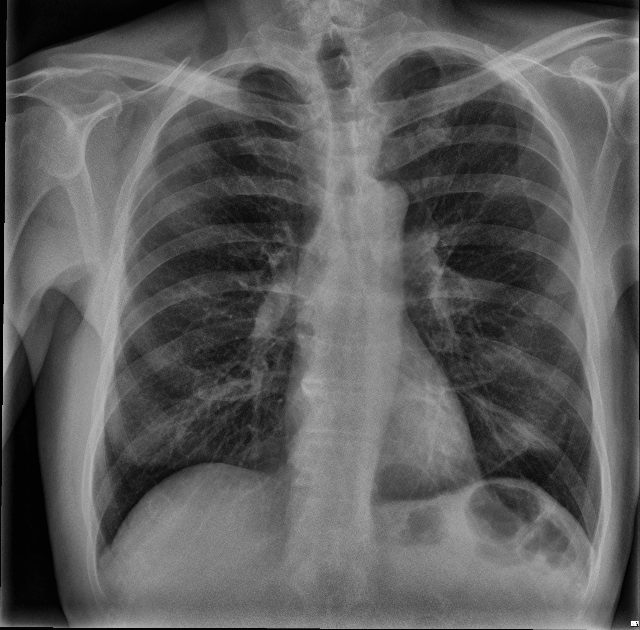}

        \label{fig:test3}
    \end{subfigure}
    \hfill
    \begin{subfigure}[b]{0.3\linewidth}
        \centering
        \includegraphics[width=0.7\linewidth]{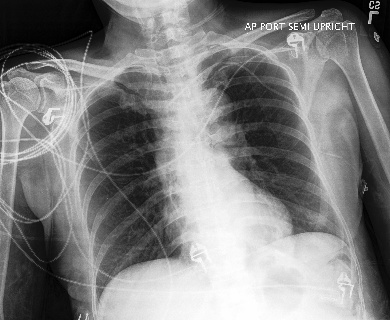}

        \label{fig:test4}
    \end{subfigure}
    \hfill
    \begin{subfigure}[b]{0.3\linewidth}
        \centering
        \includegraphics[width=0.7\linewidth]{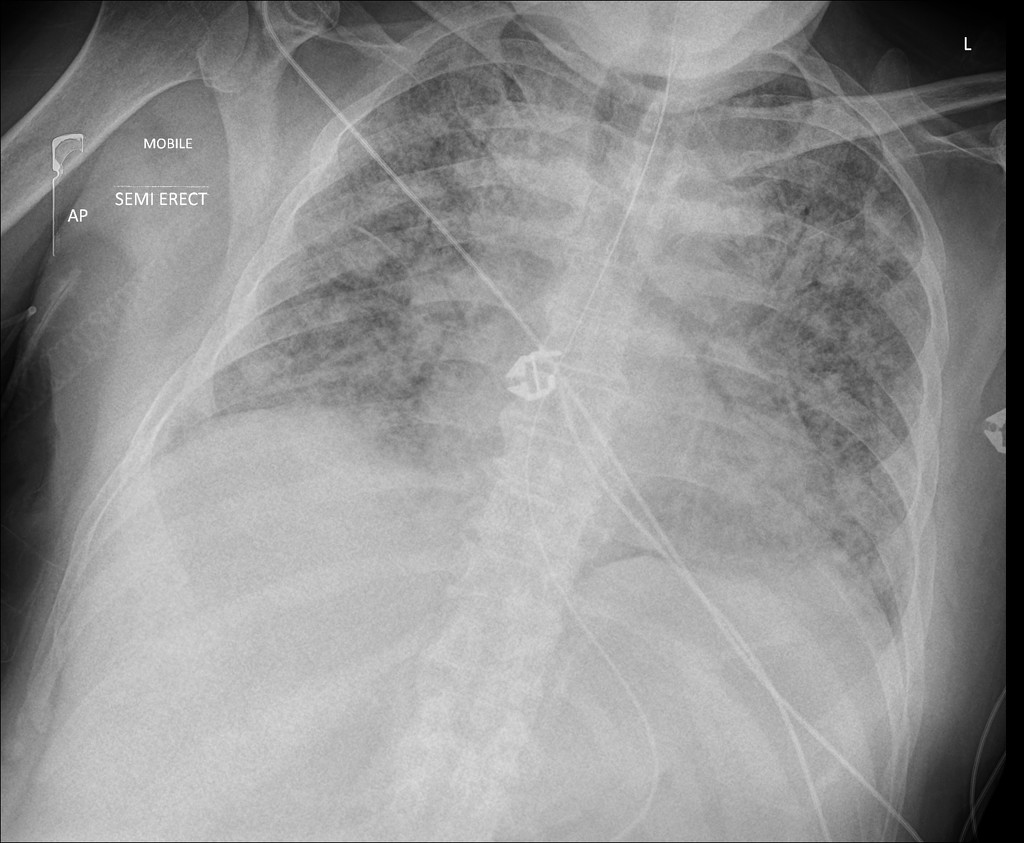}

        \label{fig:test5}
    \end{subfigure}
        \hfill
    \begin{subfigure}[b]{0.3\linewidth}
        \centering
        \includegraphics[width=0.7\linewidth]{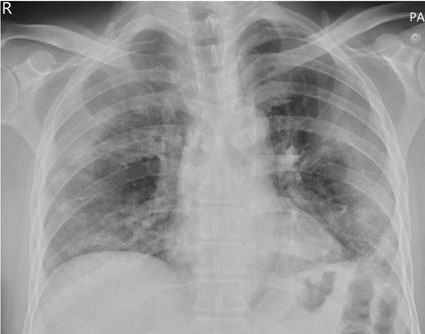}

        \label{fig:test5}
    \end{subfigure}
\caption{Randomly selected frontal CXR images from different sources}
\label{fig:normal_chest}
\end{figure}

The most common imaging technique used as the first clinical step for chest-related diseases is CXR \cite{zompatori2014overview}. Hence, more CXRs could be collected publicly than CT images. A batch of randomly selected samples from the dataset with frontal view, also known as anteroposterior (AP) or posteroanterior (PA), is shown in Fig. \ref{fig:normal_chest}.

\noindent There is another CXR imaging view called L, standing for Lateral, which is an adjunct for the main frontal view image. Lateral CXR is performed when there is diagnosis uncertainty using frontal CXR \cite{ittyachen2017forgotten}. Thus it is not as common as frontal CXR, and due to its different angle, it is excluded from our data. 


\subsection{Dataset Sources}
Since COVID-19 is a novel disease, the number of publicly available x-rays is relatively small. There are different databases, regularly updated day by day, which our dataset is constructed upon them:
\begin{enumerate}
    \item Radiopaedia \footnote{\href{https://radiopaedia.org}{https://radiopaedia.org}}: open-edit radiology resource where radiologists submit their daily cases.
    \item SIRM \footnote{\href{https://sirm.org/category/senza-categoria/covid-19/}{https://sirm.org/category/senza-categoria/covid-19/}}: the website of the Italian Society of Medical and Interventional Radiology, which has a dedicated database of COVID-19.
    \item EuroRad \footnote{\href{https://eurorad.org/}{https://eurorad.org/}}: a peer-reviewed image resource of radiological case reports.
    \item Figure1 \footnote{\href{https://figure1.com/covid-19-clinical-cases}{https://figure1.com/covid-19-clinical-cases}}: an image-based social forum that has dedicated a COVID-19 clinical cases section.
    \item COVID-19 image data collection \cite{cohen2020covid}: a GitHub repository by Dr. Cohen \textit{et al.}, which is a combination of some of the mentioned resources and other images.
    \item Twitter COVID-19 CXR dataset \footnote{\href{http://twitter.com/ChestImaging/}{http://twitter.com/ChestImaging/}}: a twitter thread of a cardiothoracic radiologist from Spain who has shared high-quality positive subjects.
    \item Peer-reviewed papers: papers which have shared their clinical images, such as \cite{jacobi2020portable} and \cite{qian2020severe}.
    \item Hannover Medical School dataset \cite{hannover2020covid}: a GitHub repository containing images from the Institute for Diagnostic and Interventional Radiology in Hannover, Germany.
    \item Social media: images collected from Instagram pages\footnote{\href{https://www.instagram.com/theradiologistpage}{https://www.instagram.com/theradiologistpage} and \href{https://www.instagram.com/radiology\_case\_reports/}{https://www.instagram.com/radiology\_case\_reports/}}.
\end{enumerate}


Four resources were considered to collect normal CXRs:
\begin{enumerate}
    \item Pediatric CXR dataset \cite{kermany2018identifying}: AP-view CXRs of children collected from Guangzhou Medical Center, including normal, bacterial pneumonia, and viral pneumonia cases.
    \item NIH CXR-14 dataset \cite{wang2017chestx}: Curated by the National Institute of Health (NIH), this large dataset has more than 100,000 images of normal chests and different abnormal lungs, including pneumonia and consolidation.
    \item Radiopaedia: Other than infected cases, healthy CXRs taken for the purpose of medical check-ups are also available in Radiopaedia.
    \item Tuberculosis Chest X-ray Image Datasets \cite{jaeger2014two}: Provided by U.S. National Library of Medicine, it has two datasets containing 406 normal x-rays.
\end{enumerate}

Currently, \numOfImages{} images of COVID-19 patients are collected in different sizes and formats. All collected images are publicly accessible in the dedicated repository\footnote{COVID-19 Chest X-ray Image Data Collection. Available on: \href{https://github.com/armiro/COVID-CXNet}{https://github.com/armiro/COVID-CXNet}}. The dataset includes 5,000 normal CXRs as well as 4,600 images of patients with CAP collected from the NIH CXR-14 dataset. 




\subsection{Preprocessing and Enhancements}

\begin{figure}[H]
    \begin{subfigure}{0.24\linewidth}
        \centering
        \includegraphics[width=\linewidth]{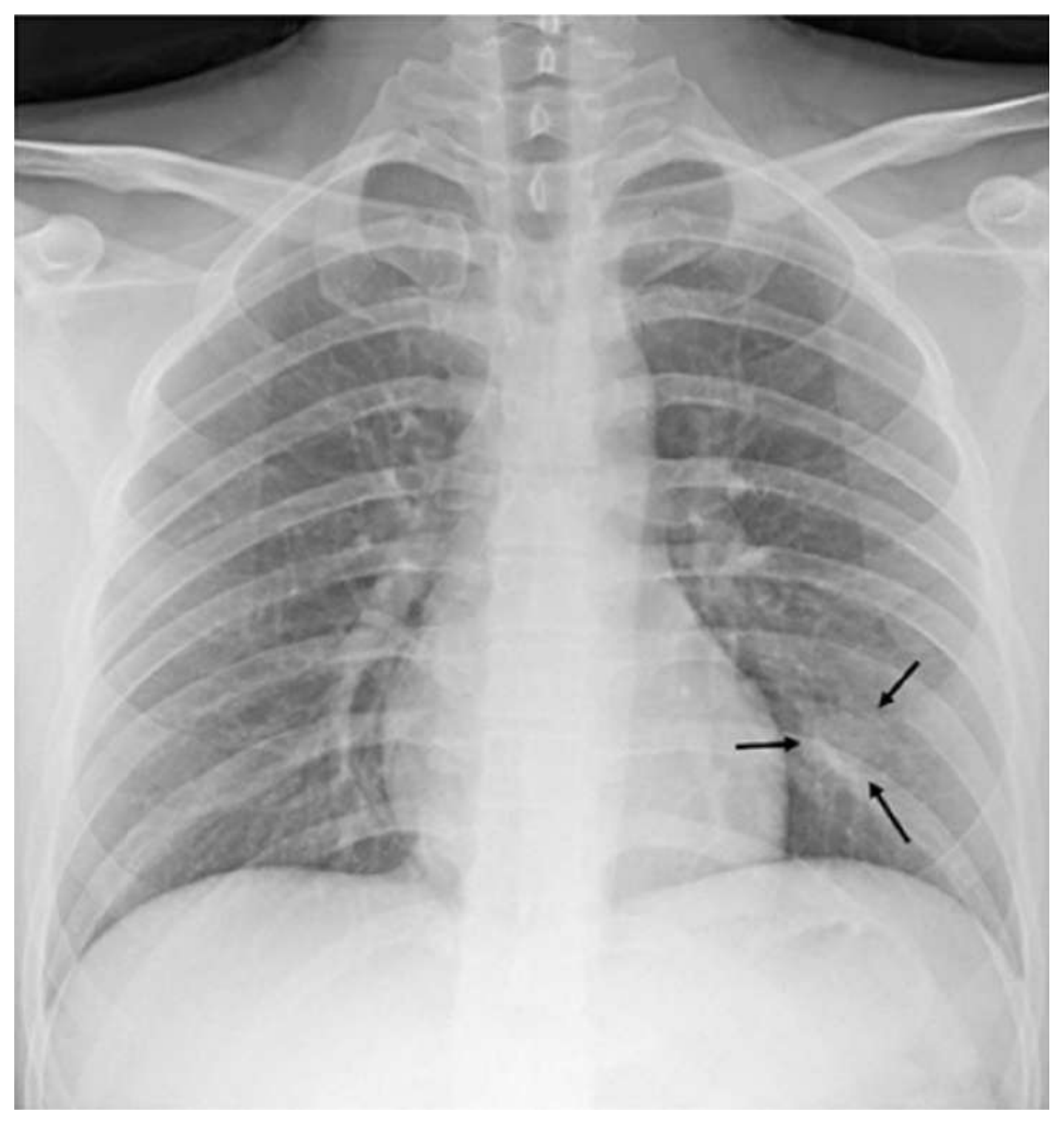}
        \caption{}
        \label{fig:img_original}
    \end{subfigure}
    \hfill
    \begin{subfigure}{0.24\linewidth}
        \centering
        \includegraphics[width=\linewidth]{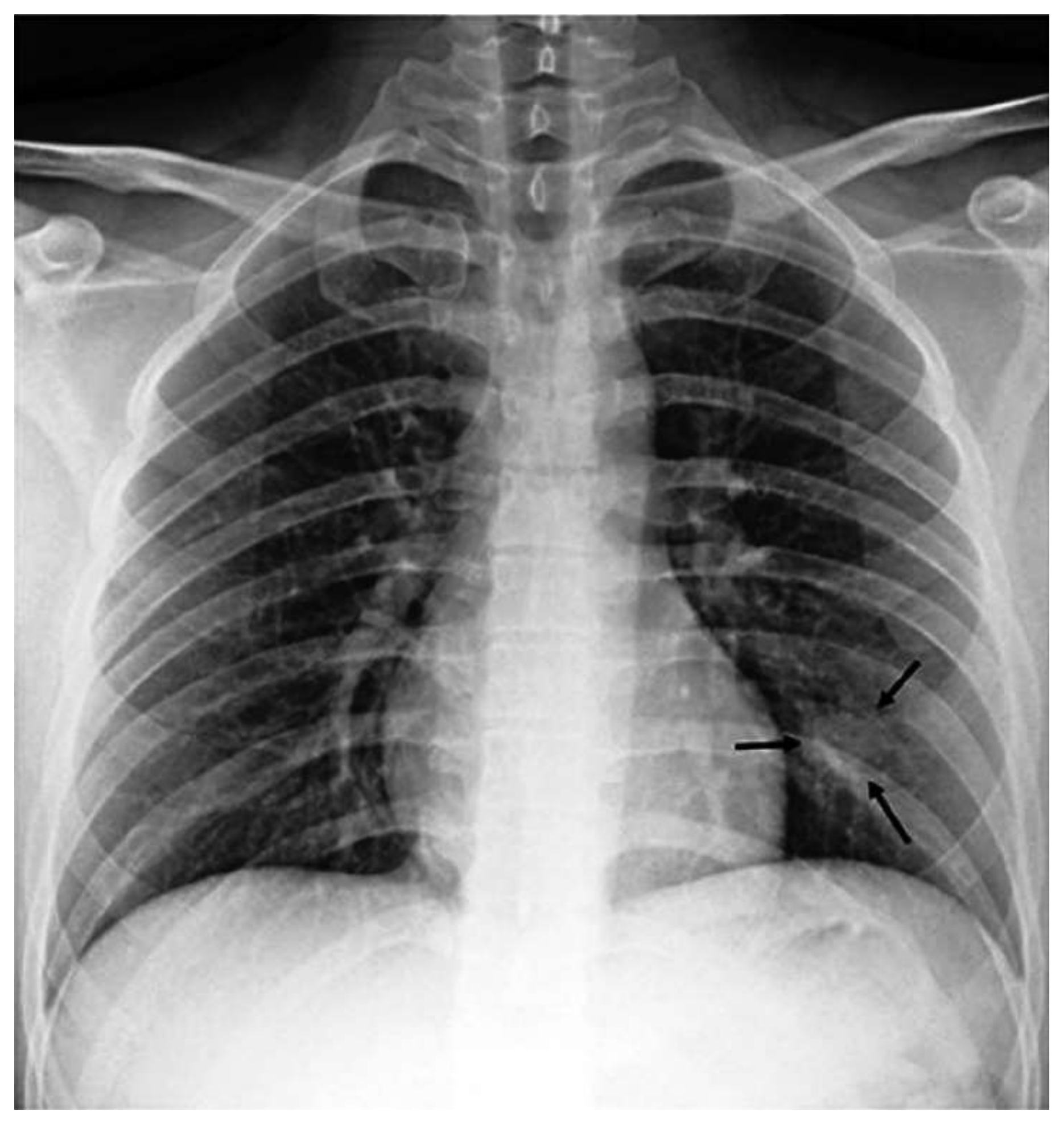}
        \caption{}
        \label{fig:img_he}
    \end{subfigure}
    \hfill
    \begin{subfigure}{0.24\linewidth}
        \centering
        \includegraphics[width=\linewidth]{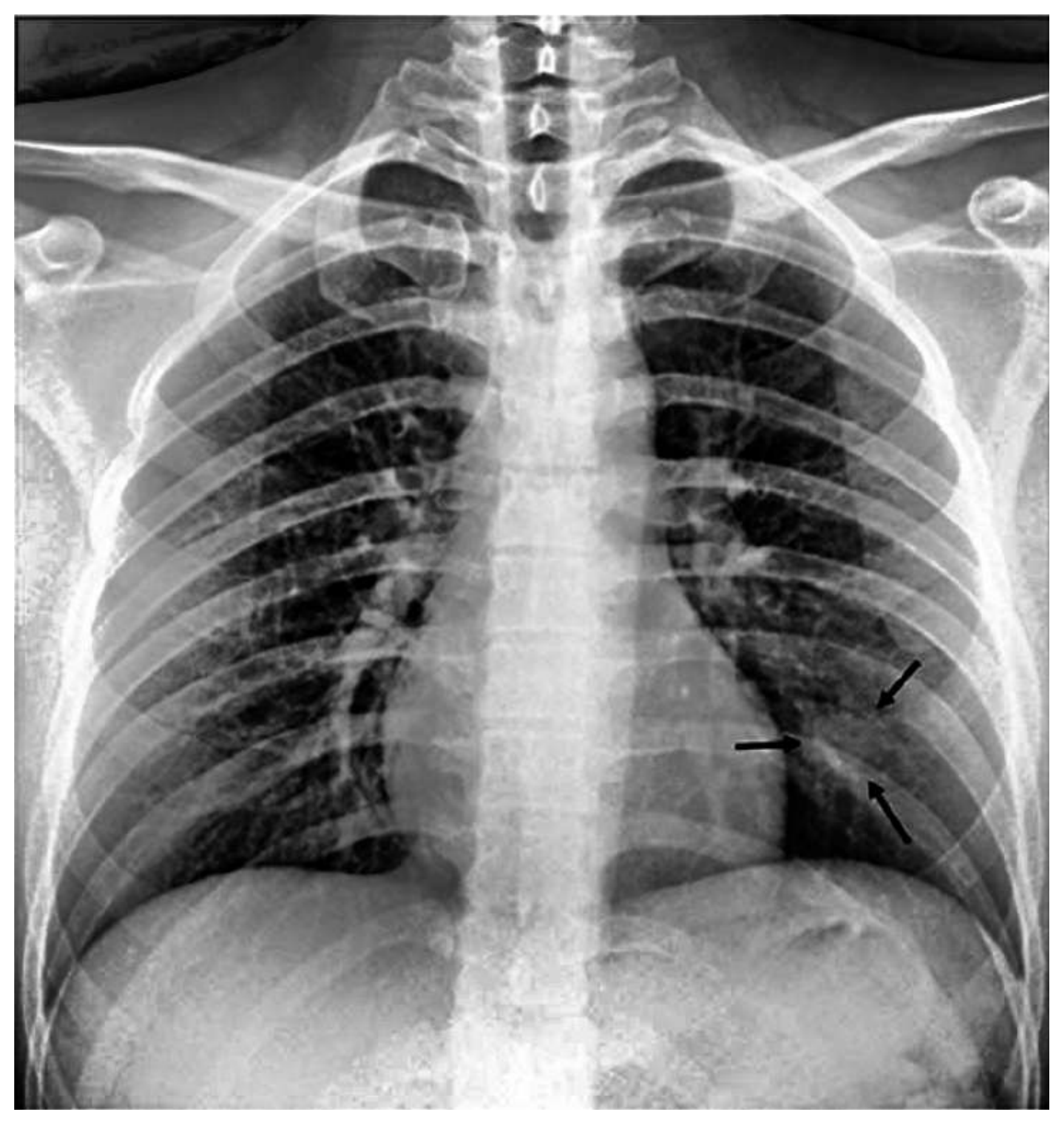}
        \caption{}
        \label{fig:img_ahe}
    \end{subfigure}
    \hfill
    \begin{subfigure}{0.24\linewidth}
        \centering
        \includegraphics[width=\linewidth]{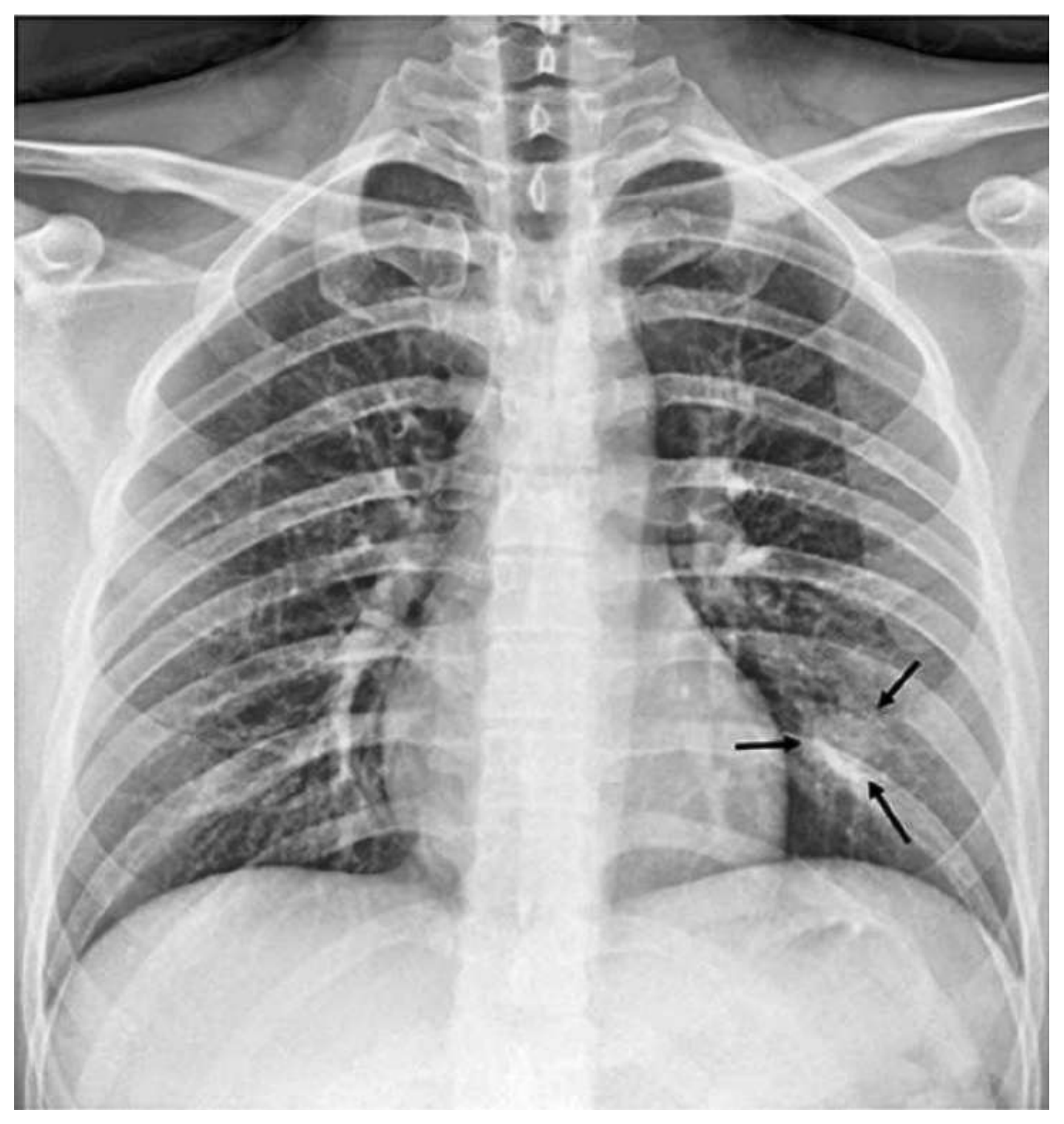}
        \caption{}
        \label{fig:img_clahe}
    \end{subfigure}

\caption{Different image enhancement methods. (a) is the main image, (b) is the image with histogram equalization (HE), (c) is adaptive histogram equalization (AHE) applied on the image, and (d) is the image with contrast limited AHE}
\label{fig:img_enhancements}
\end{figure}

Due to the small number of images in positive class, image augmentation is utilized to prevent overfitting. Images are also normalized and downsized to \((320, 320)\) to prevent resource exhaustion and decrease RAM usage. There are various image enhancement methods based on histogram equalization that increase image contrast to make non-linearities more distinguishable. Radiologists also use manual contrast improvement to diagnose mass and nodules better. An example of enhancement algorithms applied on a marker-annotated CXR is shown in Fig. \ref{fig:img_enhancements}.

\begin{figure}[H]
    \begin{subfigure}{0.3\linewidth}
        \centering
        \includegraphics[width=0.8\textwidth]{img_original.pdf}
        \caption{Main image}
        \label{fig:img_original2}
    \end{subfigure}
    \hfill
    \begin{subfigure}{0.3\linewidth}
        \centering
        \includegraphics[width=0.8\linewidth]{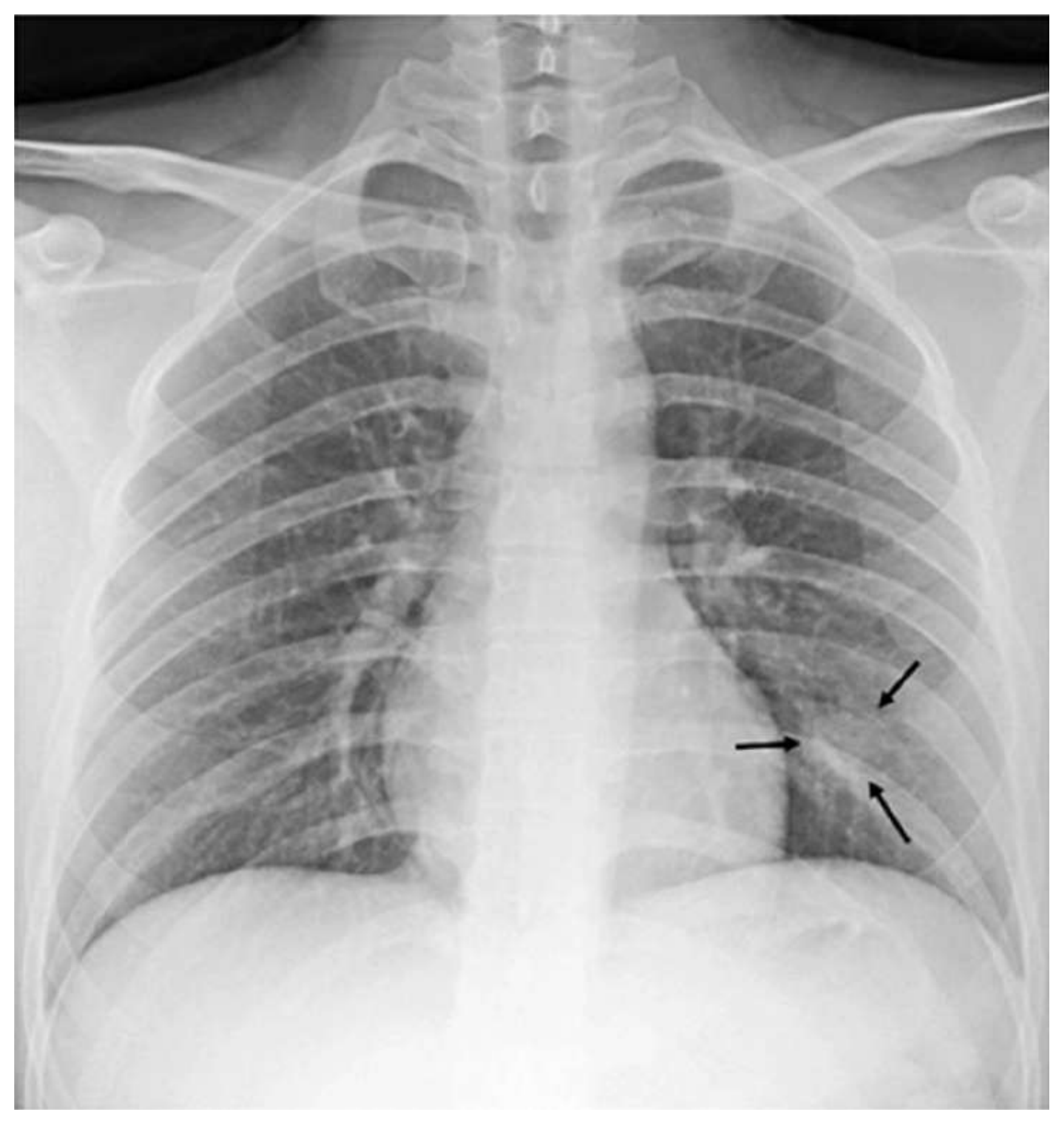}
        \caption{BEASF (\(\gamma=0.5\))}
        \label{fig:img_beasf1}
    \end{subfigure}
    \hfill
    \begin{subfigure}{0.3\linewidth}
        \centering
        \includegraphics[width=0.8\linewidth]{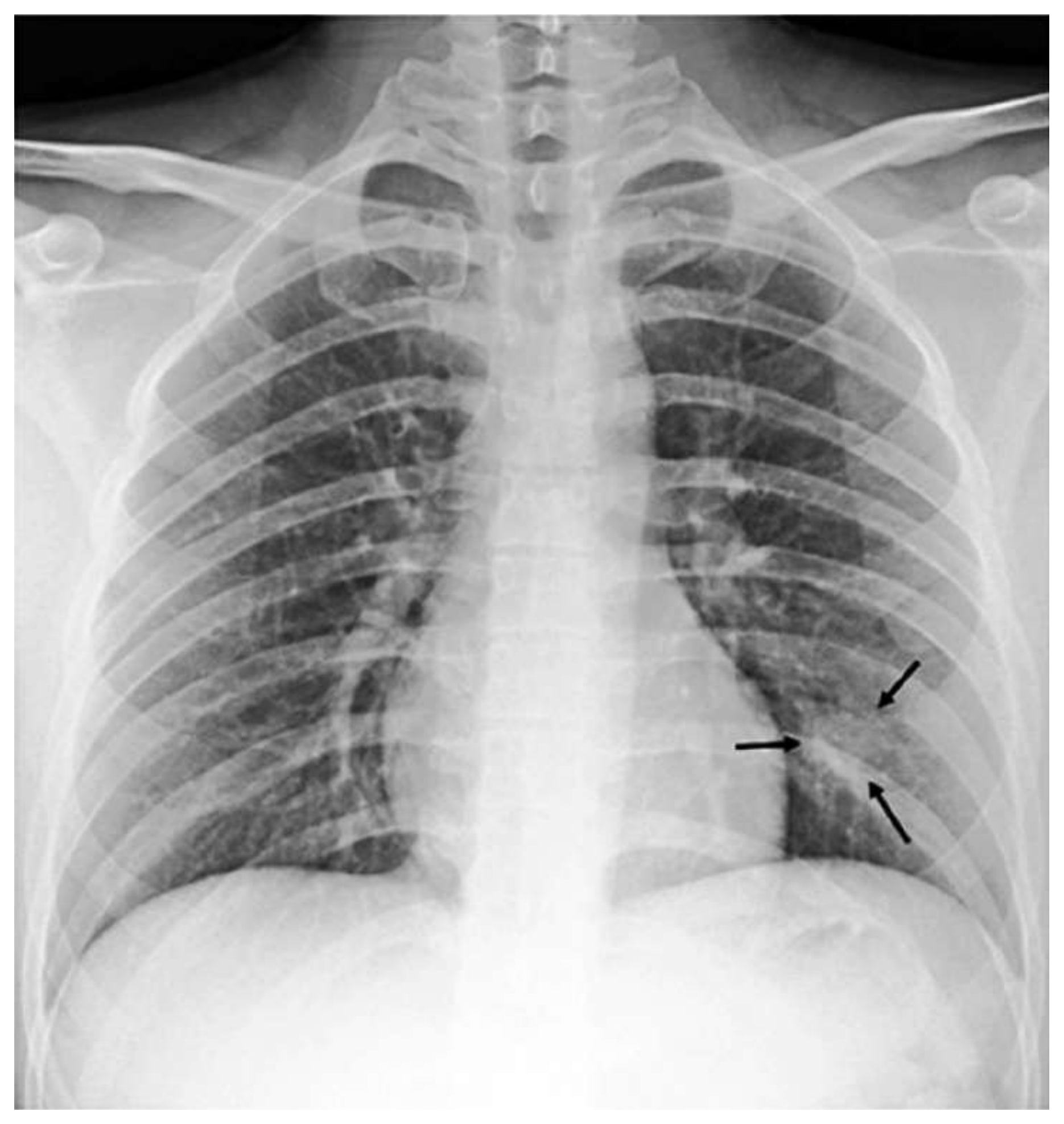}
        \caption{BEASF (\(\gamma=1.0\))}
        \label{fig:img_beasf2}
    \end{subfigure}
    \hfill
    \begin{subfigure}{0.3\linewidth}
        \centering
        \includegraphics[width=0.8\linewidth]{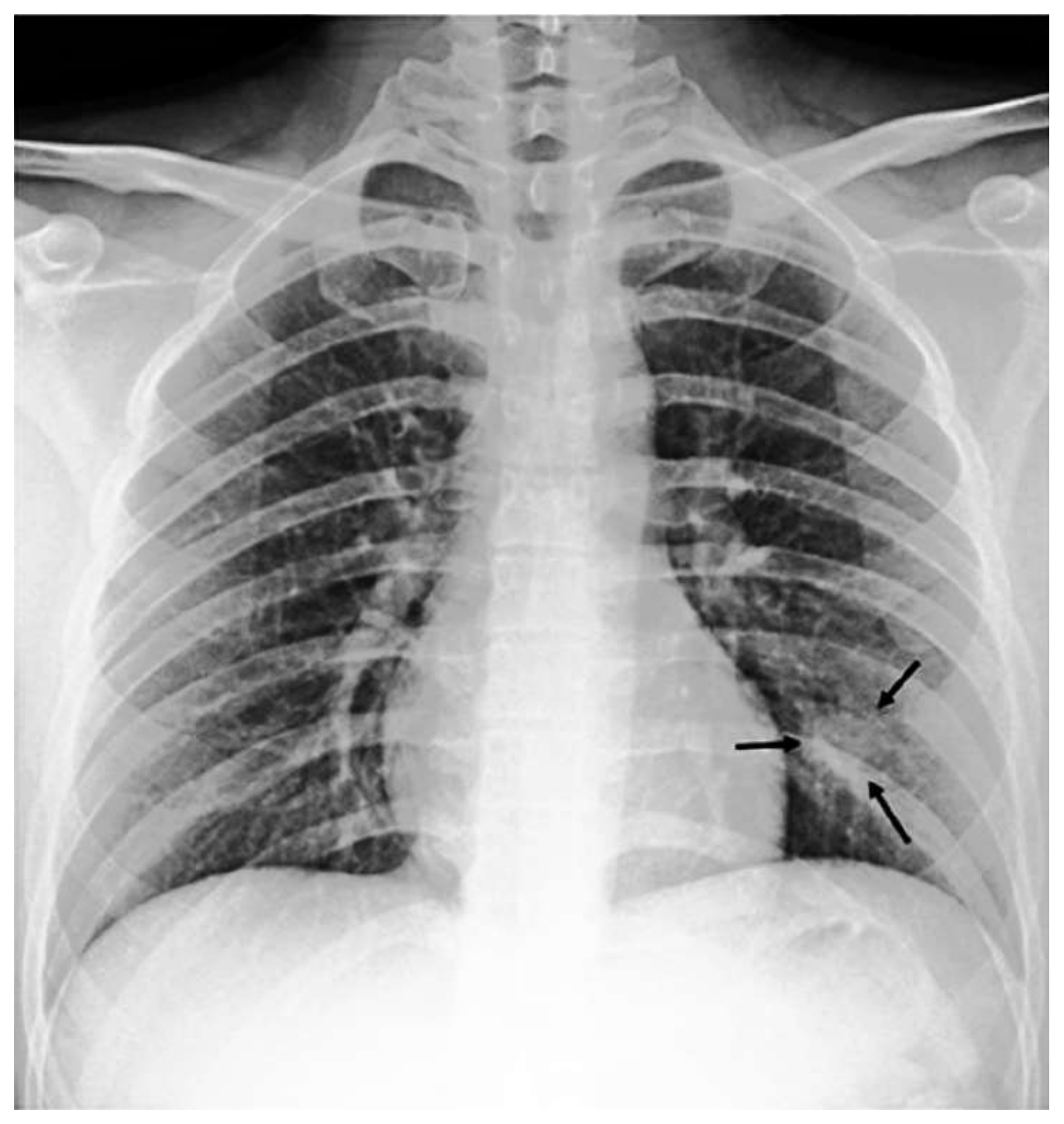}
        \caption{BEASF (\(\gamma=1.5\))}
        \label{fig:img_beasf3}
    \end{subfigure}
    \hfill
    \begin{subfigure}{0.3\linewidth}
        \centering
        \includegraphics[width=0.8\linewidth]{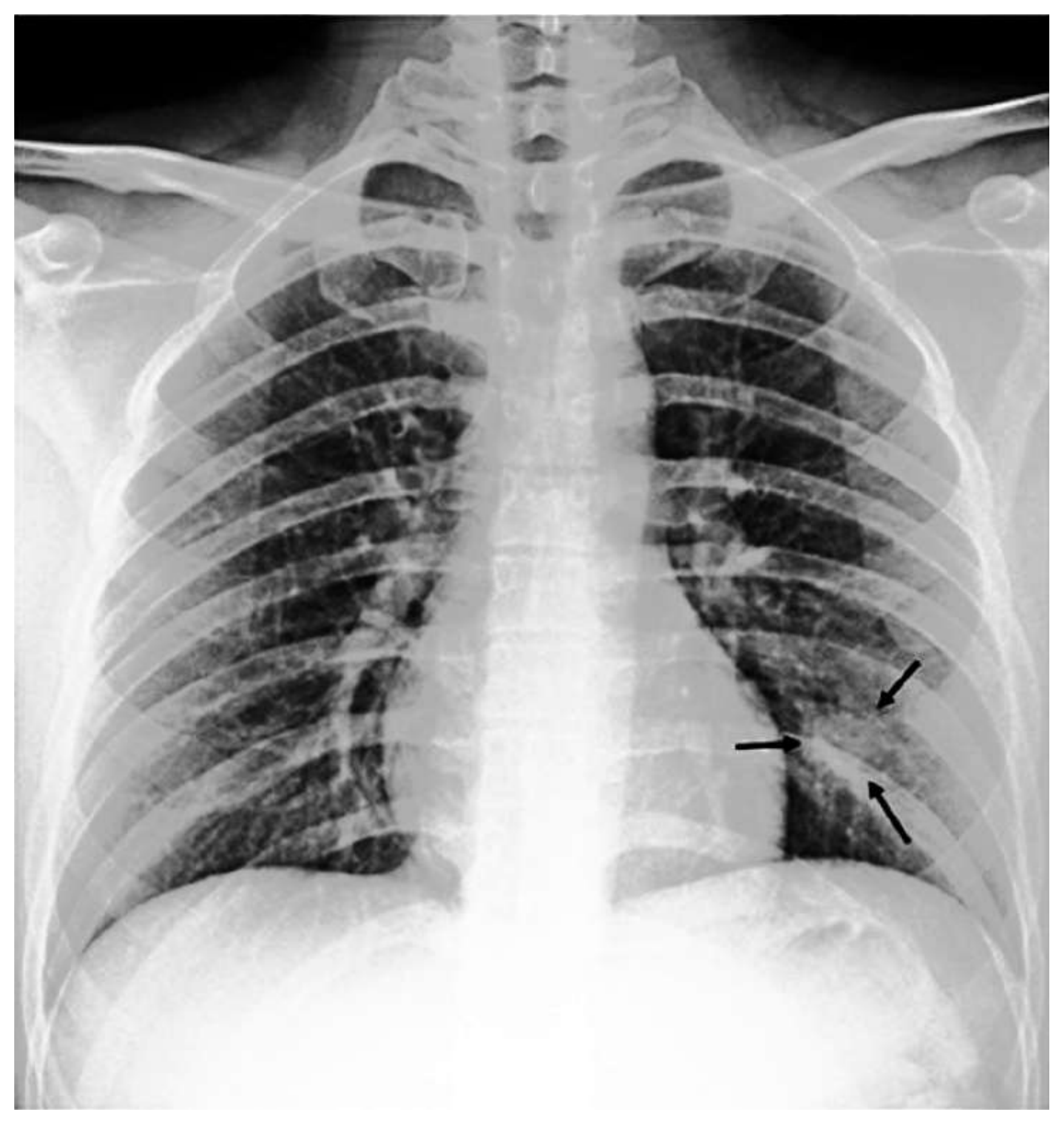}
        \caption{BEASF (\(\gamma=2.0\))}
        \label{fig:img_beasf4}
    \end{subfigure}
    \hfill
    \begin{subfigure}{0.3\linewidth}
        \centering
        \includegraphics[width=0.8\linewidth]{img_clahe.pdf}
        \caption{CLAHE}
        \label{fig:img_clahe2}
    \end{subfigure}
\caption{BEASF with different hyperparameter values compared with original image and CLAHE}
\label{fig:beasf_vs_clahe}
\end{figure}

\noindent As expected, Contrast Limited Adaptive Histogram Equalization (CLAHE) has better revealed nodular-shaped opacity related to COVID-19 pneumonia. CLAHE is one of the most popular enhancement methods in different image types \cite{pizer1987adaptive}. Another histogram equalization-based algorithm is Bi-histogram Equalization with Adaptive Sigmoid Function (BEASF) \cite{arriaga2014image}. BEASF adaptively improves image contrast based on the global mean value of the pixels. It has a hyperparameter \(\gamma\) to define the sigmoid function slope. Fig. \ref{fig:beasf_vs_clahe} depicts the output of BEASF with different \(\gamma\) values.


\noindent Although BEASF did not result in opacity detection improvement in all of the images, it could compliment the CLAHE method. Therefore, a BEASF-enhanced image with a \(\gamma=1.5\) is concatenated with CLAHE-enhanced and the main images to be fed into the model.

\begin{figure}[H]
    \centering
    \includegraphics[width=0.7\linewidth]{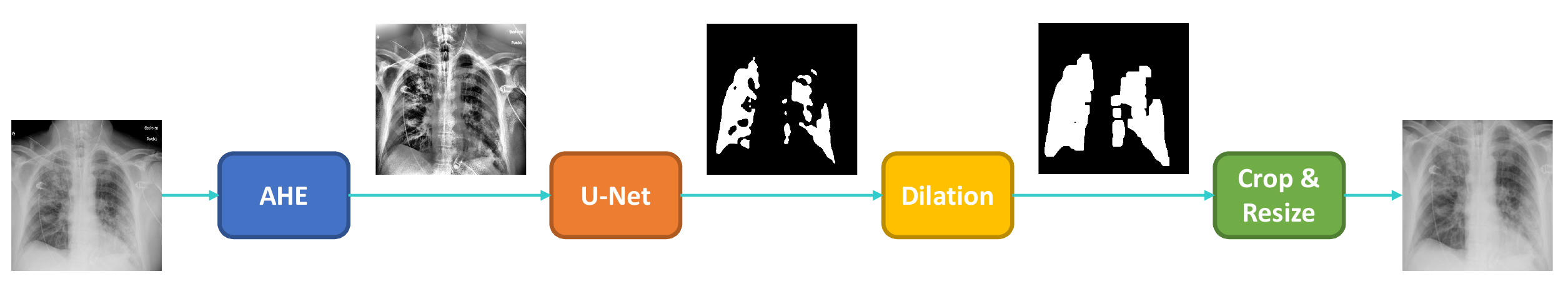}
    \caption{The segmentation approach based on the U-Net}
    \label{fig:segmentation}
\end{figure}

A U-Net based semantic segmentation \cite{ronneberger2015u} is also utilized to extract lung pixels from the body and the background. A collection of CXRs with manually segmented lung masks from Shenzhen Hospital Dataset \cite{stirenko2018chest} and Montgomery County Dataset \cite{candemir2013lung} are used for training. The diagram of the ROI extraction block is shown in Fig. \ref{fig:segmentation}.

\noindent Using model checkpoints, best weights are used to generate final masks. Afterwards, edge preservation is considered by applying dilation as well as adding margins to the segmented lung ROIs. Lung-segmented CXR is then used as the model input.

\section{Proposed Method}
In this section, model development is explained in different steps. At first, a base convolutional model is designed and trained on different portions of the dataset. Then, pretrained models based on the ImageNet dataset are discussed. Finally, a pretrained model on a similar image type is explained. 

\subsection{Base Model} \label{base_model}
\begin{figure}[H]
    \centering
    \includegraphics[width=\linewidth]{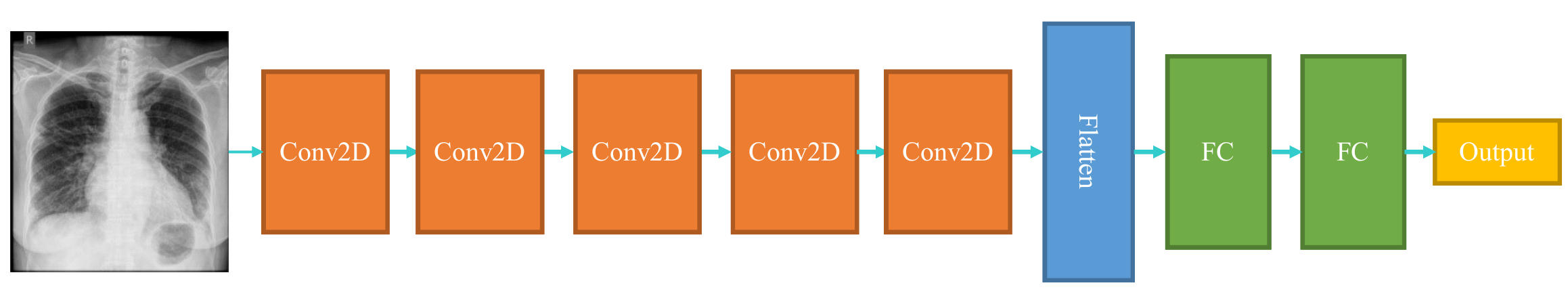}
    \caption{A high-level illustration of the base model}
    \label{fig:base_model}
\end{figure}

The base model consists of 5 convolutional layers, followed by a flatten layer and three fully connected layers. No batch normalization or pooling layers are used for this implementation stage. Fig. \ref{fig:base_model} illustrates the base model architecture.

\noindent Convolution layers have 32 filters, each of which has a kernel size of $3\times3$. The activation function is set as a rectified linear unit (ReLU), which adds non-linearity to images helping the model with better decision making. Fully connected layers have 10 neurons, and the last layer has one neuron, which demonstrates the probability of the input image belonging to the normal class (\(p=0.0\)) or pneumonia class (\(p=1.0\)). 

\subsection{Pretrained Models}
\begin{figure}[H]
    \centering
    \includegraphics[width=\linewidth]{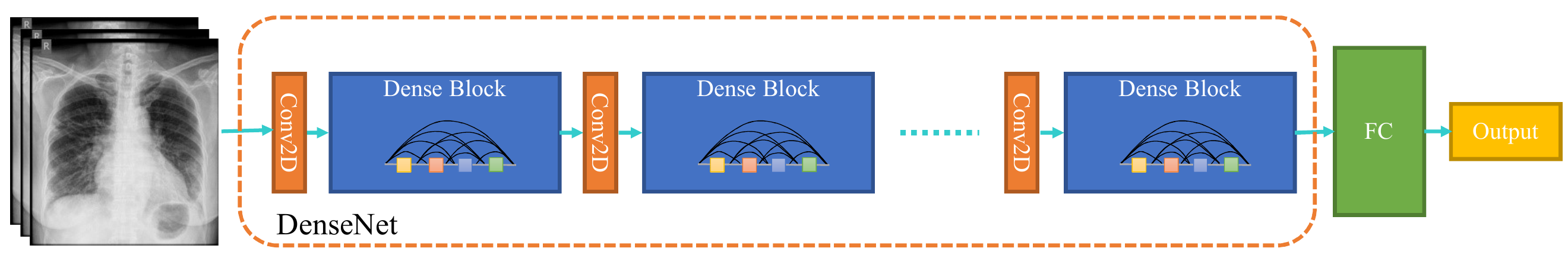}
    \caption{DenseNet-121-based model; Input shape is (224, 224, 3) and weights are from ImageNet}
    \label{fig:densenet}
\end{figure}

Transfer learning is to benefit from a pretrained model in a new classification task. Some pretrained models are trained on millions of images for many epochs and achieved high accuracy on a general task. We experimentally selected DenseNet and ResNet architectures pretrained on the ImageNet. A high-level illustration of the DenseNet-based model is shown in Fig. \ref{fig:densenet}.

\noindent It is worth mentioning that pretrained models are used for fine-tuning, training on target datasets for a small number of epochs, instead of retraining for many epochs. Since ImageNet images and labels are different from the CXR dataset, a pretrained model on the same data type should also be considered.

\begin{figure}[H]
    \centering
    \includegraphics[width=0.25\linewidth]{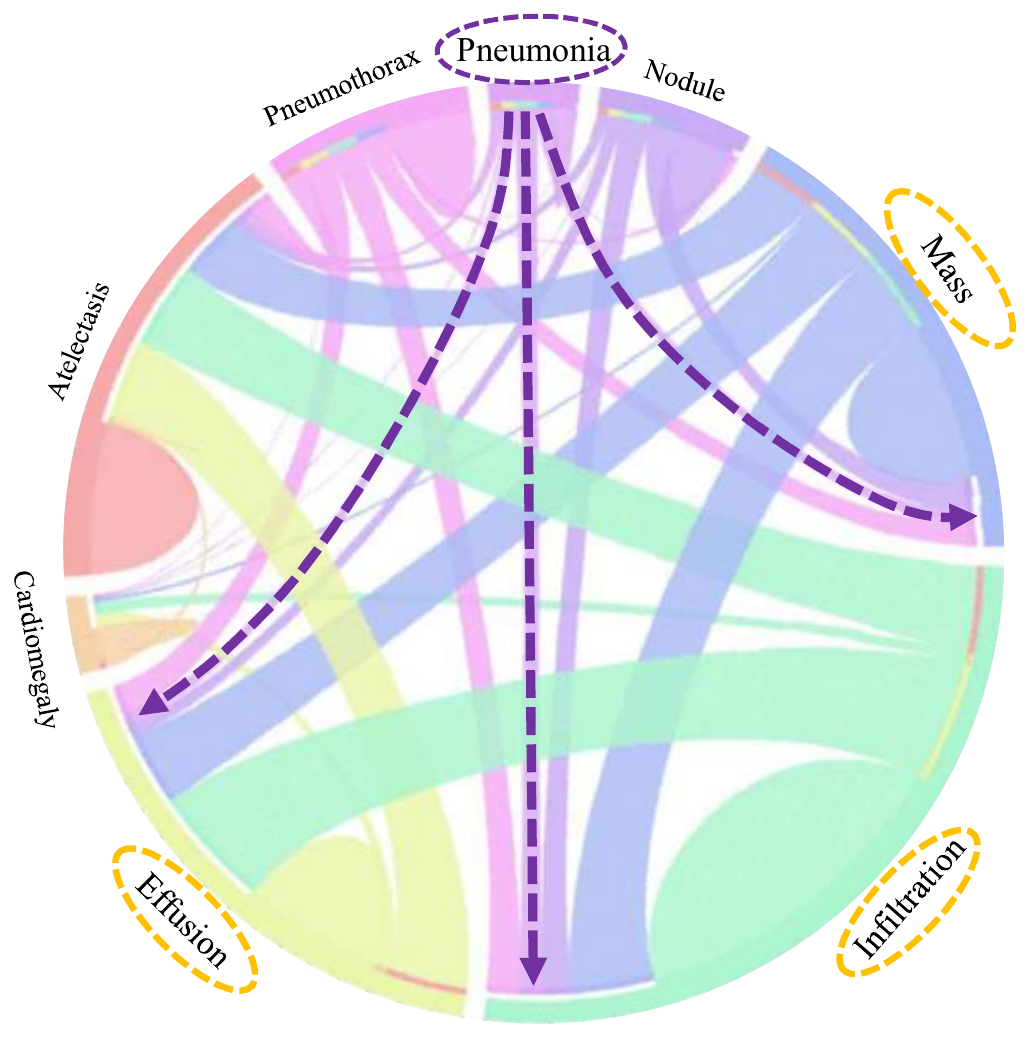}
    \caption{Co-occurrence of different CXR findings as a circular diagram by \cite{wang2017chestx}}
    \label{fig:thoracic_map}
\end{figure}

CheXNet is trained on CXR-14, a large publicly available CXR dataset with 14 different diseases such as pneumonia and edema \cite{rajpurkar2017chexnet}. CheXNet claims to have a radiologist-level diagnosis accuracy, has better performance than previous related research studies \cite{wang2017chestx}, \cite{yao2017learning}, and has simpler architecture than later approaches \cite{ranjan2018jointly}. CheXNet is based on DenseNet architecture and has been trained on frontal CXRs. It also could be used as a better option for the final model backbone. According to \cite{wang2017chestx}, pneumonia is correlated to other thoracic findings shown in Fig. \ref{fig:thoracic_map}.

\noindent Considering these correlations, we can use a combination of CheXNet output neurons to use as the classifier without any fine-tuning.

\subsection{The Proposed Model: COVID-CXNet}
\begin{figure}[H]
    \centering
    \includegraphics[width=\linewidth]{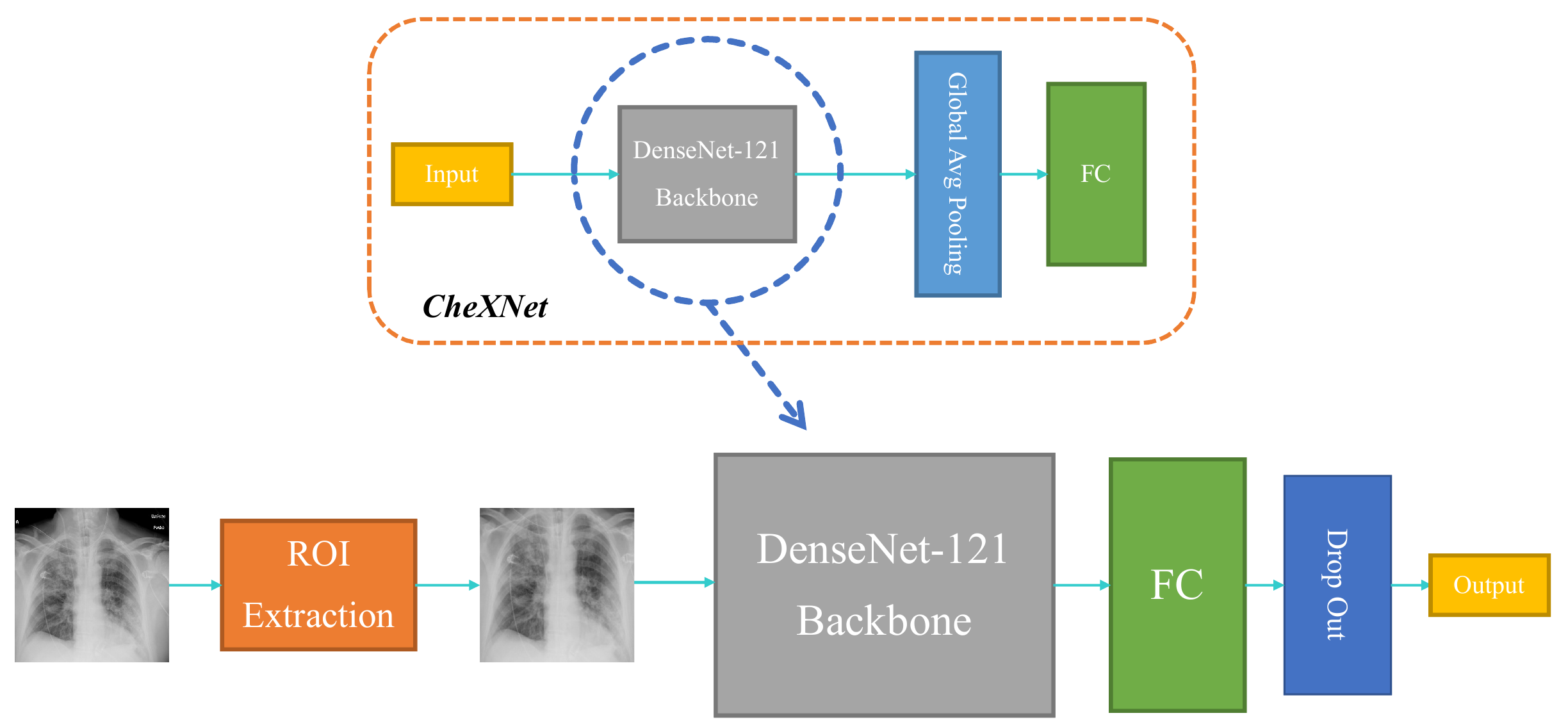}
    \caption{COVID-CXNet model architecture based on the DenseNet-121 feature extractor as the backbone}
    \label{fig:cxnet}
\end{figure}

The proposed COVID-CXNet is a CheXNet-based model, fine-tuned on the COVID-19 CXR dataset with 431 layers and $\approx$ 7M parameters. The architecture of the COVID-CXNet is presented in Fig. \ref{fig:cxnet}.

\noindent Our proposed network uses the DenseNet-121 as its backbone. There are different architectures utilized to build CheXNet, out of which the DenseNet has shown better capabilities for detecting pulmonary diseases \cite{rajpurkar2017chexnet}. COVID-CXNet has a FC layer consisting of 10 nodes followed by a dropout layer with a 0.2 dropping rate to prevent overfitting. The activation function of the last layer is changed from SoftMax to Sigmoid because the task is a binary classification. Our proposed model's advantage over the base model or other custom networks is the training speed as we are fine-tuning a backbone with pretrained weights. In comparison with other CheXNet-based models, COVID-CXNet is benefiting from a lung segmentation module and different image enhancements algorithms within the preprocessing section. Moreover, fine-tuning on a larger dataset along with several overfitting-prevention methods such as dropout layer and label smoothing will help our model outperform in terms of correctly localizing pneumonia in CXRs.

\section{Experimental Results} \label{results}
To evaluate the performance, several metrics are considered. Accuracy score is the primary metric used for statistical classification, which is required but inadequate here as we are more interested in efficiently classifying positive samples. Thus, f1-score for the positive class is also measured. The main metric here is visualization heatmaps because the small number of positive samples make the model prone to overfitting by deciding based on the wrong features. The most common manifestations of COVID-19 pneumonia in CXRs are air-space opacities in different forms, such as GGOs or consolidations. Opacities are identified as opaque regions (whiter than usual) in CXRs. They are mostly distinguished as bilateral, involving both lungs, and multifocal opacifications. Rare findings happen in the late stages of disease course, which may include pleural effusion and pneumothorax \cite{jacobi2020portable}.



\subsection{Base Model} \label{base_model_res}
\begin{figure}[H]
    \begin{subfigure}{0.5\linewidth}
        \centering
        \includegraphics[width=\linewidth]{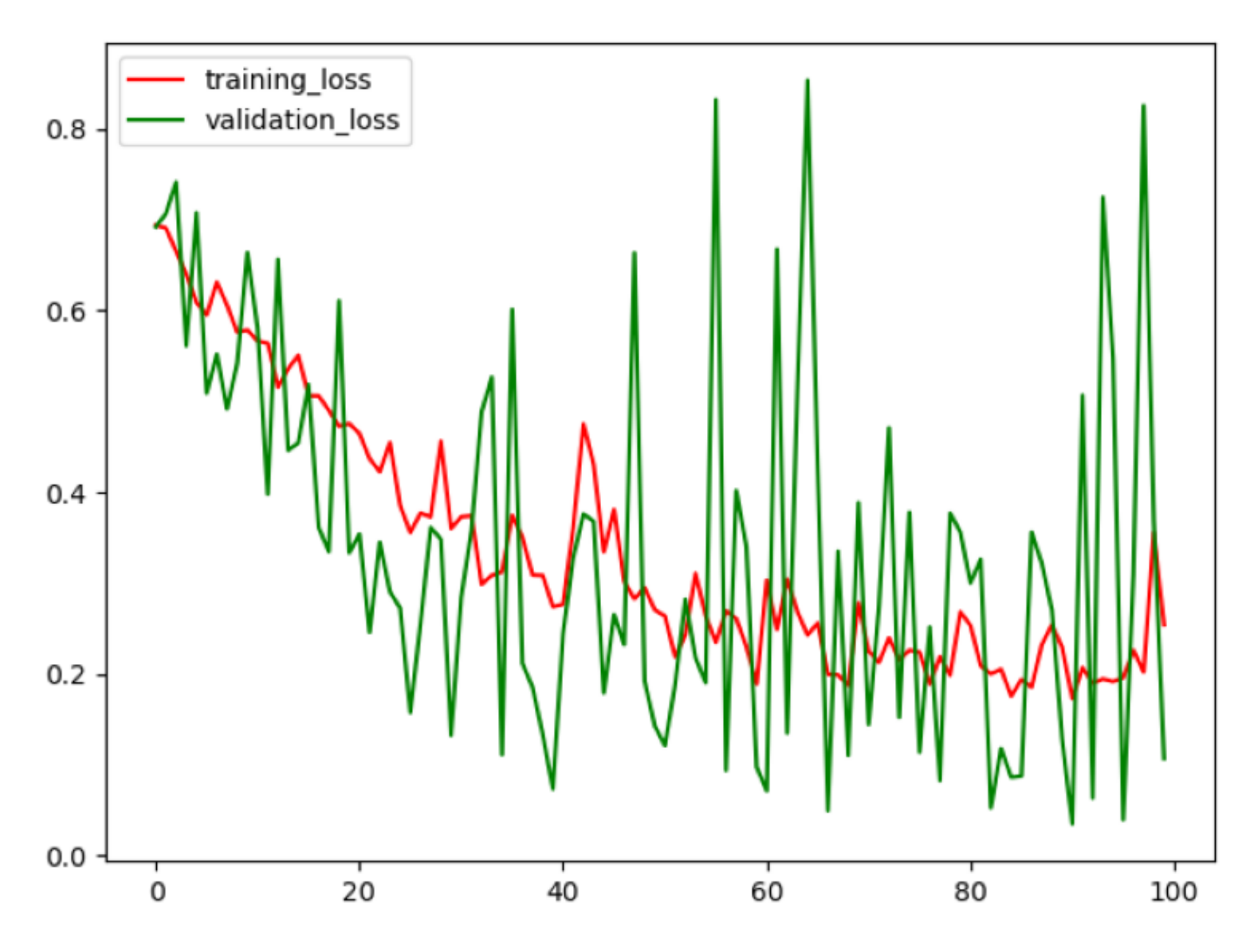}
        \caption{}
        \label{fig:v1.1}
    \end{subfigure}
    \hfill
    \begin{subfigure}{0.5\linewidth}
        \centering
        \includegraphics[width=\linewidth]{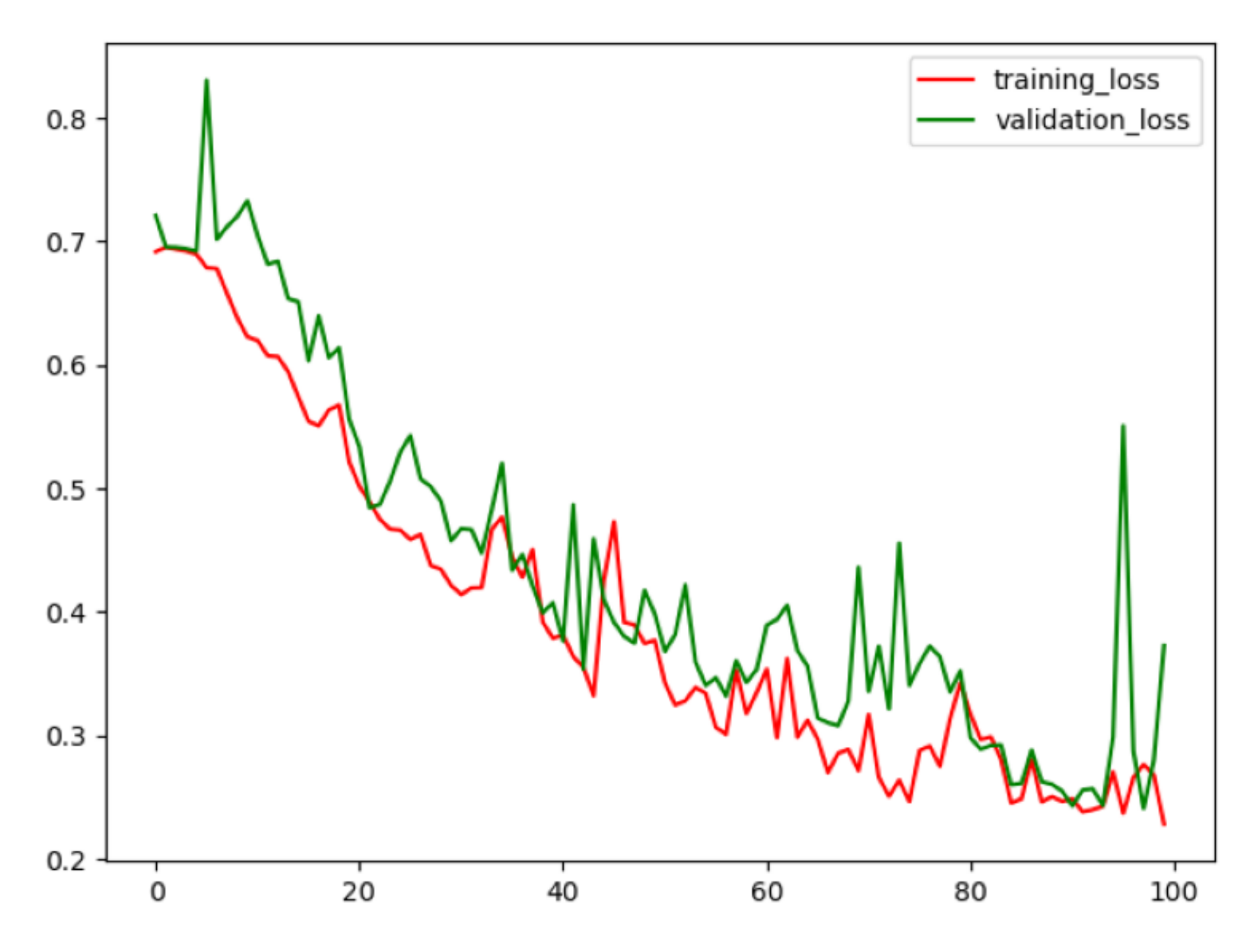}
        \caption{}
        \label{fig:v1.2}
    \end{subfigure}
\caption{Learning Curves for base model trained on (a) 450 images, and (b) 600 images}
\label{fig:training}
\end{figure}

In order to train the base model, the optimizer is set to “adam” with the optimal learning rate obtained using exponentially learning rate increasing method \cite{smith2017cyclical}. The best learning rate is where the highest decreasing slope happens ($\approx$ 0.0001). Training the base model using 300 samples resulted in an accuracy of 96.72\% on the test-set within 100 epochs. As the dataset extended, fluctuations in the curves damped gradually. With a training-set of 480 images, it reaches a reasonable accuracy on the test-set of 120 images. Model loss curves on the training-set and the validation-set (which is test-set here) on different dataset sizes are plotted in Fig. \ref{fig:training}.

\begin{figure}[H]
    \begin{subfigure}{0.33\linewidth}
        \centering
        \includegraphics[width=0.9\linewidth]{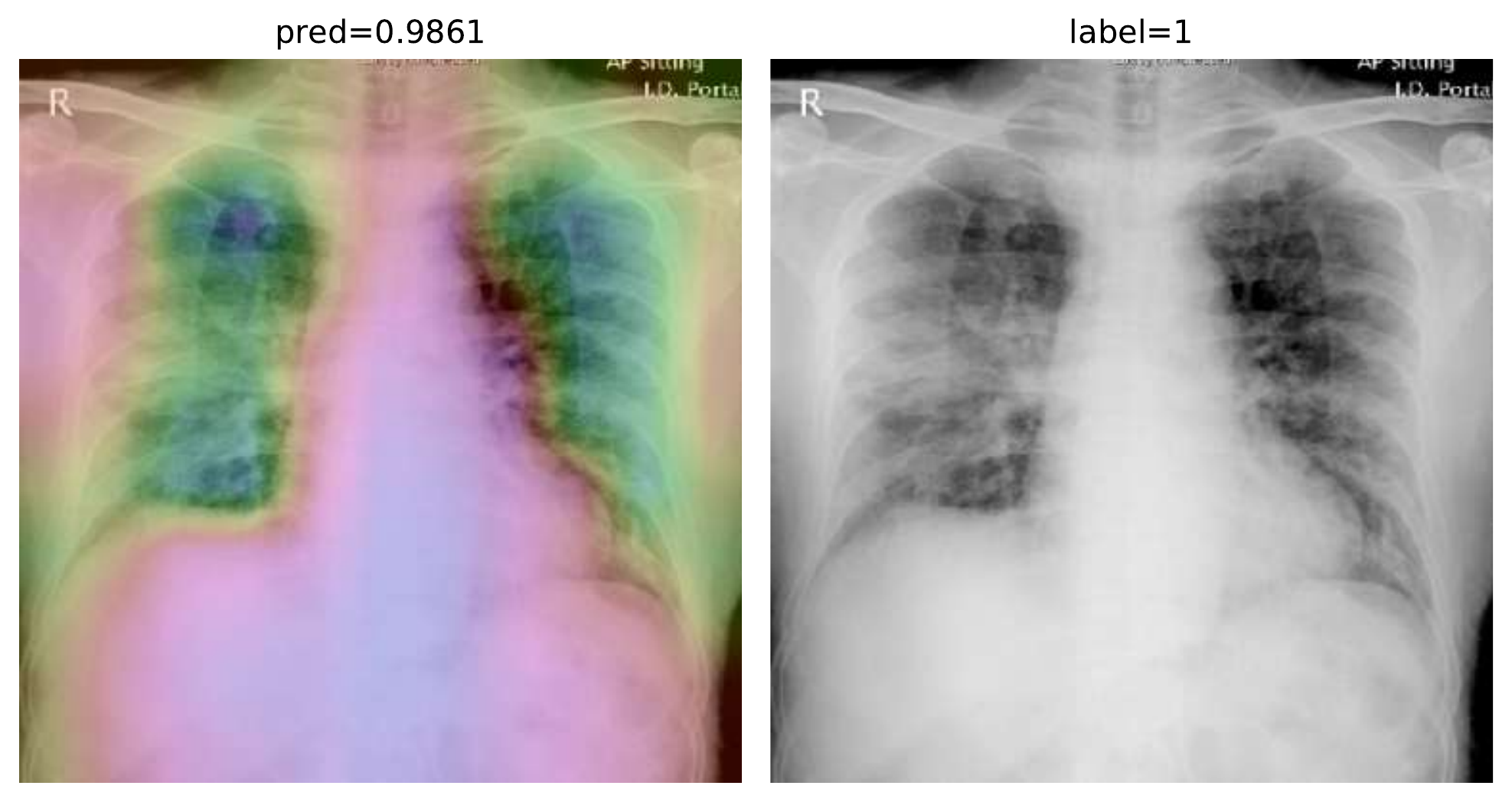}
        \label{fig:base_vis1}
    \end{subfigure}
    \begin{subfigure}{0.33\linewidth}
        \centering
        \includegraphics[width=0.9\linewidth]{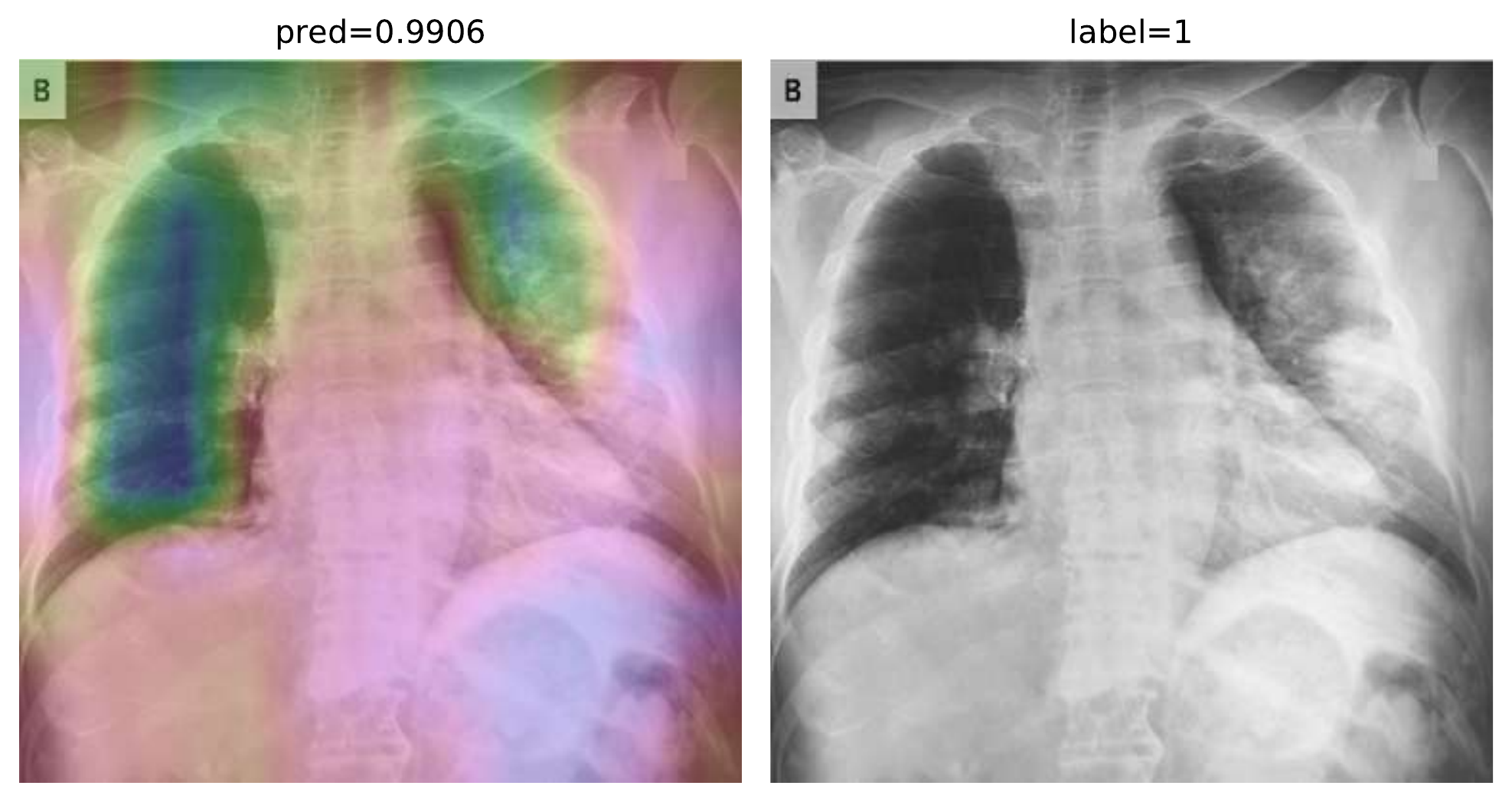}
        \label{fig:base_vis2}
    \end{subfigure}
    \begin{subfigure}{0.33\linewidth}
        \centering
        \includegraphics[width=0.9\linewidth]{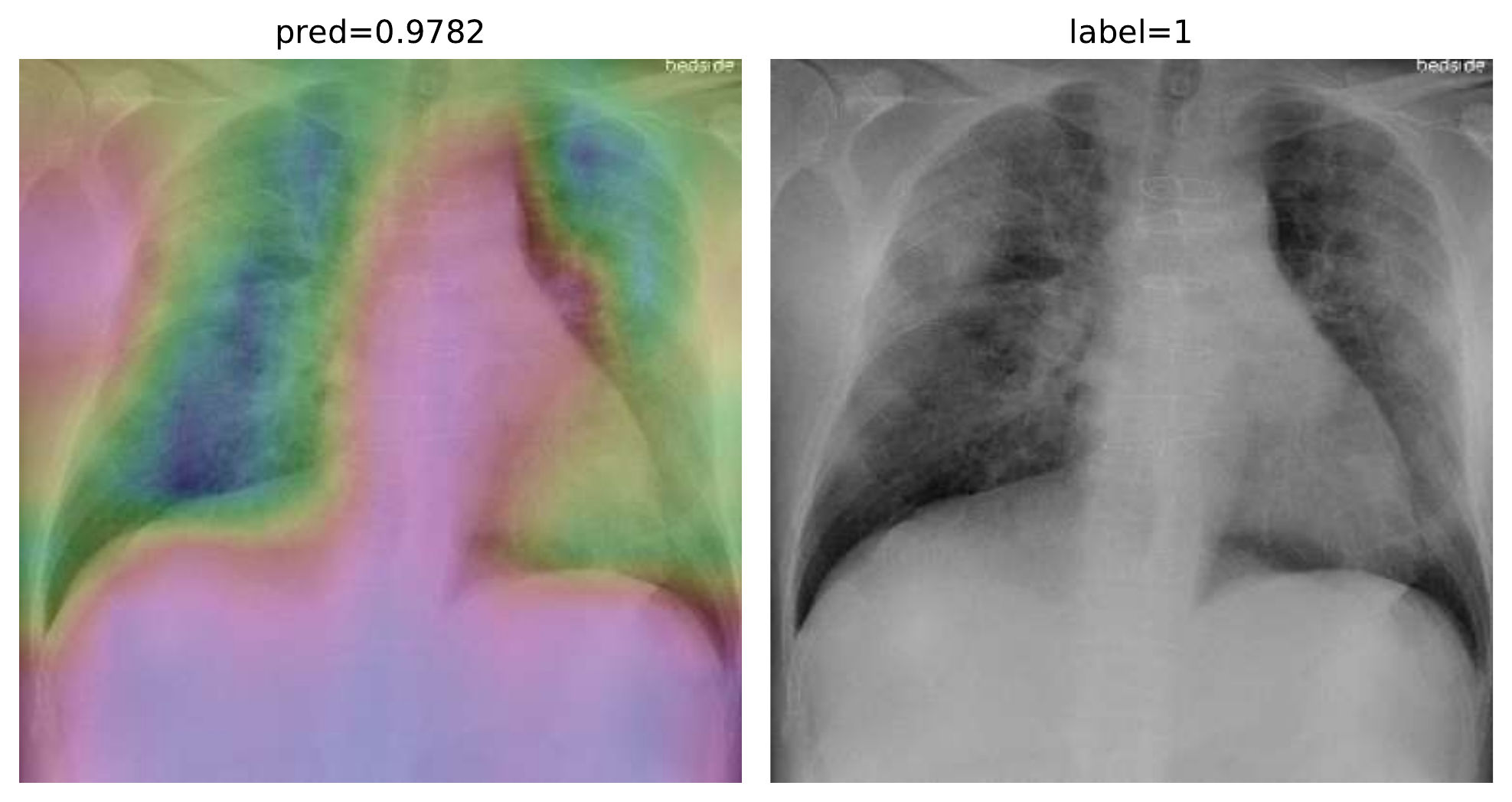}
        \label{fig:base_vis3}
    \end{subfigure}
    \begin{subfigure}{0.33\linewidth}
        \centering
        \includegraphics[width=0.9\linewidth]{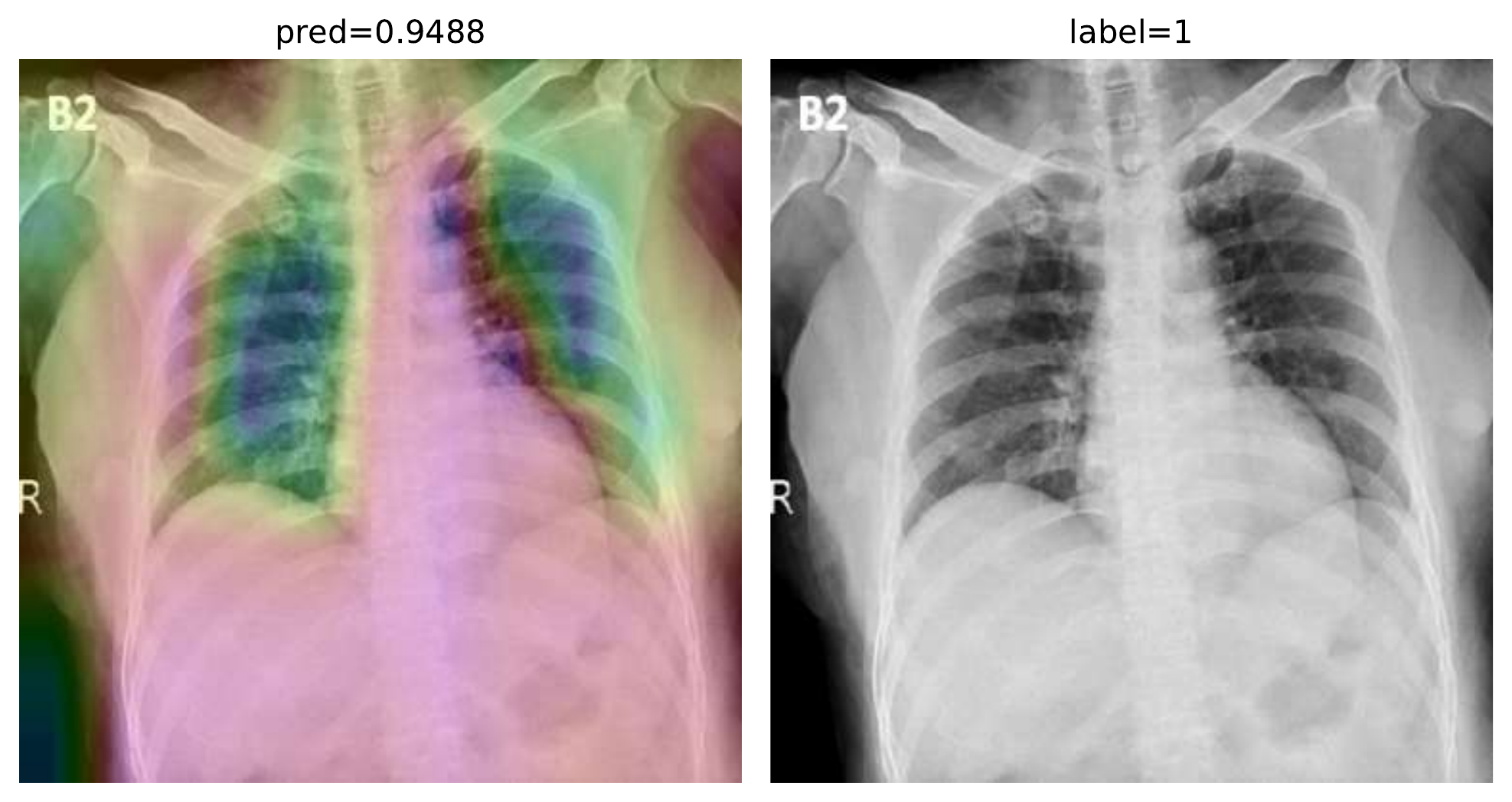}
        \label{fig:base_vis4}
    \end{subfigure}
    \begin{subfigure}{0.33\linewidth}
        \centering
        \includegraphics[width=0.9\linewidth]{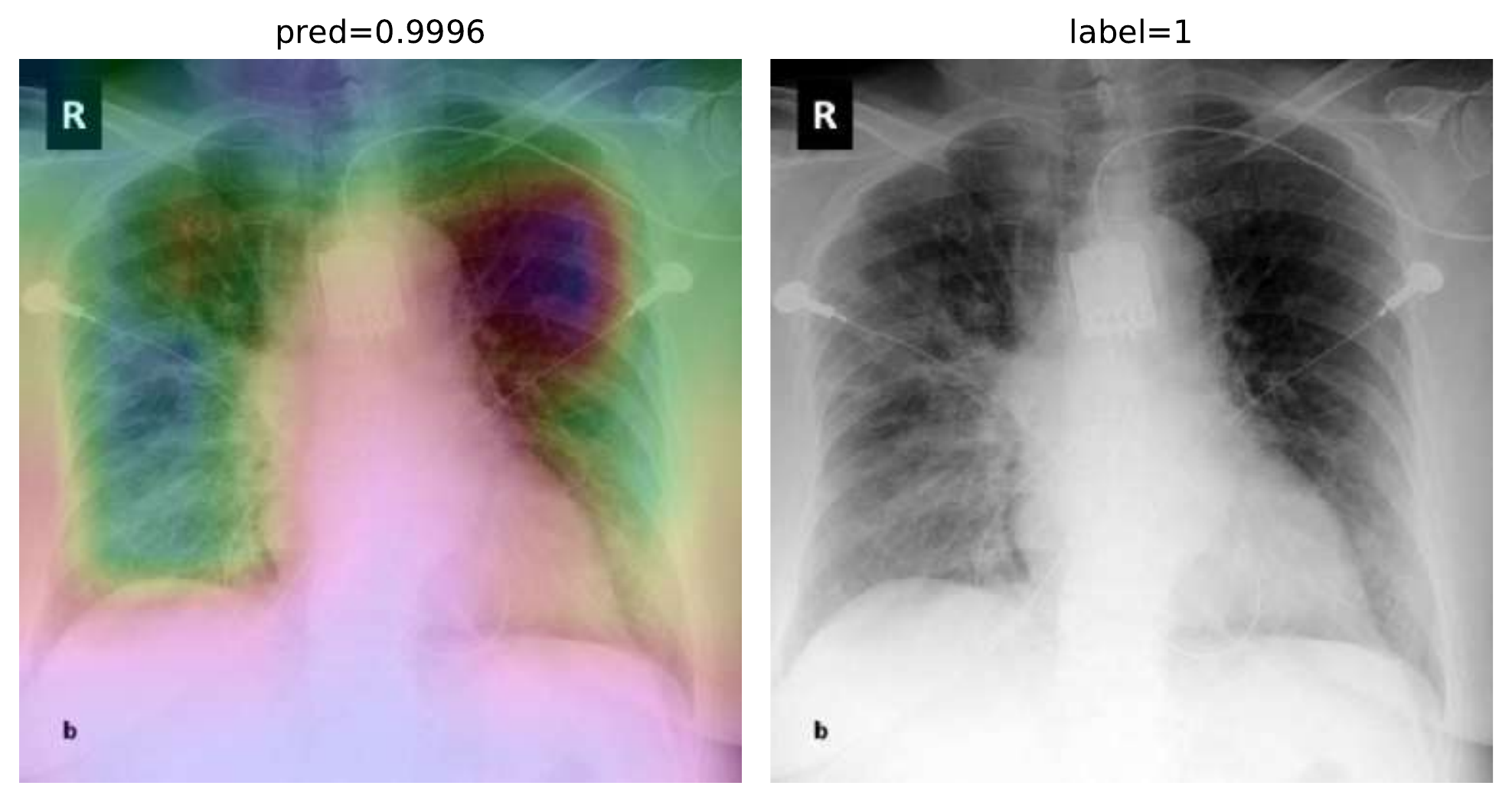}
        \label{fig:base_vis5}
    \end{subfigure}
    \begin{subfigure}{0.33\linewidth}
        \centering
        \includegraphics[width=0.9\linewidth]{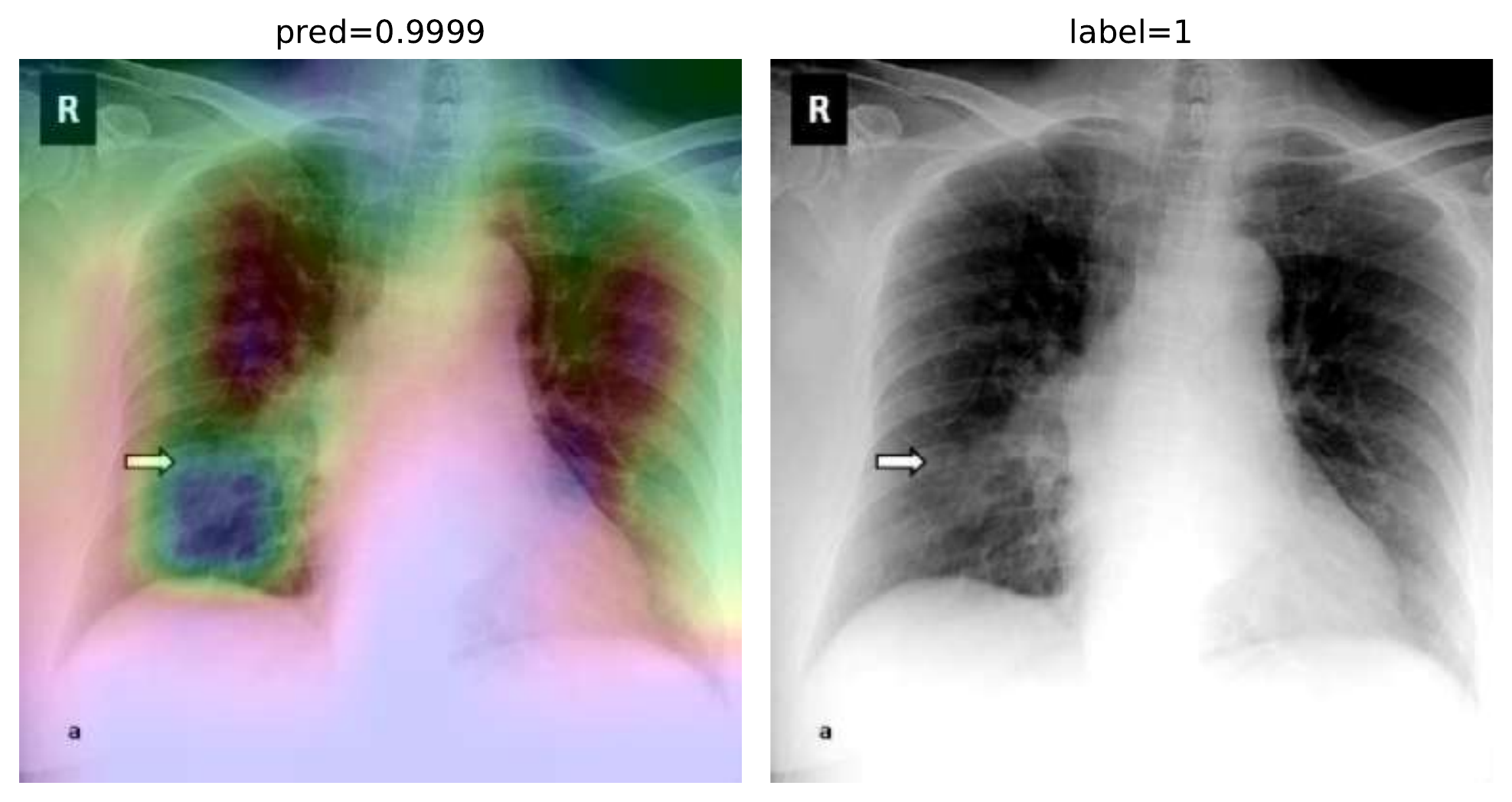}
        \label{fig:base_vis6}
    \end{subfigure}
\caption{Grad-CAM heatmaps of the base model for 6 images of positive class. Important regions are wrong in most images, while classification scores are notably high}
\label{fig:base_heatmaps}
\end{figure}

The base model hit accuracy of 96.10\%, relatively high compared to the number of images in the training-set, and complexity of pneumonia identification in CXRs. To validate the performance, CNN architecture is demystified. A popular technique for CNN visual explanation is Gradient-weighted Class Activation Mapping (Grad-CAM). Grad-CAM concept is to use the gradients of any target concept, flowing into the last convolutional layer to produce a coarse localization map with a concentration on important regions in the image for predicting the class \cite{selvaraju2017grad}. Another visualization is Local Interpretable Model-Agnostic Explanations (LIME). LIME performs local interpretability on input images, training local surrogate models on image components to find regions with the highest impact on the decision \cite{ribeiro2016should}. Grad-CAM of the base model for images of both normal and infected classes is illustrated in Fig. \ref{fig:base_heatmaps}. 

Although classification is successfully implemented with high accuracy score, extracted imaging features are wrong. One possible reason could be the fact that normal CXRs are mostly for pediatrics. To go further about the problem of the model and to prove whether it is because of the normal CXR dataset, the model was evaluated on a small external dataset of 60 images. The confusion matrix in Table \ref{base_model_table} shows that the model is not consistent regarding normal cases.

\begin{table}[H]
\renewcommand\thetable{1}
\caption{Confusion matrix of the base model on external test-set}
\label{base_model_table}
\centering
\begin{tabular}{|c|c|c|c|}
\hline
\multicolumn{2}{|c|}{\multirow{2}{*}{\textbf{Base Model}}} & \multicolumn{2}{c|}{\textbf{Predicted}} \\ \cline{3-4} 
\multicolumn{2}{|c|}{}                                     & \textit{Normal}   & \textit{COVID-19}   \\ \hline
\multirow{2}{*}{\textbf{Actual}}    & \textit{Normal}      & 21                & 9                   \\ \cline{2-4} 
                                    & \textit{COVID-19}    & 3                 & 27                  \\ \hline
\end{tabular}
\end{table}

\begin{figure}[H]
    \begin{subfigure}{0.30\linewidth}
        \centering
        \includegraphics[width=\linewidth]{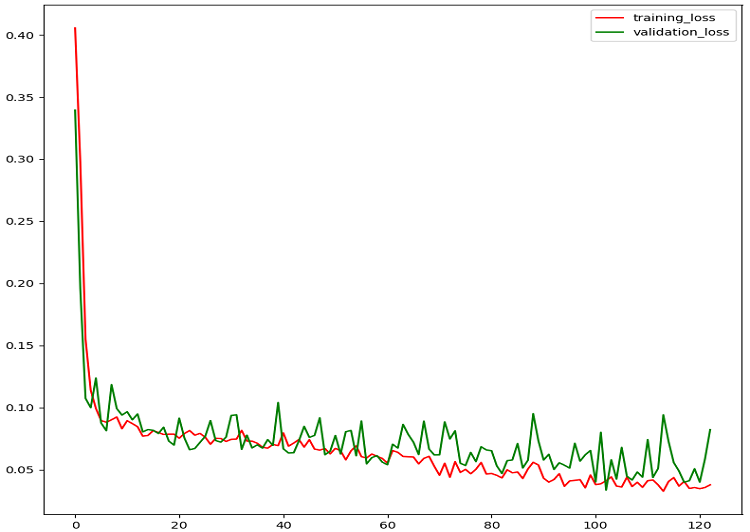}
        \caption{}
        \label{fig:v2.0_loss}
    \end{subfigure}
    \hfill
    \begin{subfigure}{0.30\linewidth}
        \centering
        \includegraphics[width=\linewidth]{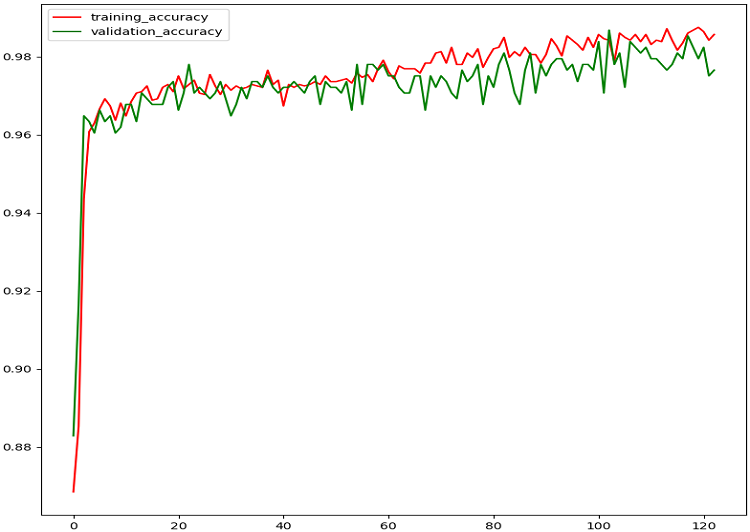}
        \caption{}
        \label{fig:v2.0_acc}
    \end{subfigure}
    \hfill
    \begin{subfigure}{0.30\linewidth}
        \centering
        \includegraphics[width=\linewidth]{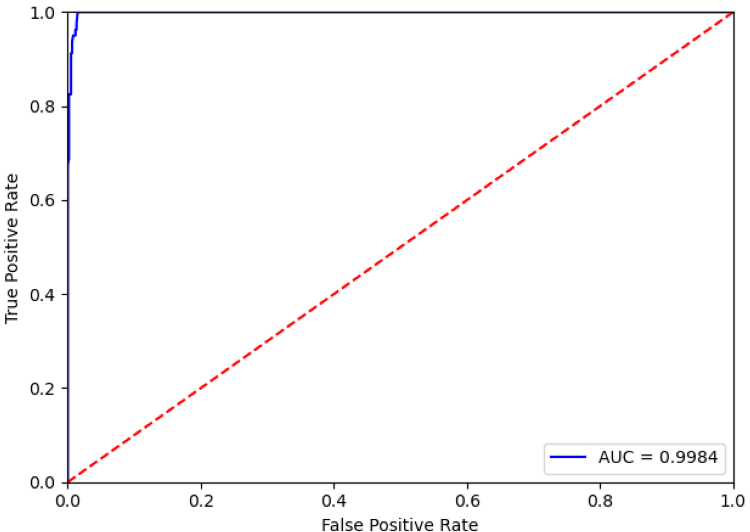}
        \caption{}
        \label{fig:v2.0_roc}
    \end{subfigure}
\caption{Results of the base model trained on 3,400 images; (a) training loss changes, (b) training accuracy score changes, and (c) receiver operating characteristic (ROC) plot}
\label{fig:v2.0_training}
\end{figure}

According to the results, recollecting CXRs from adult lungs is essential. The largest dataset containing normal cases is the NIH CXR-14 dataset, with almost 17,000 images. Then, we increased the number of normal CXRs in the dataset to prevent overfitting. The model is trained on a dataset of 3,400 images, 3,000 from normal and 400 from COVID-19 pneumonia classes. It is worth mentioning that classes are weighted in loss function calculation to deal with class imbalance. The results are presented in Fig. \ref{fig:v2.0_training} and Table \ref{base_model_table_683}.

\begin{table}[H]
\caption{Confusion matrix of the base model on the 683 samples}
\label{base_model_table_683}
\centering
\begin{tabular}{|c|c|c|c|}
\hline
\multicolumn{2}{|c|}{\multirow{2}{*}{\textbf{Base Model}}} & \multicolumn{2}{c|}{\textbf{Predicted}} \\ \cline{3-4} 
\multicolumn{2}{|c|}{}                                     & \textit{Normal}   & \textit{COVID-19}   \\ \hline
\multirow{2}{*}{\textbf{Actual}}    & \textit{Normal}      & 598                & 5                   \\ \cline{2-4} 
                                    & \textit{COVID-19}    & 4                 & 76                  \\ \hline
\end{tabular}
\end{table}

\begin{figure}[H]
    \begin{subfigure}{0.5\linewidth}
        \centering
        \includegraphics[width=\linewidth]{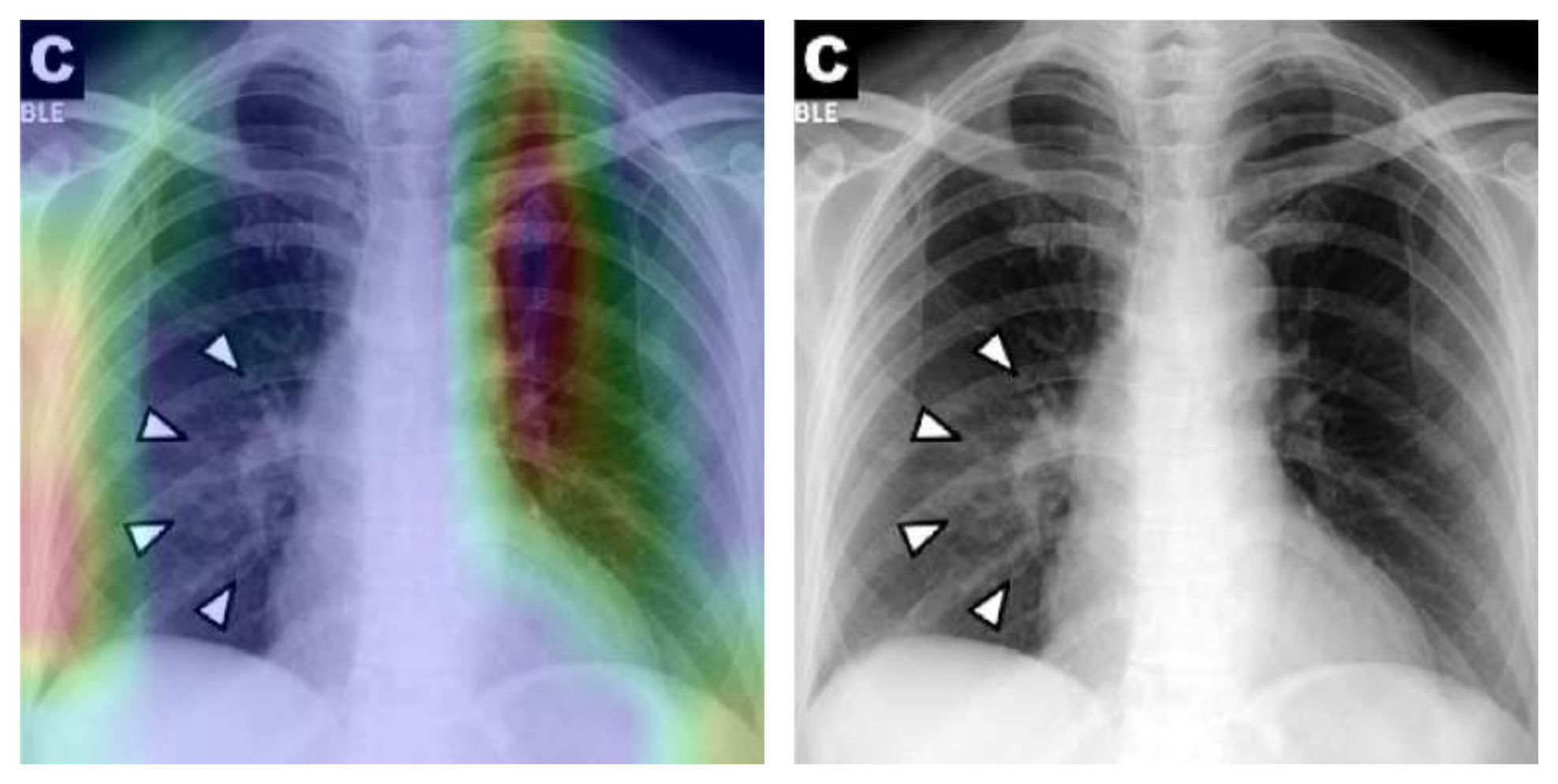}
        \caption{}
        \label{fig:3400_vis}
    \end{subfigure}
    \begin{subfigure}{0.5\linewidth}
        \centering
        \includegraphics[width=\linewidth]{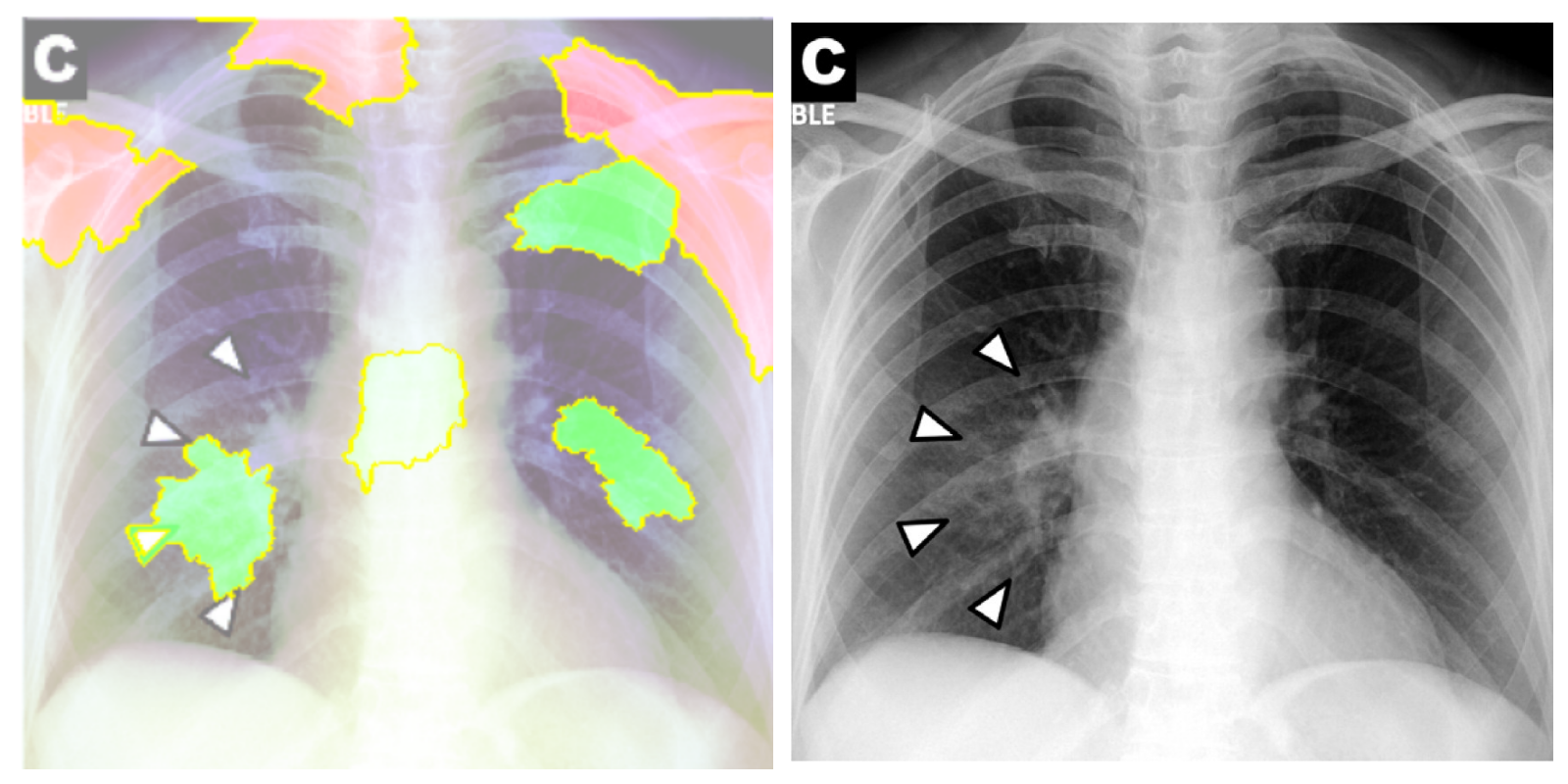}
        \caption{}
        \label{fig:3400_lime}
    \end{subfigure}
\caption{Model interpretability visualization by (a) Grad-CAM and (b) LIME image explanation}
\label{fig:3400}
\end{figure}

The base model has achieved a high area under the curve (AUC) of 0.9984 and an accuracy of 98.68\% while reaching a reasonable f-score of 0.94. Model visualization shows better performance; however, there are still various wrong regions in image explanations illustrated in Fig. \ref{fig:3400}.

In LIME explanation, green super-pixels contribute to current predicted class, and red super-pixels are most contributing to the other class \cite{ribeiro2016should}. In Grad-CAM visualization, region importance decreases from red to blue areas.

\subsection{Pretrained Models}
\begin{figure}[H]
    \centering
    \includegraphics[width=0.45\linewidth]{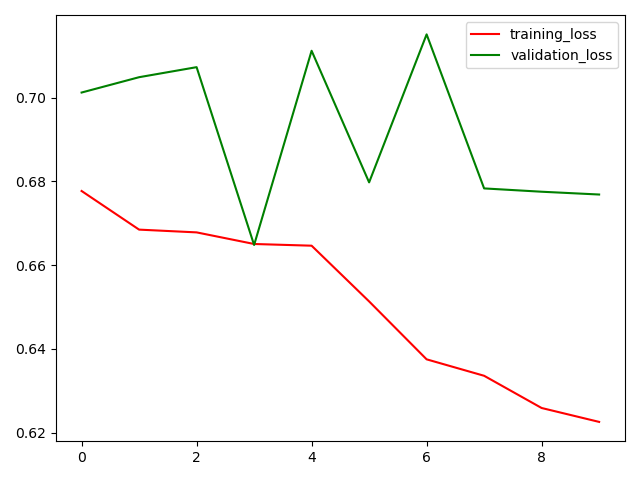}
    \caption{DenseNet-121 fine-tuning curve over 10 epochs}
    \label{fig:densenet_finetuning}
\end{figure}

DenseNet-121 is fine-tuned on the training-set containing normal pediatric cases for 20 epochs. The learning curve is plotted in Fig. \ref{fig:densenet_finetuning}.

\noindent Demonstrating bad results, the model is incapable of learning while fine-tuning only the last FC layer and freezing feature extraction layers and overfits to the data if retrained for more epochs. As expected, ImageNet categories are everyday objects which have somewhat non-similar features as pneumonia imaging patterns in CXRs. Hence, although transfer learning techniques from ImageNet-pretrained models have remarkably improved segmentation accuracy thanks to their capability of handling complex conditions, applying them for classification is still challenging due to the limited size of annotated data and a high chance of overfitting \cite{godasu2020transfer}. ResNet-50 has also produced almost the same results.

\begin{figure}[H]
    \begin{subfigure}{\linewidth}
        \centering
        \includegraphics[width=\linewidth]{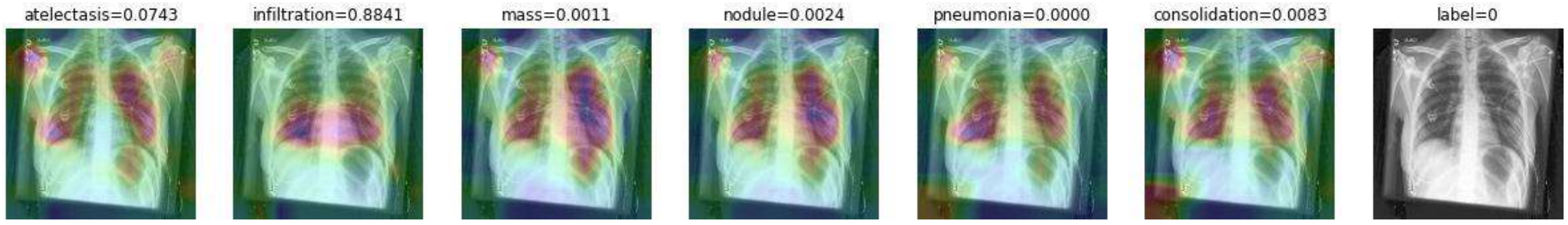}
        \caption{}
        \label{fig:chexnet_ex0}
    \end{subfigure}
    \hfill
    \begin{subfigure}{\linewidth}
        \centering
        \includegraphics[width=\linewidth]{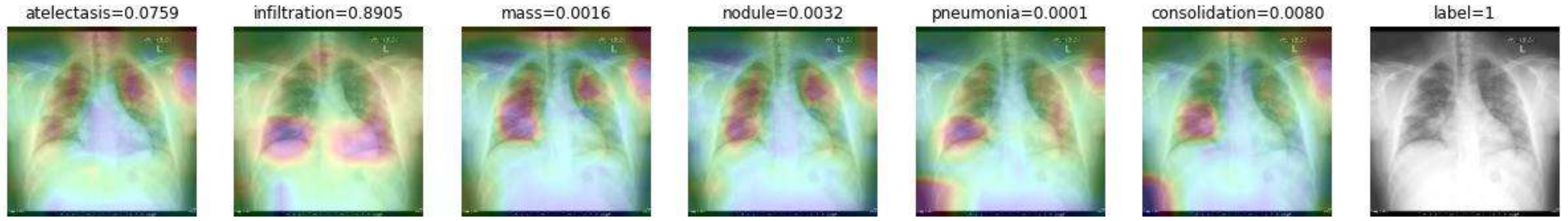}
        \caption{}
        \label{fig:chexnet_ex1}
    \end{subfigure}
\caption{CheXNet probabilities of different classes for (a) a COVID-19 positive case, and (b) a normal case}
\label{fig:chexnet_results}
\end{figure}

Regarding CheXNet pretrained model, we first probe to see if it is capable of correctly classifying COVID-19 pneumonia with no further improvements. Fig. \ref{fig:chexnet_results} shows results for two sample CXRs from both classes.

\noindent Extracted heatmaps reveal that CheXNet correctly marks chest lobes to determine each class probability. The output of each class is slightly higher in positive cases for most of the diseases as well. Some of the drawbacks are extremely high predictions for infiltration in most of the dataset images, getting stuck in regions outside lung boundaries and predominantly in corners, and missing some of the opacities particularly in lower lobes.

\subsection{COVID-CXNet}
\begin{figure}[H]
    \begin{subfigure}{0.5\linewidth}
        \centering
        \includegraphics[width=\linewidth]{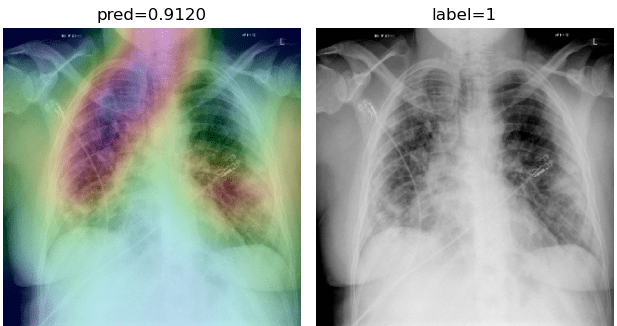}
        \label{fig:v1_ex1}
    \end{subfigure}
    \begin{subfigure}{0.5\linewidth}
        \centering
        \includegraphics[width=\linewidth]{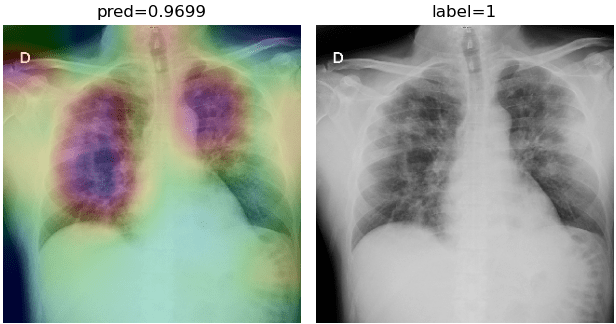}
        \label{fig:v1_ex2}
    \end{subfigure}
    \begin{subfigure}{0.5\linewidth}
        \centering
        \includegraphics[width=\linewidth]{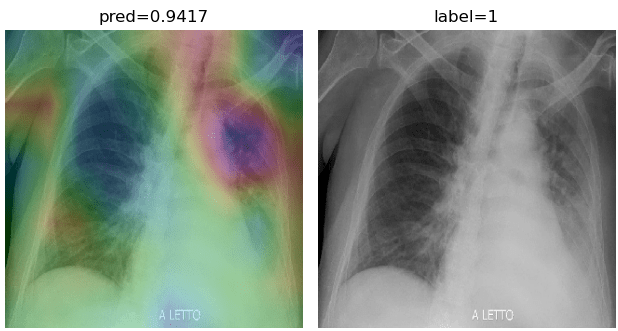}
        \label{fig:v1_ex3}
    \end{subfigure}
    \begin{subfigure}{0.5\linewidth}
        \centering
        \includegraphics[width=\linewidth]{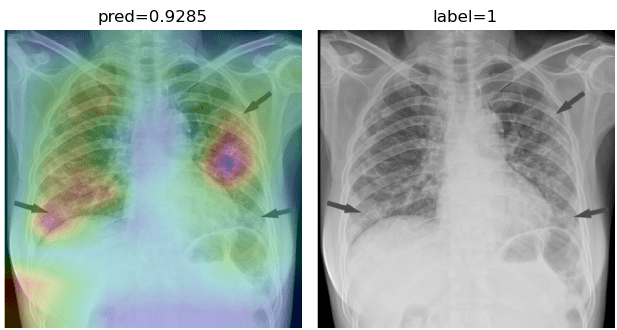}
        \label{fig:v1_ex4}
    \end{subfigure}
\caption{Grad-CAM visualization of the proposed model over sample cases}
\label{fig:cxnet_v1_heatmaps}
\end{figure}

To overcome the aforementioned issues and force the model's attention to the correct regions of interest (ROIs), we introduce the COVID-CXNet. Our model is initialized with the pretrained weights from CheXNet. A dataset of 3,628 images, 3,200 normal CXRs and 428 COVID-19 CXRs, are divided into 80\% as training-set and 20\% as test-set. Batch size is set to 16, rather than 32 in previous models, regarding memory constraints. Grad-CAMs of the COVID-CXNet for random images are plotted in Fig. \ref{fig:cxnet_v1_heatmaps}.

\noindent More Grad-CAMs are available in Appendix \ref{appendix: B}. Heatmaps are more accurate than previous models, while an accuracy of 99.04\% and an f-score of 0.96 are achieved. Table \ref{cxnet_cm} is the confusion matrix of the proposed model.

\begin{table}[H]
\caption{Confusion matrix of COVID-CXNet}
\label{cxnet_cm}
\centering
\begin{tabular}{|c|c|c|c|}
\hline
\multicolumn{2}{|c|}{\multirow{2}{*}{\textbf{COVID-CXNet}}} & \multicolumn{2}{c|}{\textbf{Predicted}} \\ \cline{3-4} 
\multicolumn{2}{|c|}{}                                     & \textit{Normal}   & \textit{COVID-19}   \\ \hline
\multirow{2}{*}{\textbf{Actual}}    & \textit{Normal}      & 641                & 4                   \\ \cline{2-4} 
                                    & \textit{COVID-19}    & 3                 & 78                  \\ \hline
\end{tabular}
\end{table}

\begin{figure}[H]
    \begin{subfigure}{0.33\linewidth}
        \centering
        \includegraphics[width=0.85\linewidth]{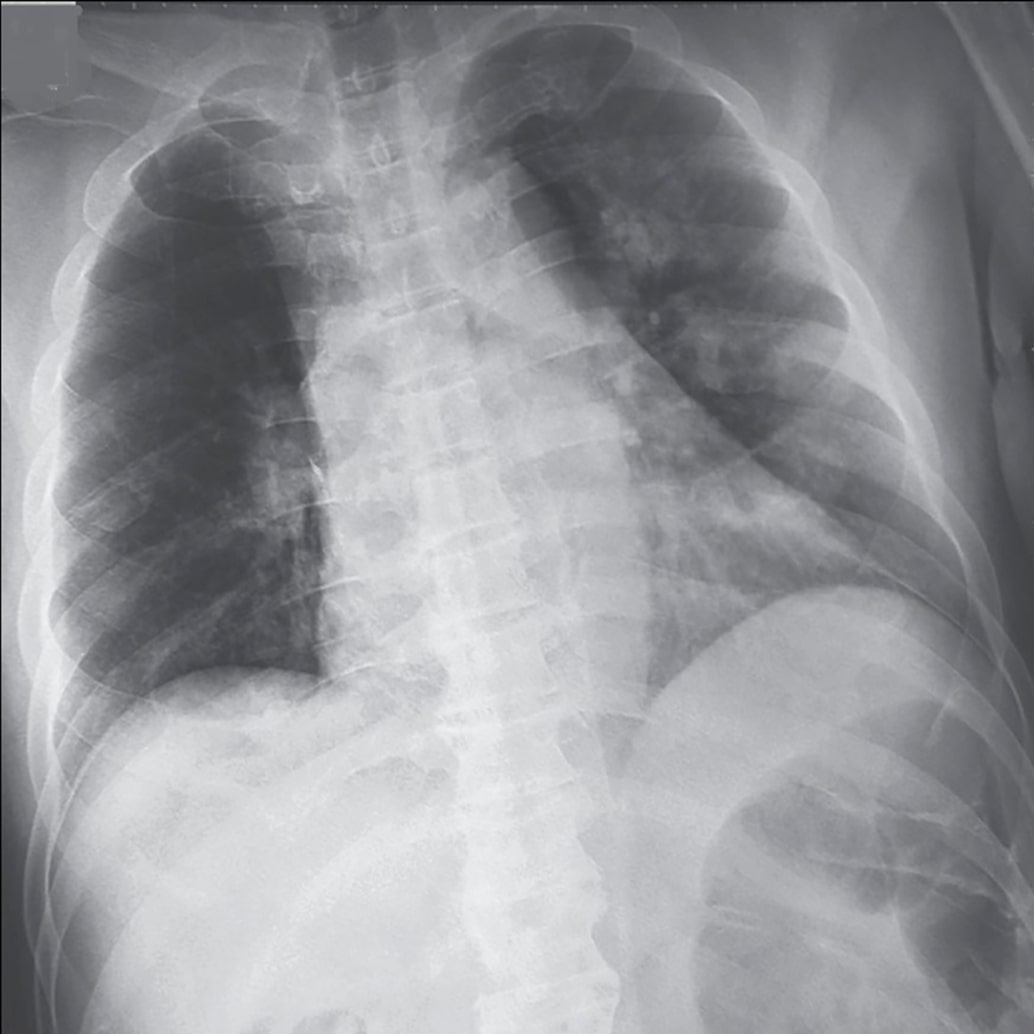}
        \caption{}
        \label{fig:comp_img}
    \end{subfigure}
    \hfill
    \begin{subfigure}{0.33\linewidth}
        \centering
        \includegraphics[width=0.85\linewidth]{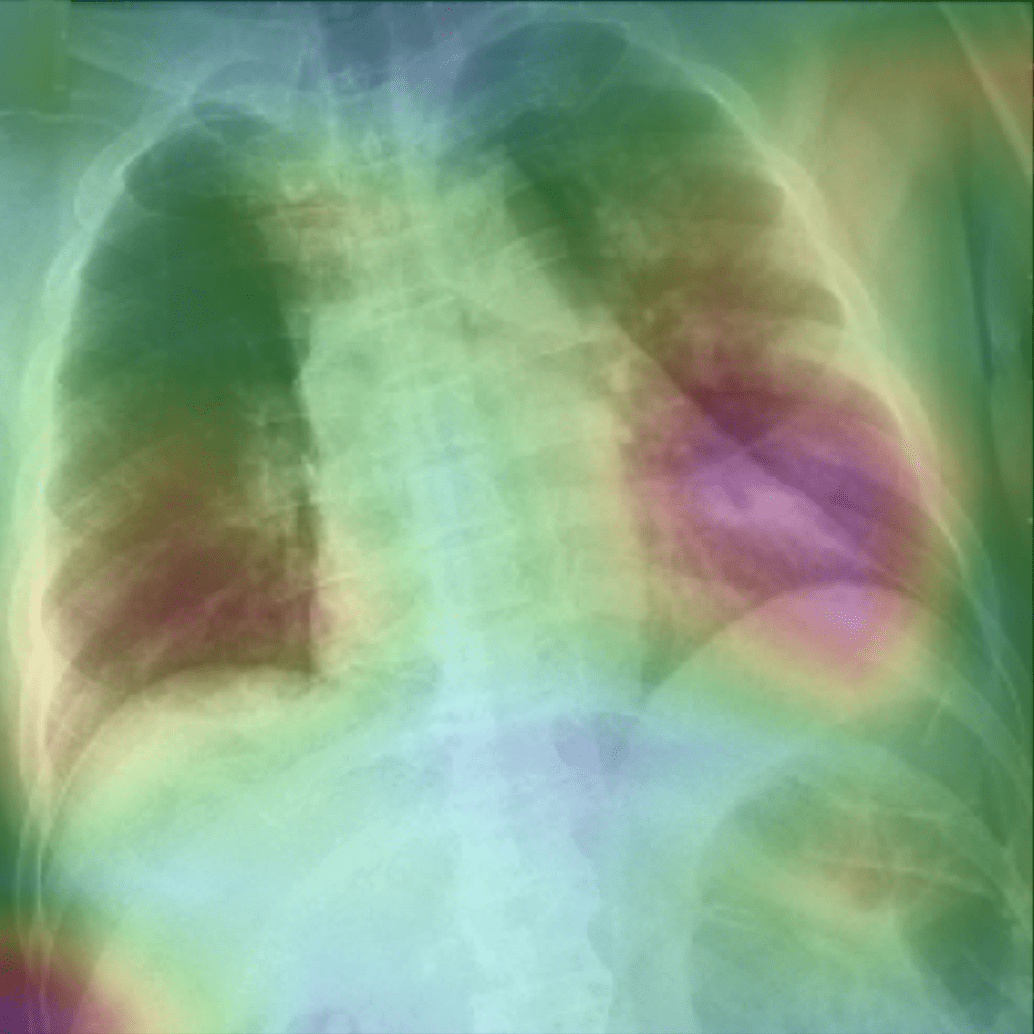}
        \caption{}
        \label{fig:comp_chexnet}
    \end{subfigure}
    \hfill
    \begin{subfigure}{0.33\linewidth}
        \centering
        \includegraphics[width=0.85\linewidth]{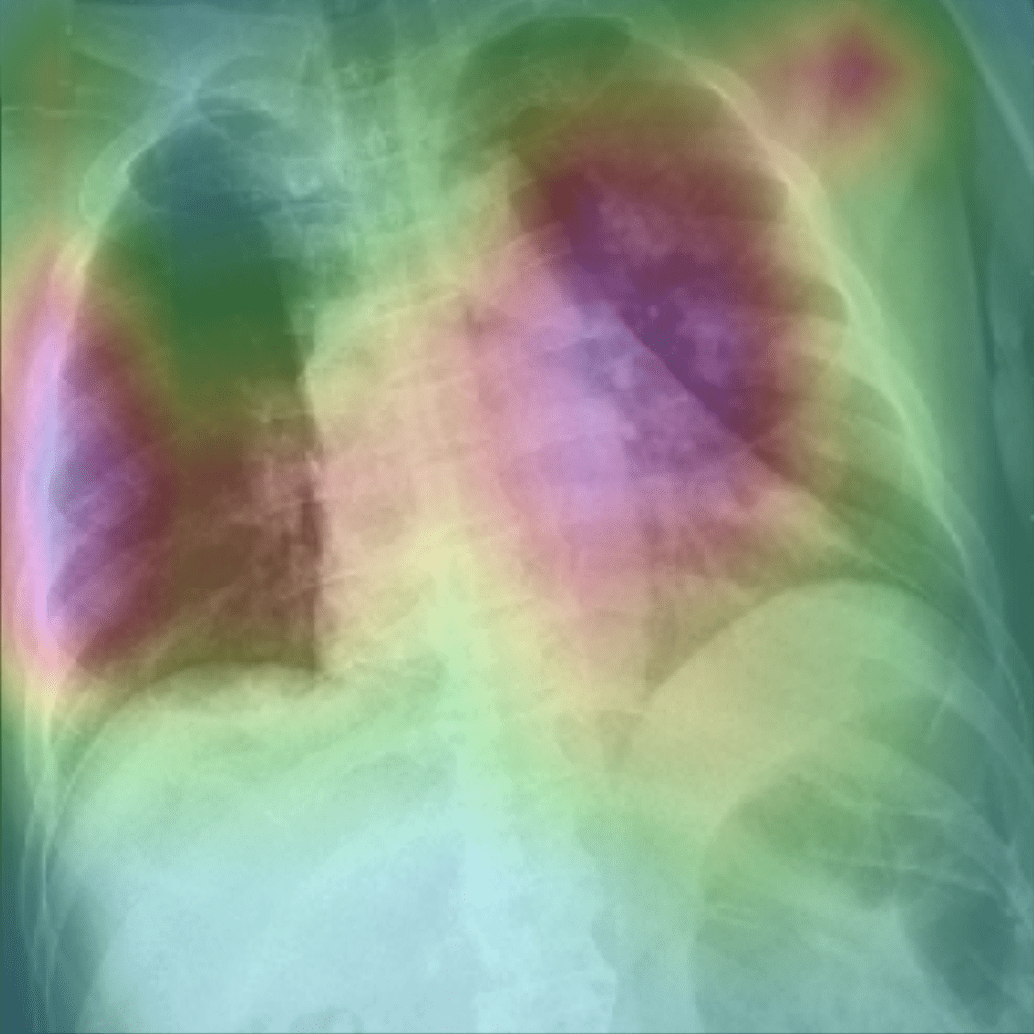}
        \caption{}
        \label{fig:comp_cxnet}
    \end{subfigure}
\caption{Comparison between the CheXNet and the proposed model; (a) is the image with patchy opacities in the upper left zone, (b) and (c) are heatmaps of the CheXNet and the proposed COVID-CXNet, respectively.}
\label{fig:chexnet_vs_cxnet}
\end{figure}

\noindent Proposed CheXNet-based model is capable of correctly classifying images. In many cases, it can localize pneumonia findings more precisely than the CheXNet. An example is illustrated in Fig. \ref{fig:chexnet_vs_cxnet}.

\begin{figure}[H]
    \begin{subfigure}[b]{0.5\linewidth}
        \centering
        \includegraphics[width=0.7\linewidth]{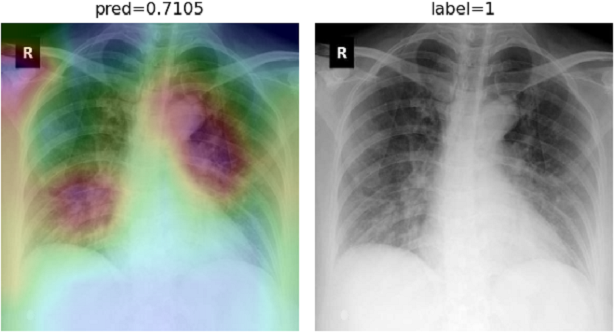}
        \caption{}
        \label{fig:no_text_ex0}
    \end{subfigure}
    \hfill
    \begin{subfigure}[b]{0.5\linewidth}
        \centering
        \includegraphics[width=0.7\linewidth]{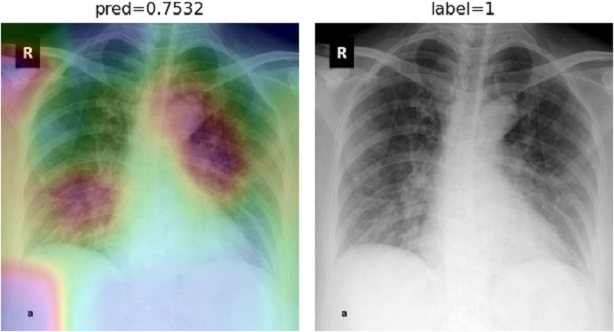}
        \caption{}
        \label{fig:with_text_ex0}
    \end{subfigure}
    \hfill
    \begin{subfigure}[b]{0.5\linewidth}
        \centering
        \includegraphics[width=0.7\linewidth]{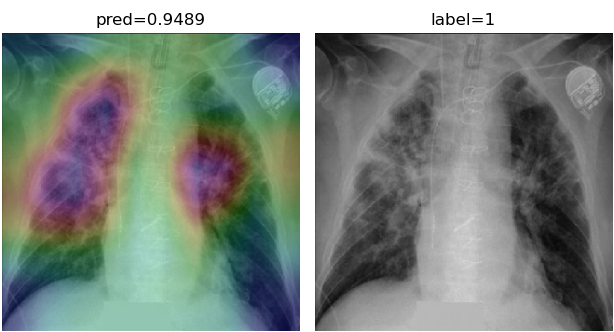}
        \caption{}
        \label{fig:no_text_ex1}
    \end{subfigure}
    \hfill
    \begin{subfigure}[b]{0.5\linewidth}
        \centering
        \includegraphics[width=0.7\linewidth]{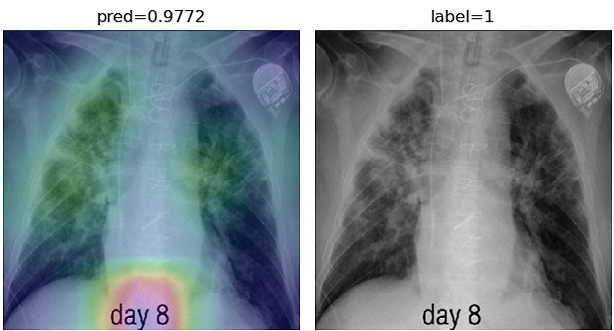}
        \caption{}
        \label{fig:with_text_ex1}
    \end{subfigure}
\caption{Text removal effect on model results. Images on the right have dates and signs, which are concealed in the images on the left.}
\label{fig:text_removal_effect}
\end{figure}

Fig. \ref{fig:chexnet_vs_cxnet} shows a CXR with an infiltrate in the upper lobe of the left hemithorax \cite{phan2020importation}; while CheXNet missed the region of pneumonia, the proposed model correctly uncovered the infiltration area. One concern about COVID-CXNet results is that it has pointed into other irrelevant regions, even outside the lungs. The same problem happens when there are frequently-appeared texts and signs, such as dates, present in the image. Fig. \ref{fig:text_removal_effect} shows how text removal can improve model efficiency.

\begin{figure}[H]
    \begin{subfigure}{0.5\linewidth}
        \centering
        \includegraphics[width=0.7\linewidth]{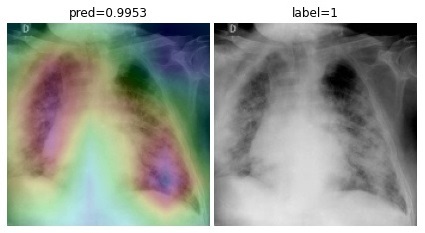}
        \label{fig:v2_ex1}
    \end{subfigure}
    \begin{subfigure}{0.5\linewidth}
        \centering
        \includegraphics[width=0.7\linewidth]{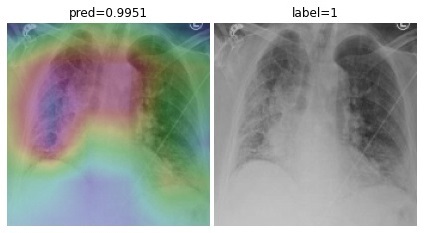}
        \label{fig:v2_ex2}
    \end{subfigure}
    \begin{subfigure}{0.5\linewidth}
        \centering
        \includegraphics[width=0.7\linewidth]{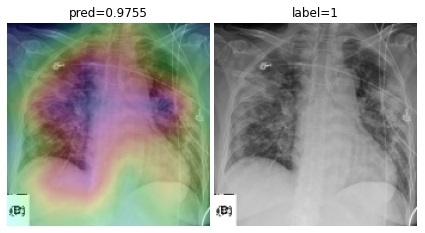}
        \label{fig:v2_ex3}
    \end{subfigure}
    \begin{subfigure}{0.5\linewidth}
        \centering
        \includegraphics[width=0.7\linewidth]{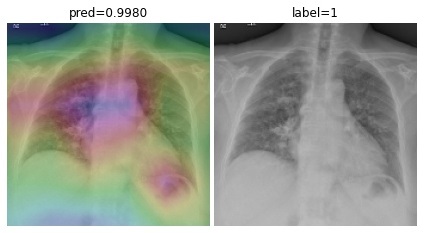}
        \label{fig:v2_ex4}
    \end{subfigure}
\caption{Grad-CAM visualization of the proposed model, trained with lung-segmented CXRs, over sample cases.}
\label{fig:cxnet_v2_heatmaps}
\end{figure}

While text removal methods can prevent overfitting, we can simply force the model to look into the lungs in order to address both problems in one effort. To accomplish this task, a U-Net based segmentation illustrated in Fig. \ref{fig:segmentation} is applied to the input images before enhancements. Visualization results for COVID-CXNet with the ROI-segmentation block are shown in Fig. \ref{fig:cxnet_v2_heatmaps}.

A figure with more Grad-CAMs is attached in Appendix \ref{appendix: C}. From Fig. \ref{fig:cxnet_v2_heatmaps}, it can be observed that COVID-CXNet with ROI-segmentation has delivered superior performance regarding the localization of pneumonia features. Worthwhile to mention that image augmentation is expanded by adding zoom-in, zoom-out, and brightness adjustment. Label smoothing is also applied to the loss function. 


The proposed method has shown a negligible drop in metric scores; accuracy is decreased by 0.42\%, and f-score is declined by 0.02. This decrease is a result of training with a larger dataset and accurately segmented ROIs, which means it has become more robust against unseen samples. There is a trade-off between catching good features and higher metric scores; while better features result in a more generalized model, high metric scores may indicate overfitting. 


As an extra step, we expanded COVID-CXNet for multiclass classification between normal, COVID-19 pneumonia (CP), and non-COVID pneumonia to examine its performance regarding the differentiation between two types of pneumonia. CP is often appeared with bilateral findings, whereas non-COVID pneumonia or CAP mostly has unilateral consolidations. Since most images are collected from the CXR-14 dataset, a histogram matching is applied to adjust histograms according to a base image. The output layer is changed to have three neurons with the SoftMax activation function. Confusion matrix is shown in Table \ref{multi_cxnet_cm}.

\begin{table}[H]
\caption{Confusion matrix of multiclass COVID-CXNet}
\label{multi_cxnet_cm}
\centering
\begin{tabular}{|c|c|c|c|c|}
\hline
\multicolumn{2}{|c|}{\multirow{2}{*}{\textbf{COVID-CXNet}}} & \multicolumn{3}{c|}{\textbf{Predicted}}      \\ \cline{3-5} 
\multicolumn{2}{|c|}{}                                     & \textit{Normal} & \textit{CAP} & \textit{CP} \\ \hline
\multirow{3}{*}{\textbf{Actual}}     & \textit{Normal}     & 671             & 44           & 9           \\ \cline{2-5} 
                                     & \textit{CAP}        & 205             & 451          & 16          \\ \cline{2-5} 
                                     & \textit{CP}         & 6               & 12           & 126         \\ \hline
\end{tabular}
\end{table}

\begin{figure}[H]
    \begin{subfigure}{0.33\linewidth}
        \centering
        \includegraphics[width=\linewidth]{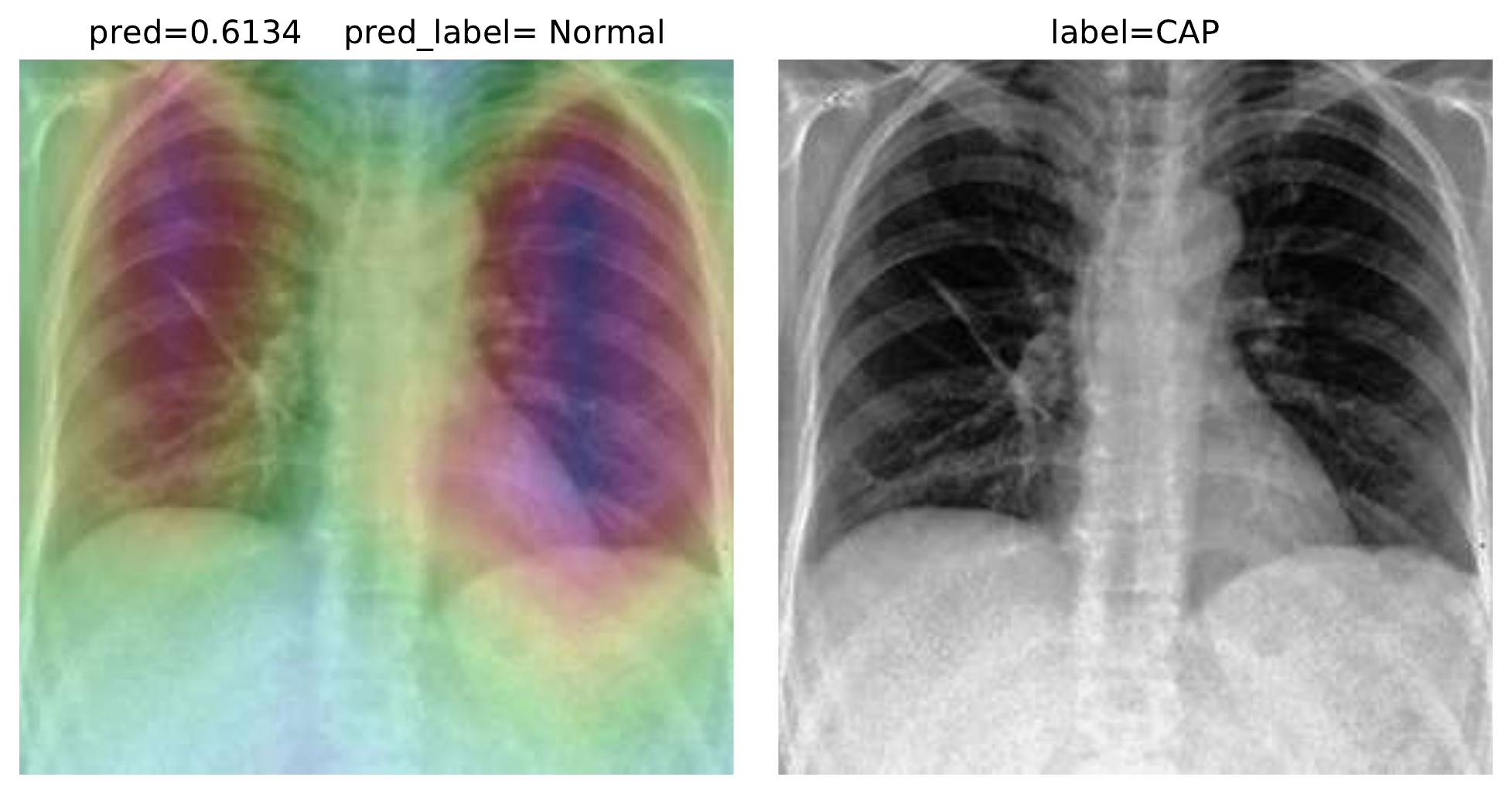}
        \label{fig:multiclass1}
    \end{subfigure}
    \begin{subfigure}{0.33\linewidth}
        \centering
        \includegraphics[width=\linewidth]{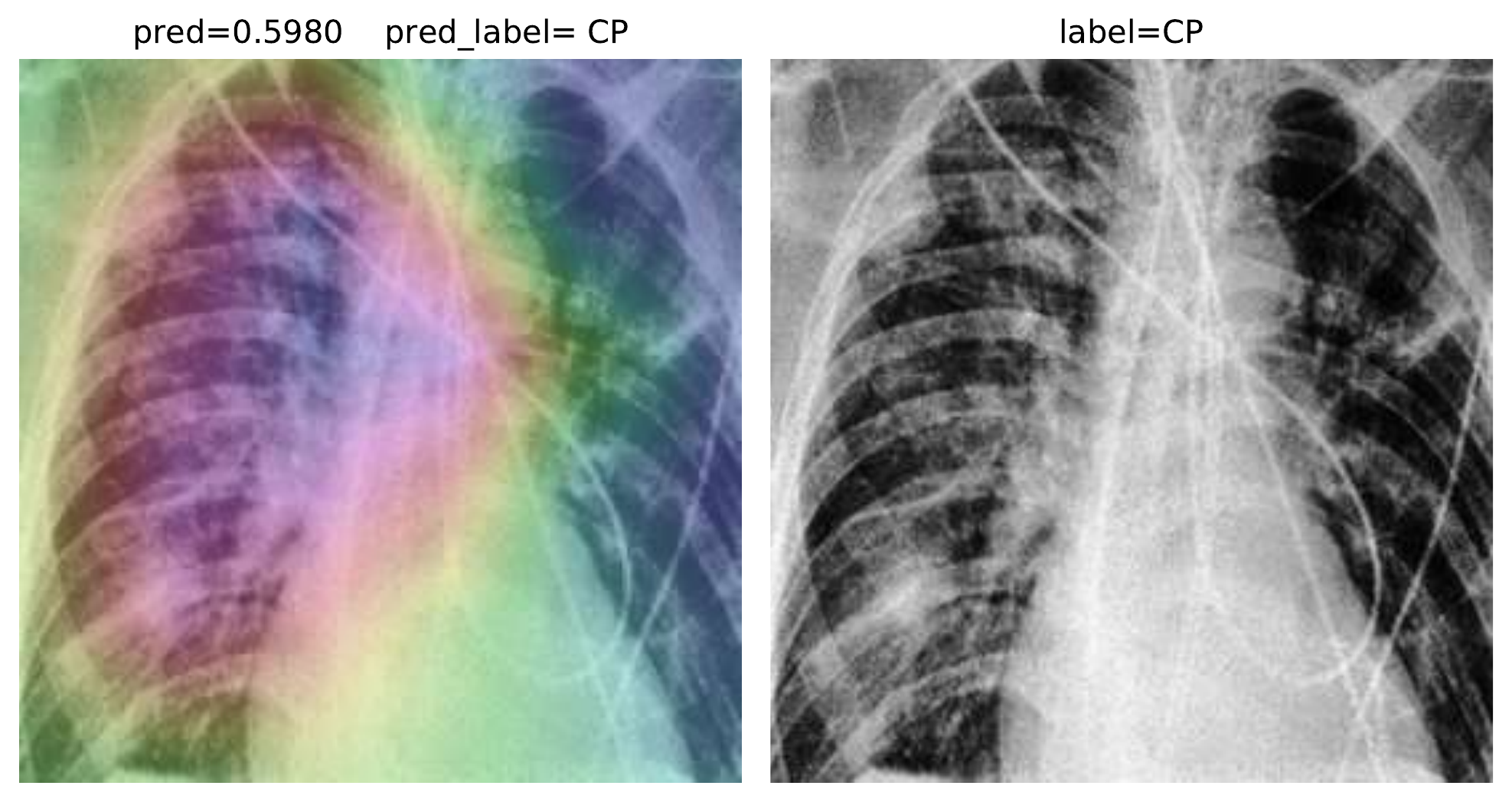}
        \label{fig:multiclass2}
    \end{subfigure}
    \begin{subfigure}{0.33\linewidth}
        \centering
        \includegraphics[width=\linewidth]{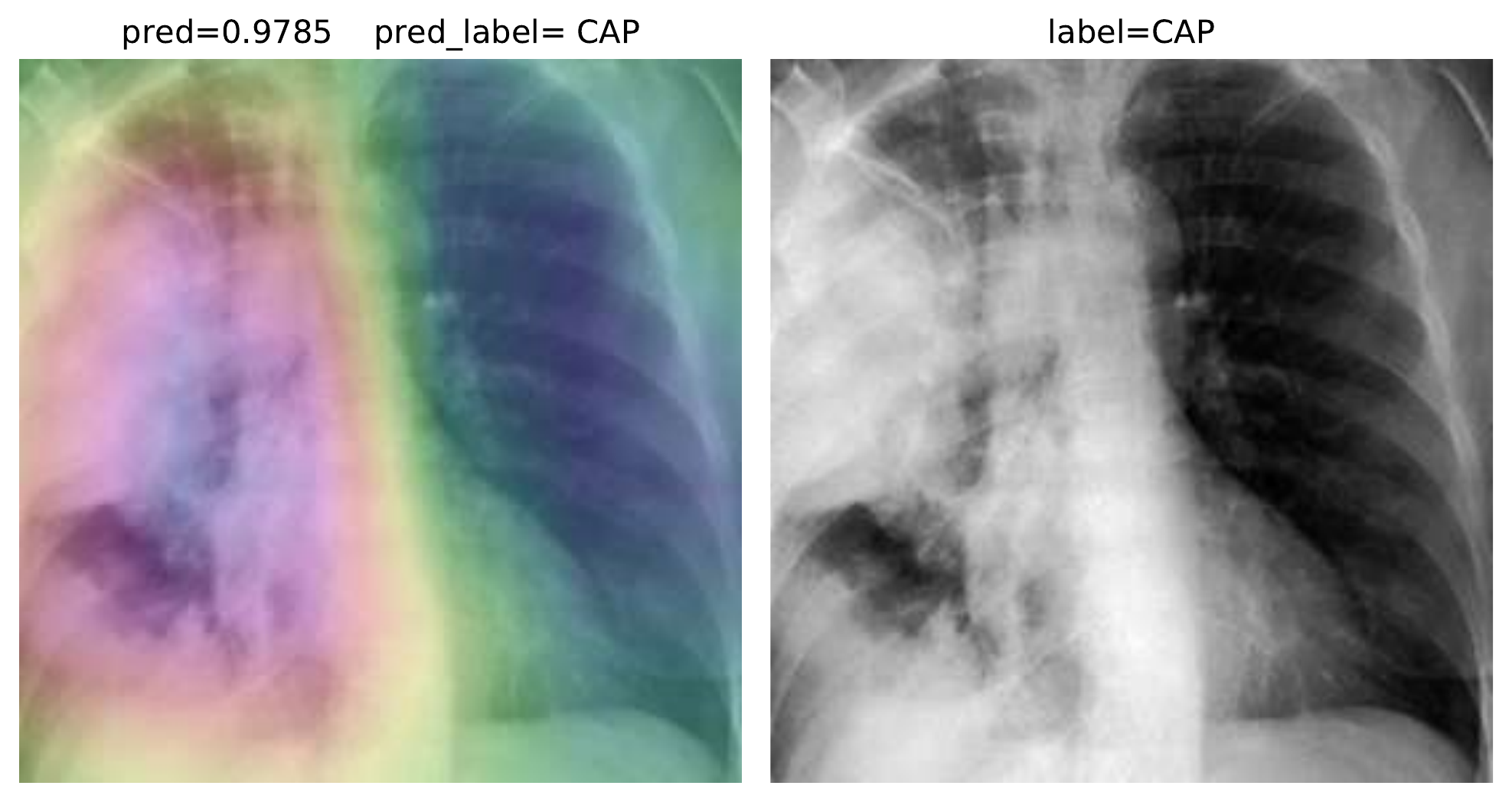}
        \label{fig:multiclass3}
    \end{subfigure}
    \begin{subfigure}{0.33\linewidth}
        \centering
        \includegraphics[width=\linewidth]{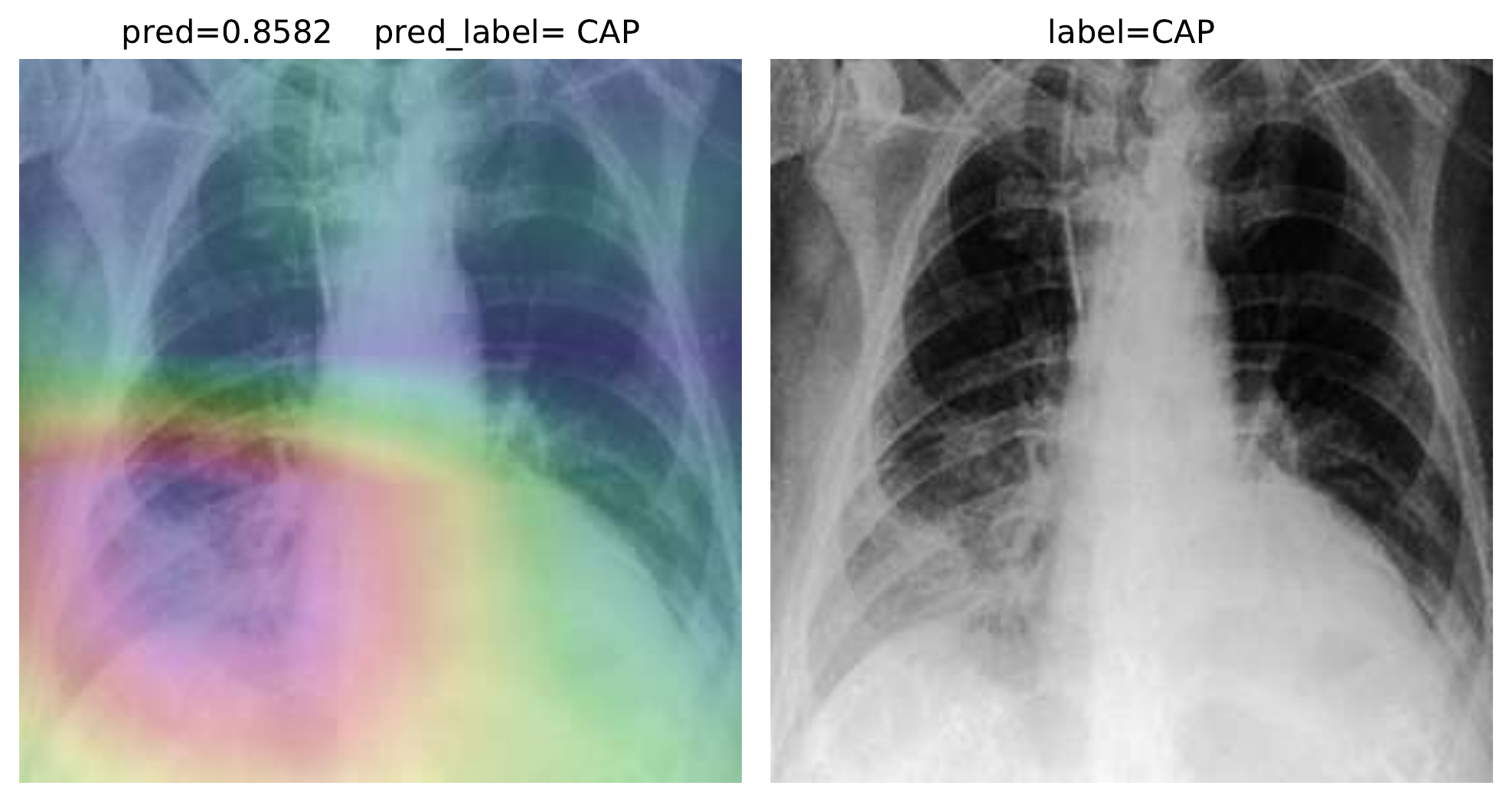}
        \label{fig:multiclass4}
    \end{subfigure}
    \begin{subfigure}{0.33\linewidth}
        \centering
        \includegraphics[width=\linewidth]{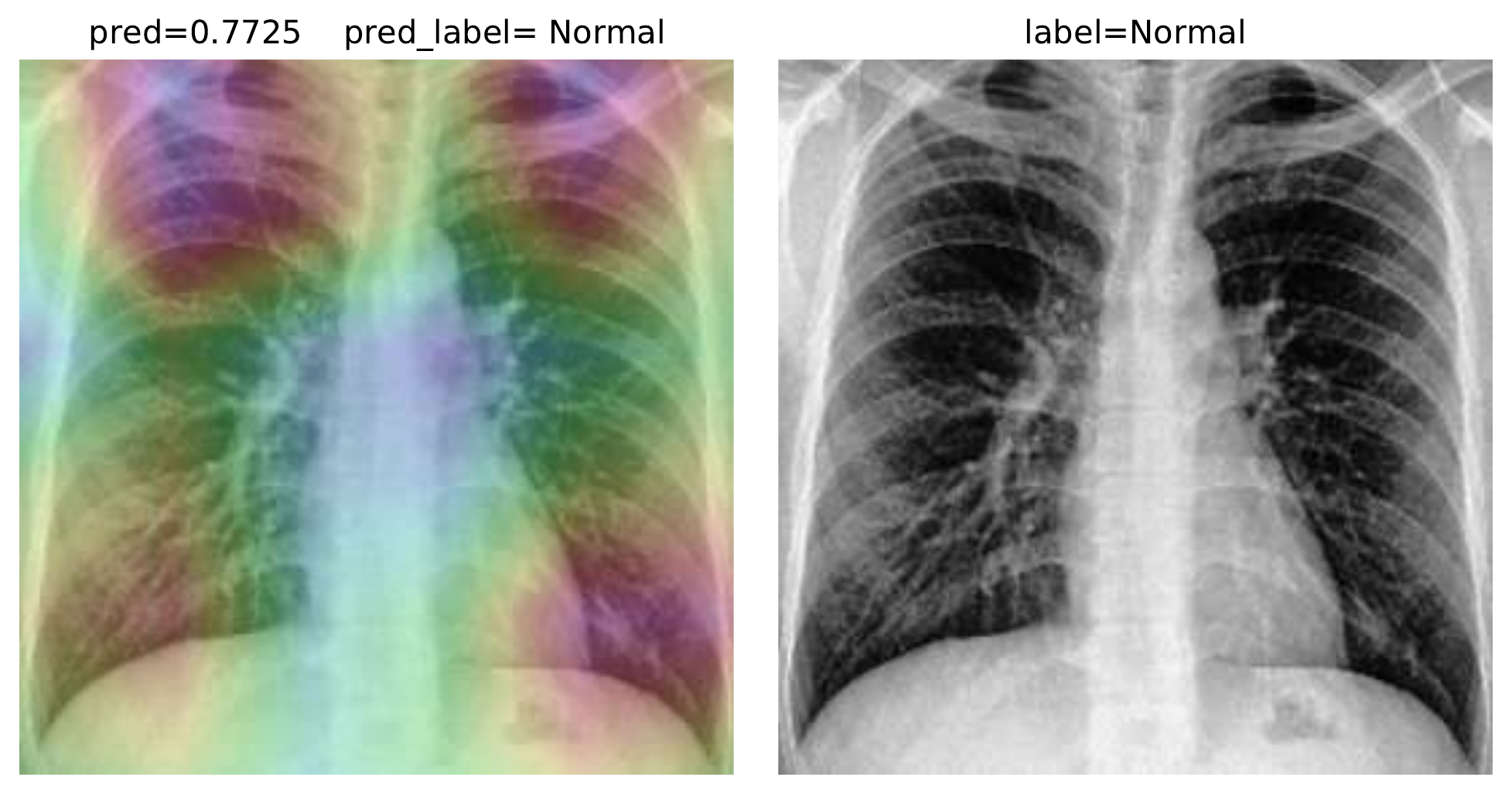}
        \label{fig:multiclass5}
    \end{subfigure}
    \begin{subfigure}{0.33\linewidth}
        \centering
        \includegraphics[width=\linewidth]{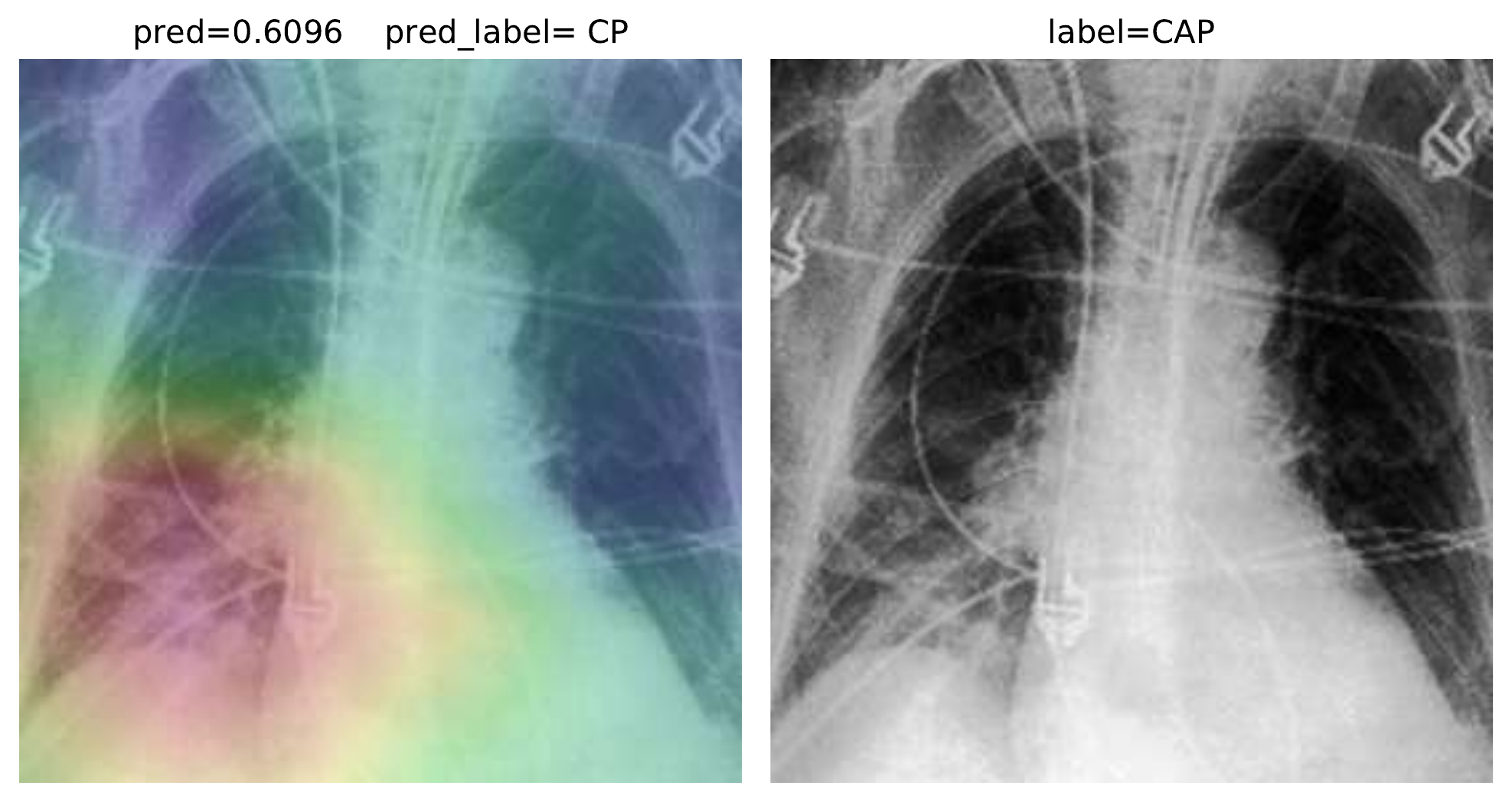}
        \label{fig:multiclass6}
    \end{subfigure}

\caption{COVID-CXNet multiclass classification visualization results}
\label{fig:multiclass_heatmaps}
\end{figure}

\noindent Accuracy score is 81.04\%, with f-scores of 0.85 and 0.76 for CP and CAP classes, respectively. In a number of cases, especially in the first stages of virus progression, CP has unilateral findings. Also, CAP may cause bilateral consolidations. Therefore, some cases are expected to be misclassified between CP and CAP. From the confusion matrix, it could be seen that a relatively high number of images are misclassified between CAP and normal. A potential reason for this issue is considered to be related to wrong labeling. Besides, some CAP CXRs are from patients with early-stage disease development. To confirm the model performance, Grad-CAMs are plotted in Fig. \ref{fig:multiclass_heatmaps}.

The model is properly looking at one lobe for detecting CAP and both lobes for CAP and normal images. There are some wrong labels, nevertheless. A figure containing more visualizations is found in Appendix \ref{appendix: D}. To further enhance statistical scores, a hierarchical approach is implemented. In the first level, we classify images into normal and pneumonia classes. In the second level, pneumonia images are categorized into CP and CAP. Final confusion matrix is illustrated in Table \ref{hierarchical_cxnet_cm}.

\begin{table}[hpbt]
\caption{Confusion matrix of hierarchical multiclass COVID-CXNet}
\label{hierarchical_cxnet_cm}
\centering
\begin{tabular}{|c|c|c|c|c|}
\hline
\multicolumn{2}{|c|}{\multirow{2}{*}{\textbf{COVID-CXNet}}} & \multicolumn{3}{c|}{\textbf{Predicted}}      \\ \cline{3-5} 
\multicolumn{2}{|c|}{}                                     & \textit{Normal} & \textit{CAP} & \textit{CP} \\ \hline
\multirow{3}{*}{\textbf{Actual}}     & \textit{Normal}     & 689             & 32           & 3           \\ \cline{2-5} 
                                     & \textit{CAP}        & 143             & 524          & 5          \\ \cline{2-5} 
                                     & \textit{CP}         & 3               & 11           & 130         \\ \hline
\end{tabular}
\end{table}

A slight improvement is observed. Overall accuracy is 87.21\%, while f-scores are 0.92 for CP class and 0.85 for CAP class. Although visualizations and metrics demonstrate promising performance, the effect of dataset bias is non-negligible for wrong predictions between CAP and normal images. A dataset of 7,700 images containing 700 CP, 3,500 CAP and 3,500 normal CXRs is used in both multiclass approaches.

\section{Discussions}
Throughout this study, several model architectures are introduced and applied to different amounts of images. Bias to pediatric CXRs and lung segmentation module are also addressed in different models. A comparison between these models is shown in Table \ref{tab:version_comparison}.

\begin{table}[H]
    \centering
    \caption{Comparison between investigated models throughout this paper (F-score is provided for CP class only)}
\begin{tabular}{|c|c|c|c|c|c|c|c|}
\hline
\textbf{Model}         & \begin{tabular}[c]{@{}c@{}}\textbf{Dataset} \\ \textbf{Size}\end{tabular} & \begin{tabular}[c]{@{}c@{}}\textbf{Pediatric} \\ \textbf{Bias}\end{tabular} & \begin{tabular}[c]{@{}c@{}}\textbf{Lung} \\ \textbf{Segmentation}\end{tabular} & \begin{tabular}[c]{@{}c@{}}\textbf{Num} \\ \textbf{Epochs}\end{tabular} & \begin{tabular}[c]{@{}c@{}}\textbf{Accuracy} \\ \textbf{Score (avg)}\end{tabular} &
\begin{tabular}[c]{@{}c@{}}\textbf{Confidence} \\
\textbf{Intervals}\end{tabular} &
\begin{tabular}[c]{@{}c@{}}\textbf{F-score} \\ \textbf{(avg)}\end{tabular} \\ \hline
Base Model v1          & 600                & Yes               & No                & 100                & 96.10\%           & 91.40 -- 97.10      & 0.82                \\ \hline
Base Model v2          & 3,400               & No               & No                 & 120                & 98.68\%          & 94.00 -- 99.30         & 0.94                \\ \hline
COVID-CXNet v1         & 3,628               & No                & No                & 10                 & 99.04\%           & 96.43 -- 99.54    & 0.96                \\ \hline
COVID-CXNet v2         & 3,628               & No                & Yes               & 10                 & 98.62\%           & 97.14 -- 99.28    & 0.94                \\ \hline
\begin{tabular}[c]{@{}c@{}}Flat Multiclass \\ COVID-CXNet \end{tabular} & 7,700               & No                & Yes               & 30                 & 81.04\%           & 80.25 -- 84.86    & 0.85                
     \\ \hline
\begin{tabular}[c]{@{}c@{}}Hierarchical Multiclass \\ COVID-CXNet \end{tabular} & 7,700               & No                & Yes               & 10+20                 & 87.21\%           & 86.03 -- 89.12    & 0.92                \\ \hline
\end{tabular}
    \label{tab:version_comparison}
\end{table}

Accuracy score ranges are achieved by running models ten different times. With the expansion of the dataset, confidence intervals shrink, and metric scores slightly decline while pneumonia symptoms localization improves. Furthermore, the proposed model is compared to other research studies discussed in section \ref{related_works} regarding several criteria, such as dataset size and f-score. The comparison is illustrated in Table \ref{tab:related_works_comparison}.

\begin{table}[H]
    \centering
    \caption{Comparison between proposed model and related works}
\begin{tabular}{|c|c|c|c|c|c|c|}
\hline
\textbf{Model}                                                              & \begin{tabular}[c]{@{}c@{}}
\textbf{Num CP} \\ \textbf{Images}\end{tabular} 
&
\begin{tabular}[c]{@{}c@{}} \textbf{Pediatric} \\ \textbf{Bias} \end{tabular}
& \begin{tabular}[c]{@{}c@{}} \textbf{Pretrained} \\ \textbf{Weights} \end{tabular}
& \textbf{Architecture}                                                     & \begin{tabular}[c]{@{}c@{}}\textbf{CP Class} \\ \textbf{F-score}\end{tabular} 
& \begin{tabular}[c]{@{}c@{}}\textbf{Lung} \\ \textbf{Segmentation}\end{tabular} \\ 
\hline
Zhang \textit{et al} \cite{zhang2020covid}               & 100   & No                  & \begin{tabular}[c]{@{}c@{}}Yes, \\ ImageNet \end{tabular}               & ResNet                                                                          & 0.72                & No                \\ \hline
Li \textit{et al} \cite{li2020covidmobilexpert}          & 179   & No                  & \begin{tabular}[c]{@{}c@{}}Yes, \\ ImageNet \end{tabular}               & \begin{tabular}[c]{@{}c@{}}Multiple\\ Best: ShuffleNetV2\end{tabular}           & -                   & No                \\ \hline

Khan \textit{et al} \cite{khan2020coronet}          & 284   & Yes                  & \begin{tabular}[c]{@{}c@{}}Yes, \\ ImageNet \end{tabular}              
& Xception           & 0.9560                   & No                \\ \hline

Rajaraman \textit{et al} \cite{rajaraman2020iteratively} & 313   & No                  & \begin{tabular}[c]{@{}c@{}}Yes, \\ ImageNet \end{tabular}               & \begin{tabular}[c]{@{}c@{}}Multiple (Ensemble)\\ Best: InceptionV3\end{tabular} & 0.9841              & Yes, U-Net        \\ \hline
Wang and Wang \cite{wang2020covid}                                        & 358         & No            & No                          & \begin{tabular}[c]{@{}c@{}}COVID-Net \\ (Custom Architecture)\end{tabular}      & 0.9480              & No                \\ \hline
Mangal \textit{et al} \cite{mangal2020covidaid}          & 358    & Yes                 & \begin{tabular}[c]{@{}c@{}}Yes, \\ CheXNet \end{tabular}                & DenseNet                                                                        & 0.9230              & No                \\ \hline
Proposed Model                                                                             & 700       & No              & \begin{tabular}[c]{@{}c@{}}Yes, \\ CheXNet \end{tabular}                 & DenseNet                                                                        & 0.9221                & Yes, U-Net        \\ \hline
\end{tabular}
    \label{tab:related_works_comparison}
\end{table}

While other models have higher f-scores, they have different issues. For example, CoroNet \cite{khan2020coronet} has a small dataset in which most images are from pediatrics. The pruned ensemble method \cite{rajaraman2020iteratively} also suffers from overparameterization and a lack of proper visualization discussion in the paper. Considering the significant number of parameters of the model and their small dataset, Grad-CAM visualizations must be investigated to a certain extent. Our proposed model is the only study to use a lung segmentation module with a CheXNet-based fine-tuning on a relatively large dataset of CP CXRs. Besides, our model benefits from different approaches, such as label smoothing and hierarchical strategy, to enhance its performance and prevent overfitting.

\subsection{Dataset}
\textit{\textbf{Data}}: While DL-based methods have demonstrated promising results in medical image classification, a significant challenge is data shortage. Since COVID-19 diagnosis using CXR images is recently becoming an interesting topic, accessible data is limited. Data augmentation is an essential method of coping with data shortcomings. However, designing a pneumonia detection model still needs much more CXR data. Most COVID-19 pneumonia detection articles have very small datasets. Although \cite{wang2017chestx} has claimed to introduce the largest open-access dataset of 385 CXRs, to the best of our knowledge, our dataset of \numOfImages{} COVID-19 CXRs has the most number of images.



\textit{\textbf{Pediatric Bias}}: Many research papers have benefited from databases built upon the proposed dataset by \cite{kermany2018identifying}. Children have different pulmonary anatomy. Hence, developed models based on normal pediatric and adult pneumonia images are highly vulnerable to the "right decision with wrong reason" problem. Besides, previous studies have proved that using various datasets containing images from different hospitals improves the pneumonia-detection results \cite{tilve2020pneumonia}. To prevent pediatric bias, we not only collected normal CXRs from different sources. Furthermore, COVID-19 pneumonia CXRs were collected from 9 different sources to improve cross-dataset robustness.

In the future, other information regarding patient status can be used alongside x-rays. Clinical symptoms can remarkably help radiologists in COVID-19 differential diagnosis. Metadata could be concatenated with the input CXR and be fed into the model to help increase its decision certainty. Providing the metadata, it is possible to have more detailed predictions, e.g. the chance of patient survival, based on clinical symptoms and the severity of pneumonia features presented by CXR.

\subsection{Architecture}
\textit{\textbf{Metrics}}: In section \ref{base_model} we introduced a simple base model. The purpose was to show how it can achieve very high accuracy scores. Digging into explainability, the model revealed wrong features responsible for its excellent metric scores. Therefore, high accuracy scores of sophisticated models from a small number of CXRs, which have high texture complexity, are tricky. Investigation of model performance based upon confusion matrices and accuracy scores could not be usually validated unless demonstrating appropriate localization of imaging features.

\textit{\textbf{Transfer Learning}}: Using pretrained models with ImageNet weights, some studies such as \cite{rajaraman2020iteratively} showed acceptable heatmaps, but only a few images were visualized. While these pretrained models may help, having small dataset sizes suggested us to fine-tune models previously trained on similar data. Our CheXNet-based model shows better performance over ImageNet-pretrained models, while not hindered by problems like overparameterization. Besides, lung segmentation was also performed by a U-Net based architecture previously trained on similar frontal CXRs. Among studies conducted, there was only one article to use CheXNet as its backbone \cite{mangal2020covidaid}, which applied the model on a fewer number of images and without lung segmentation as its image preprocessing procedure.

\textit{\textbf{CheXNet}}: CheXNet is trained on a very large dataset of CXRs and has been used for transfer learning by some other thoracic disease identification studies \cite{almuhayar2019classification, sze2019tchexnet}. However, it has its own deficiencies, such as individual sample variability as a result of data ordering changes \cite{zech2019individual} and vulnerability to adversarial attacks \cite{finlayson2018adversarial}. Enhancing thoracic abnormality detection in CXRs using CheXNet requires the development of ensemble models, which is currently prone to overfitting due to the number of images from COVID-19 positive patients.


\section{Conclusion}
In this paper, we firstly collected a dataset of CXR images from normal lungs and COVID-19 infected patients. The constructed dataset is made from images of different datasets from multiple hospitals and radiologists and is the largest public dataset to the best of our knowledge. Next, we designed and trained an individual CNN and also investigated the results of ImageNet-pretrained models. Then, a DenseNet-based model is designed and fine-tuned with weights initially set from the CheXNet model. Comparing model visualization over a batch of samples as well as accuracy scores, we denoted the significance of Grad-CAM heatmaps and its priority to be considered the primary model validation metric. Finally, we discussed several points like data shortage and the importance of transfer learning for tackling similar tasks. A final CP class f-score of 0.94 for binary classification and 0.85 for three-class classification are achieved. The proposed model development procedure is visualization-oriented as it is the best method to confirm its generalization as a medical decision support system.

\section*{Data and Code Availability}
A dataset of COVID-19 positive CXR images, used in this study, as well as source codes and the pretrained network weights are hosted on a public repository on GitHub to help accelerate further research studies.
\textit{\href{https://github.com/armiro/COVID-CXNet}{https://github.com/armiro/COVID-CXNet}}

\bibliography{main.bib}
\bibliographystyle{IEEEtran}

\newpage
\appendix
\section*{Appendices}

\section{More Grad-CAMs of the COVID-CXNet Model} \label{appendix: B}

\begin{figure}[H]
    \begin{subfigure}{0.4\linewidth}
        \centering
        \includegraphics[width=\linewidth]{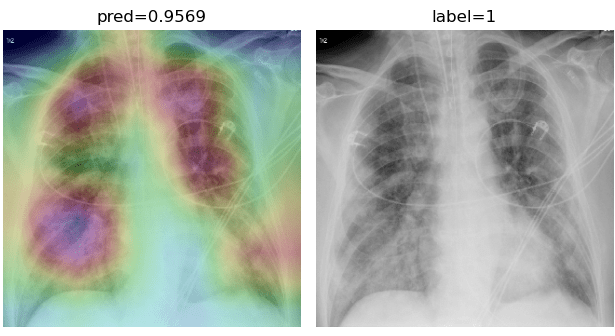}
        \label{fig:cx1_1}
    \end{subfigure}
    \hfill
    \begin{subfigure}{0.4\linewidth}
        \centering
        \includegraphics[width=\linewidth]{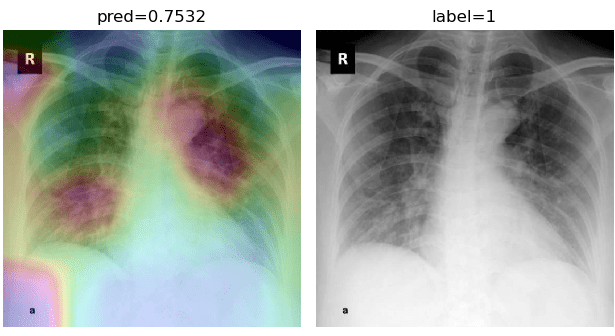}
        \label{fig:cx1_2}
    \end{subfigure}
    \hfill
    \begin{subfigure}{0.4\linewidth}
        \centering
        \includegraphics[width=\linewidth]{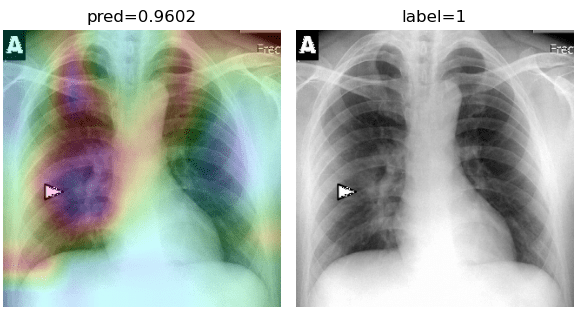}
        \label{fig:cx1_3}
    \end{subfigure}
    \hfill
    \begin{subfigure}{0.4\linewidth}
        \centering
        \includegraphics[width=\linewidth]{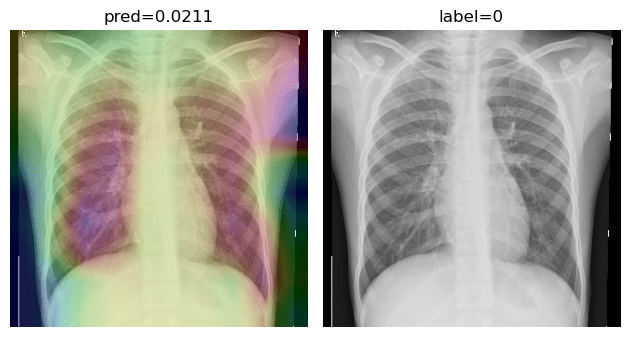}
        \label{fig:cx1_4}
    \end{subfigure}
    \hfill
    \begin{subfigure}{0.4\linewidth}
        \centering
        \includegraphics[width=\linewidth]{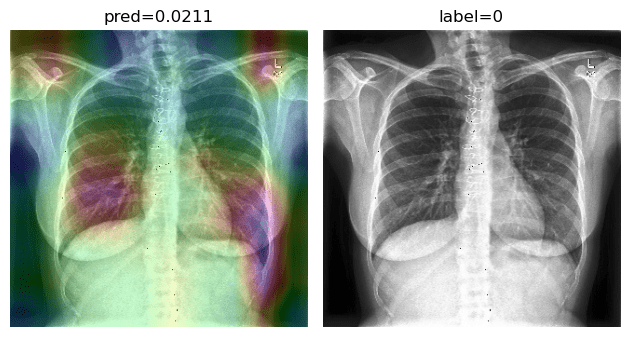}
        \label{fig:cx1_5}
    \end{subfigure}
    \hfill
    \begin{subfigure}{0.4\linewidth}
        \centering
        \includegraphics[width=\linewidth]{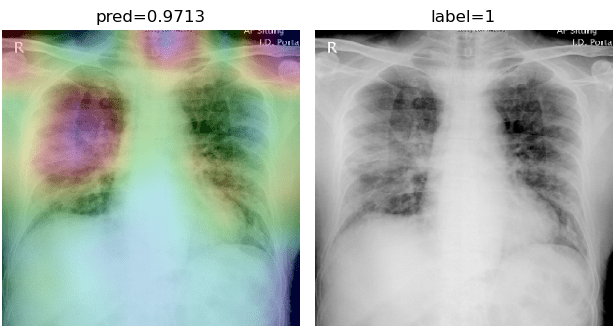}
        \label{fig:cx1_6}
    \end{subfigure}
\caption{Grad-CAMs from COVID-CXNet}
\label{fig:more_roi_cxnet_1}
\end{figure}

\newpage
\section{More Grad-CAMs of the COVID-CXNet Model with Lung Segmentation Preprocessing} \label{appendix: C}

\begin{figure}[H]
    \begin{subfigure}{0.4\linewidth}
        \centering
        \includegraphics[width=\linewidth]{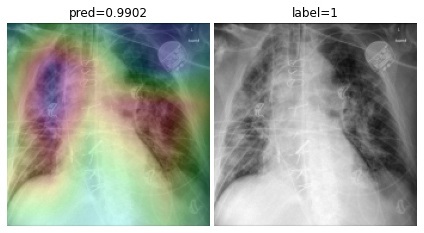}
        \label{fig:cx2_1}
    \end{subfigure}
    \hfill
    \begin{subfigure}{0.4\linewidth}
        \centering
        \includegraphics[width=\linewidth]{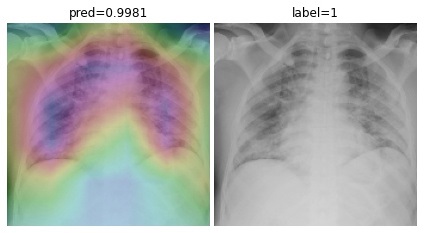}
        \label{fig:cx2_2}
    \end{subfigure}
    \hfill
    \begin{subfigure}{0.4\linewidth}
        \centering
        \includegraphics[width=\linewidth]{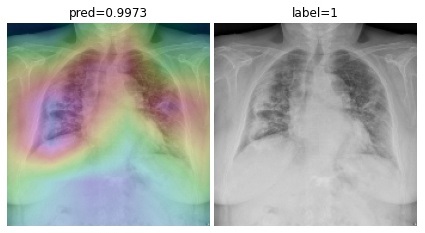}
        \label{fig:cx2_3}
    \end{subfigure}
    \hfill
    \begin{subfigure}{0.4\linewidth}
        \centering
        \includegraphics[width=\linewidth]{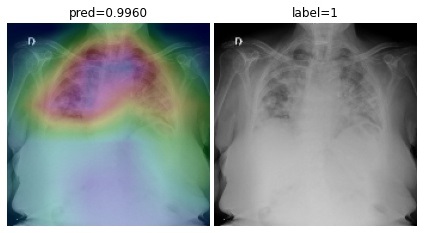}
        \label{fig:cx2_4}
    \end{subfigure}
    \hfill
    \begin{subfigure}{0.4\linewidth}
        \centering
        \includegraphics[width=\linewidth]{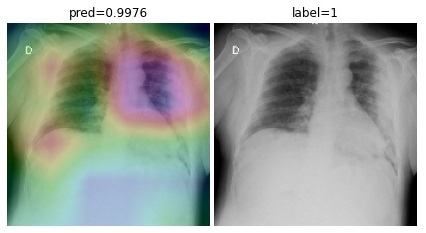}
        \label{fig:cx2_5}
    \end{subfigure}
    \hfill
    \begin{subfigure}{0.4\linewidth}
        \centering
        \includegraphics[width=\linewidth]{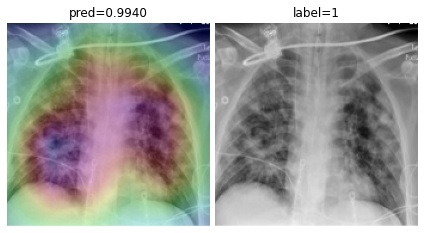}
        \label{fig:cx2_6}
    \end{subfigure}
    \hfill
    \begin{subfigure}{0.4\linewidth}
        \centering
        \includegraphics[width=\linewidth]{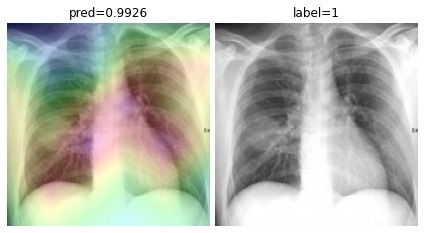}
        \label{fig:cx2_7}
    \end{subfigure}
    \hfill
    \begin{subfigure}{0.4\linewidth}
        \centering
        \includegraphics[width=\linewidth]{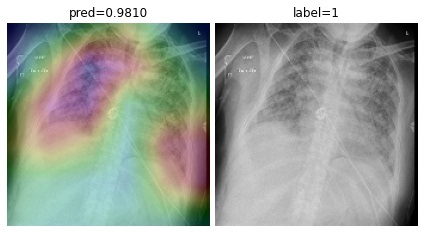}
        \label{fig:cx2_8}
    \end{subfigure}
\caption{Grad-CAMs from COVID-CXNet with lung segmentation module}
\label{fig:more_roi_cxnet_2}
\end{figure}

\newpage
\section{More Grad-CAMs of the Multiclass COVID-CXNet from Different Classes} \label{appendix: D}

\begin{figure}[H]
    \begin{subfigure}{0.4\linewidth}
        \centering
        \includegraphics[width=\linewidth]{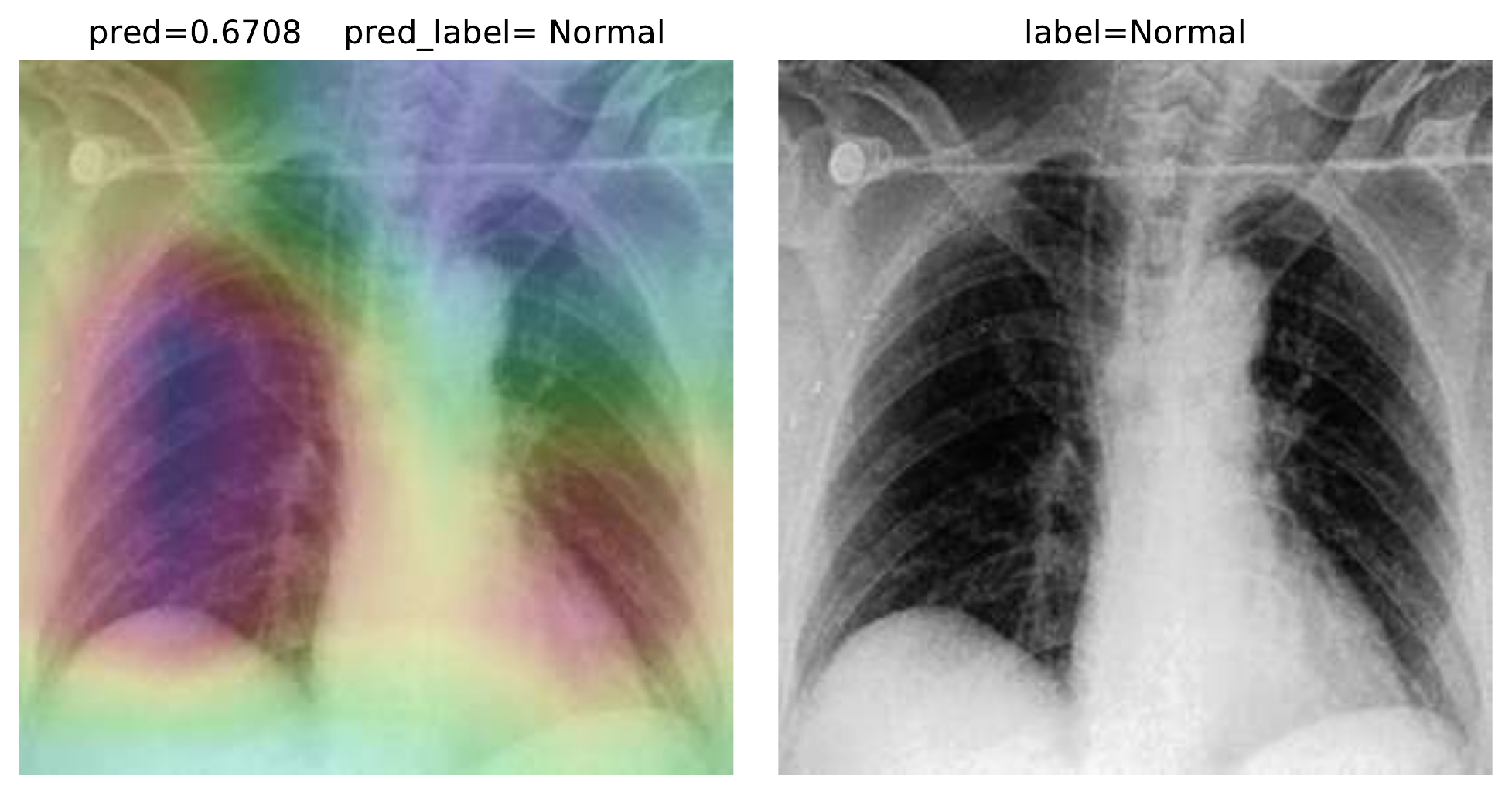}
        \label{fig:multiclass7}
    \end{subfigure}
    \begin{subfigure}{0.4\linewidth}
        \centering
        \includegraphics[width=\linewidth]{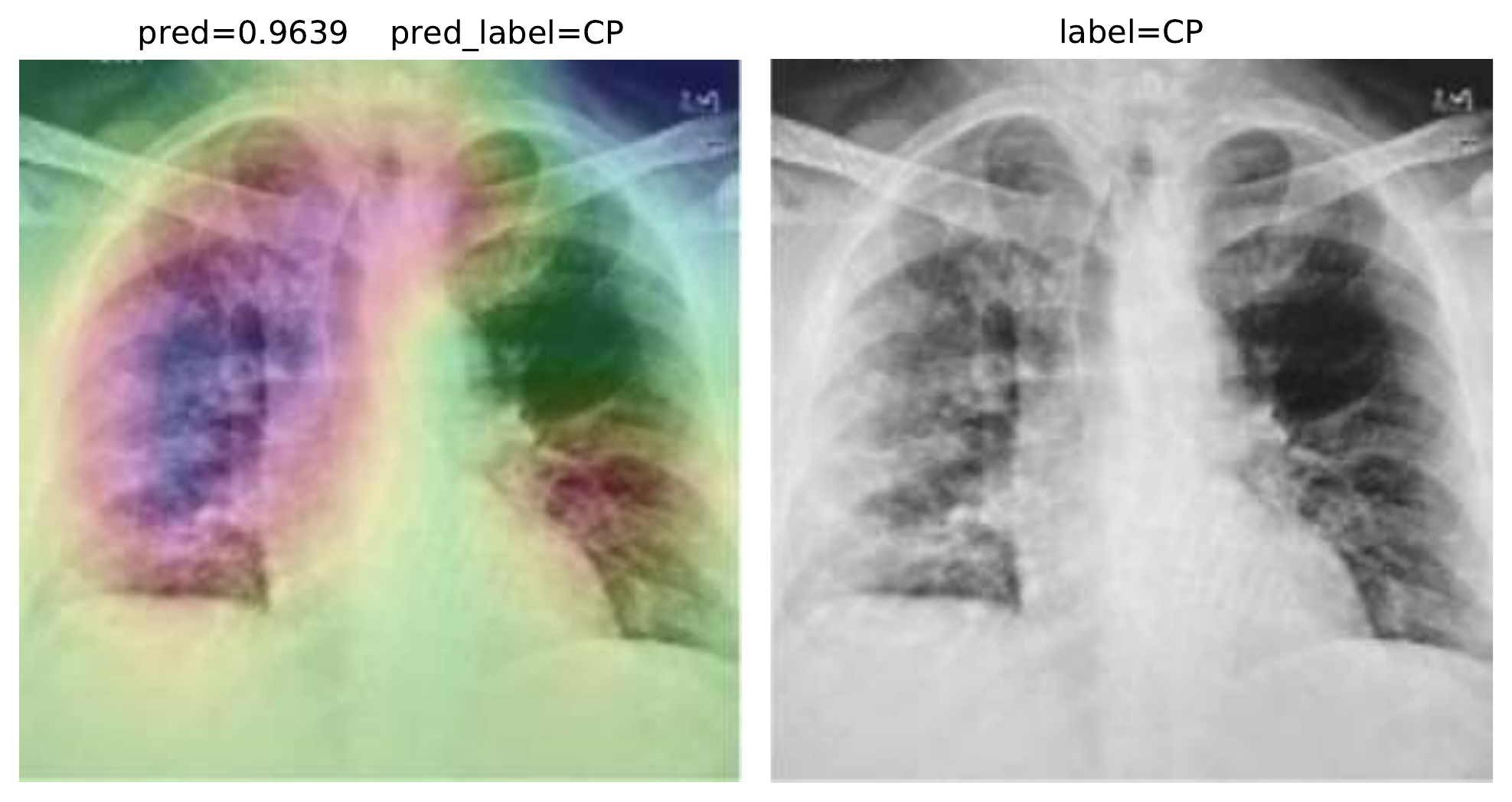}
        \label{fig:multiclass8}
    \end{subfigure}
    \begin{subfigure}{0.4\linewidth}
        \centering
        \includegraphics[width=\linewidth]{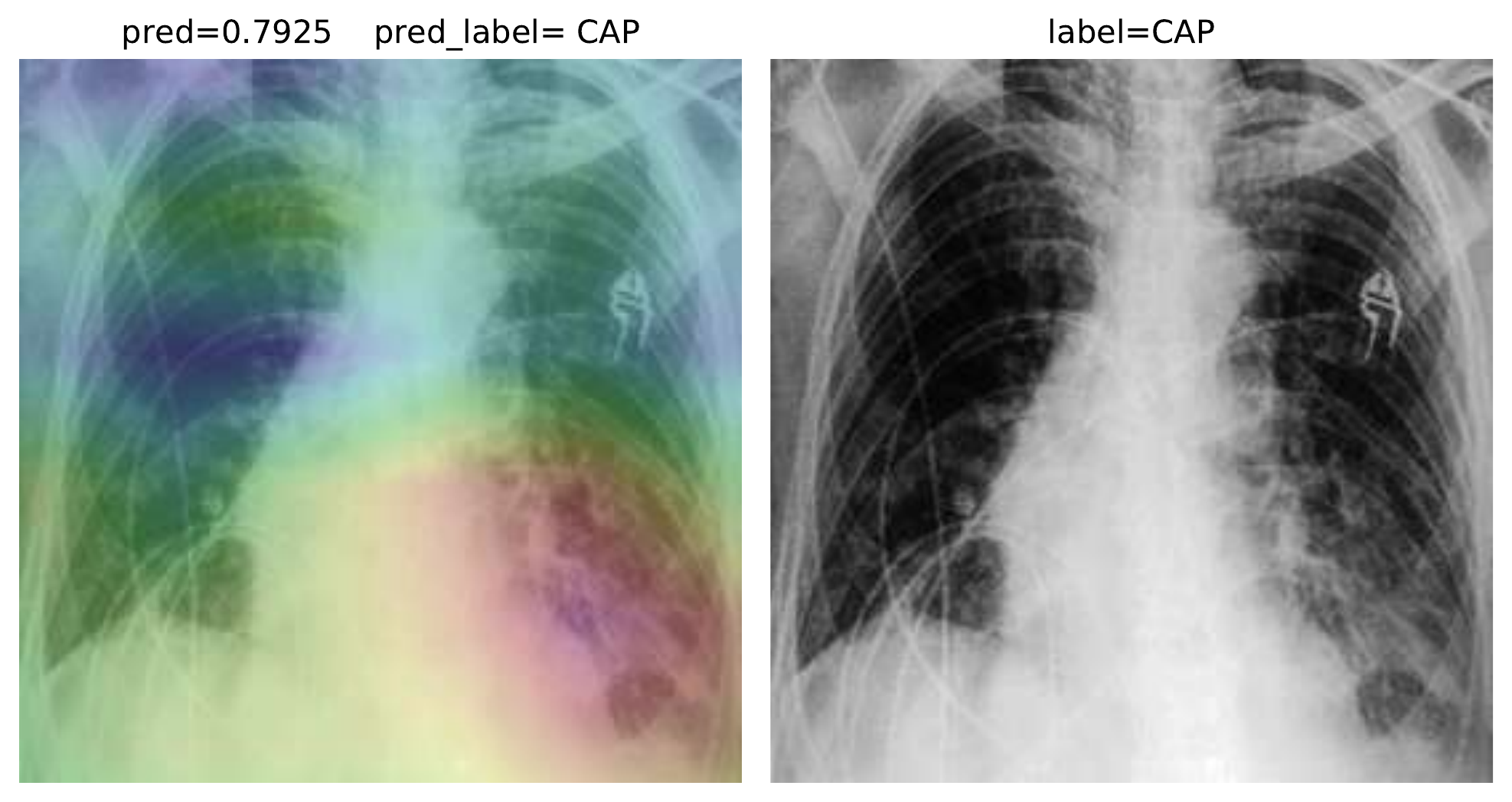}
        \label{fig:multiclass9}
    \end{subfigure}
    \begin{subfigure}{0.4\linewidth}
        \centering
        \includegraphics[width=\linewidth]{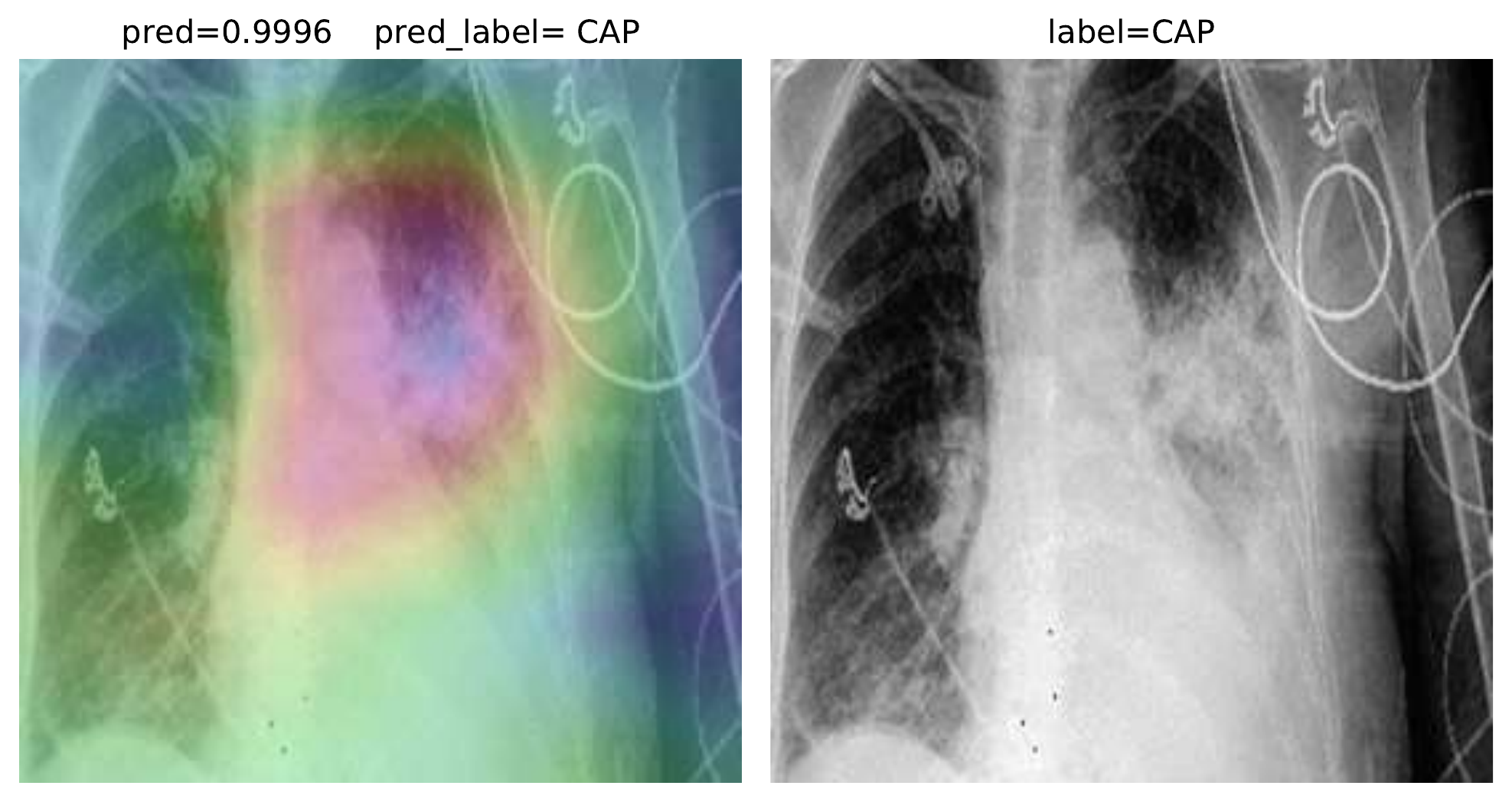}
        \label{fig:multiclass10}
    \end{subfigure}
    \begin{subfigure}{0.4\linewidth}
        \centering
        \includegraphics[width=\linewidth]{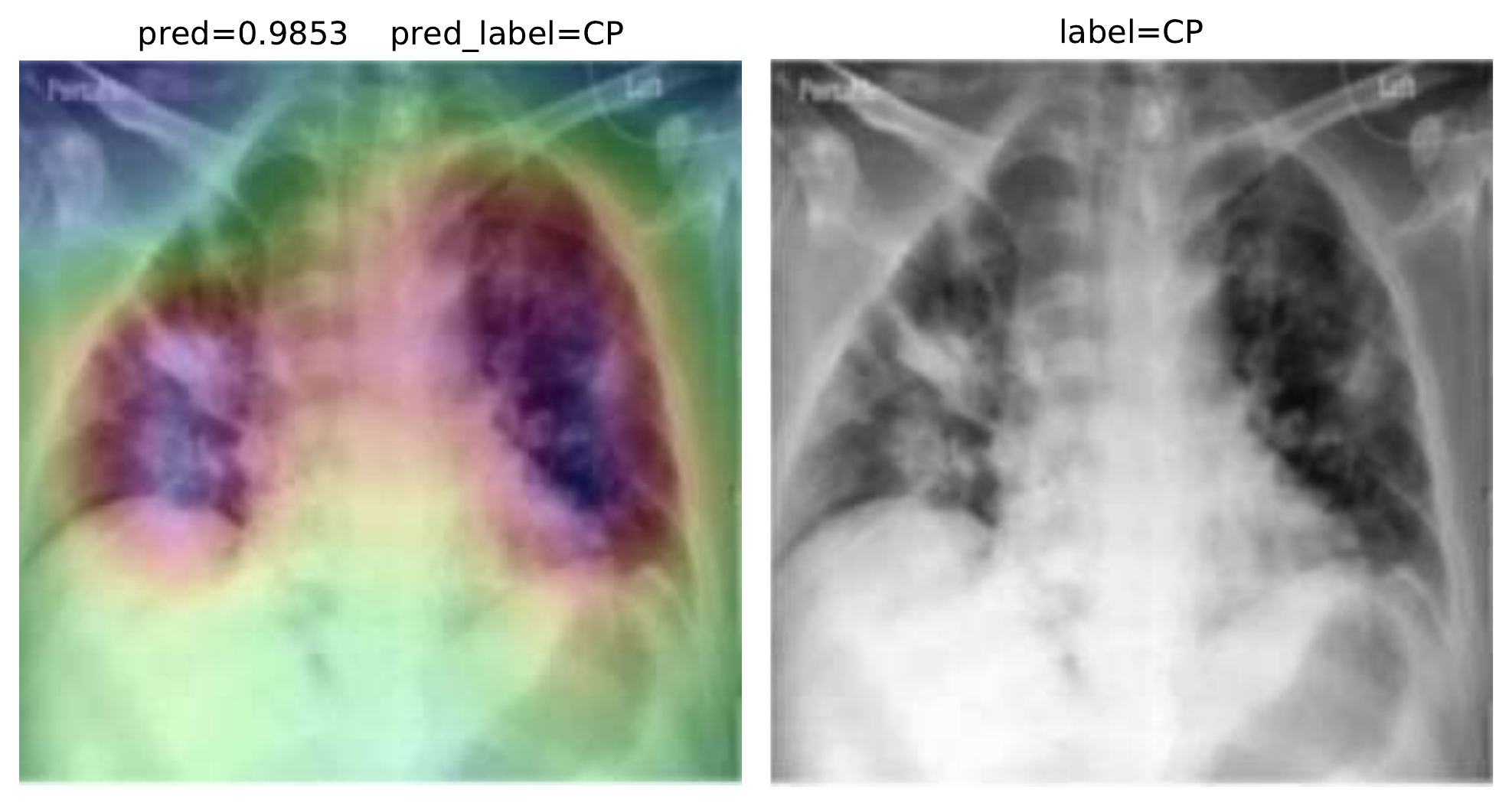}
        \label{fig:multiclass11}
    \end{subfigure}
    \begin{subfigure}{0.4\linewidth}
        \centering
        \includegraphics[width=\linewidth]{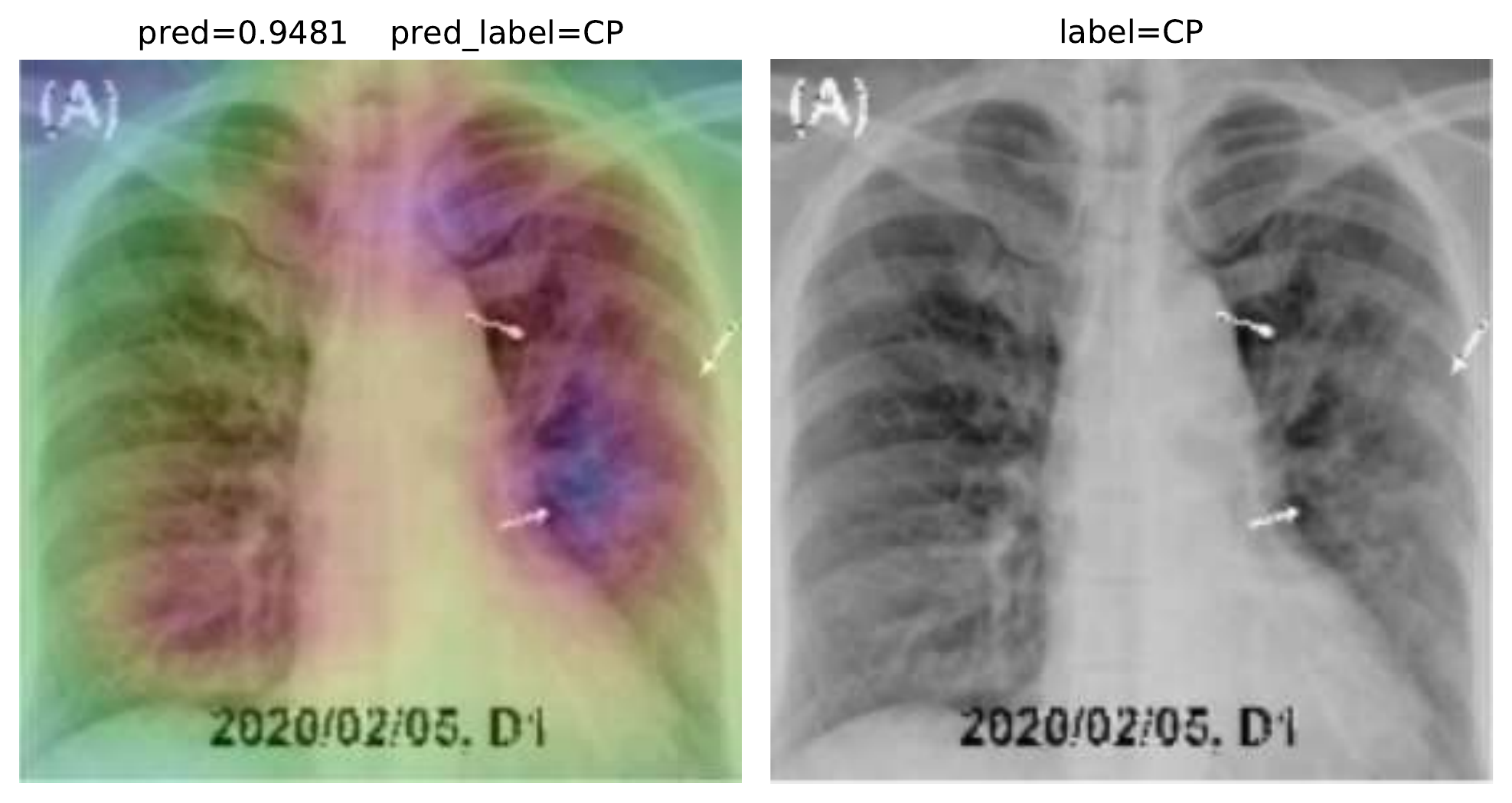}
        \label{fig:multiclass12}
    \end{subfigure}
    \hfill
    \begin{subfigure}{0.4\linewidth}
        \centering
        \includegraphics[width=\linewidth]{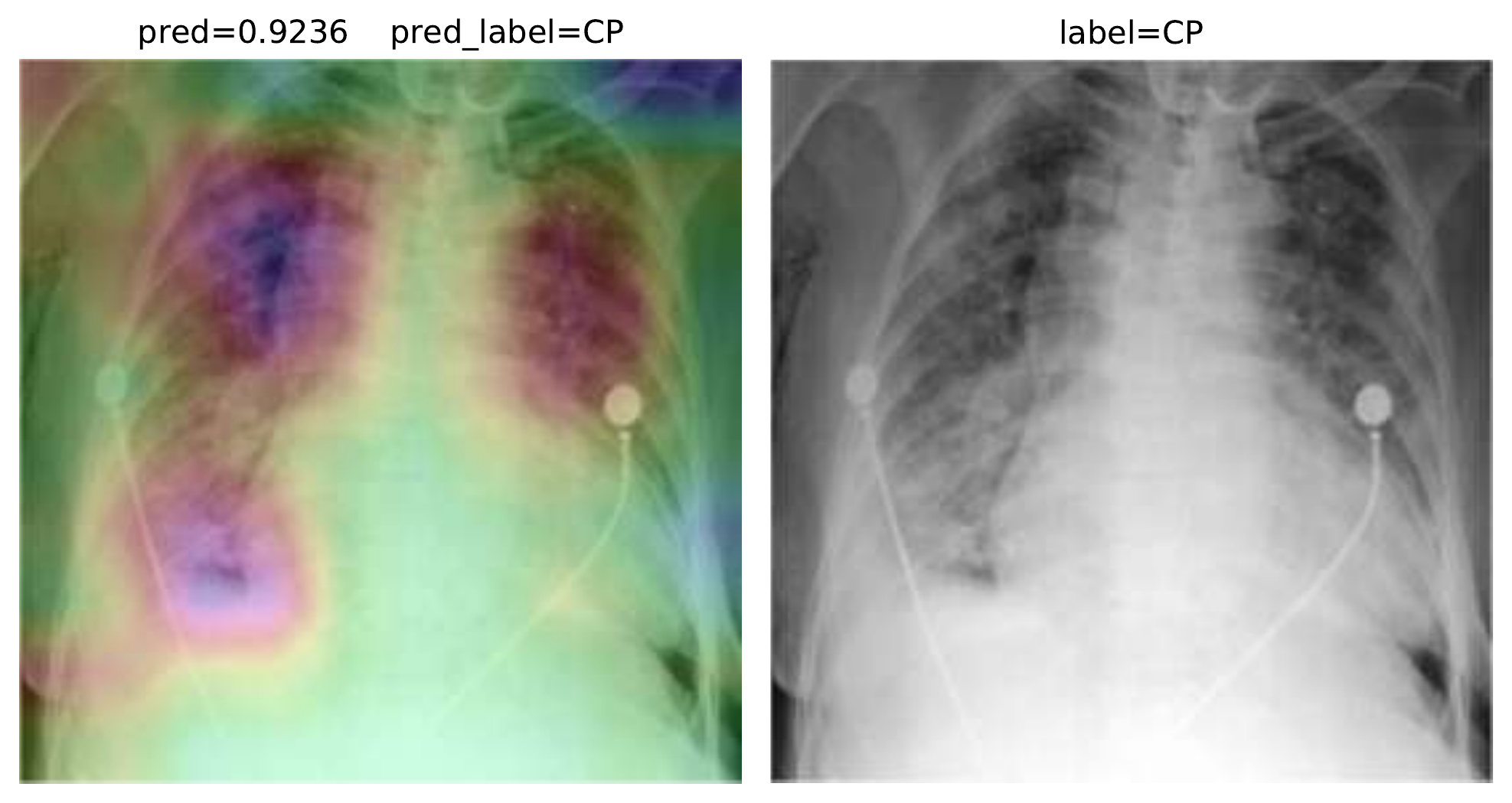}
        \label{fig:multiclass13}
    \end{subfigure}
    \hfill
    \begin{subfigure}{0.4\linewidth}
        \centering
        \includegraphics[width=\linewidth]{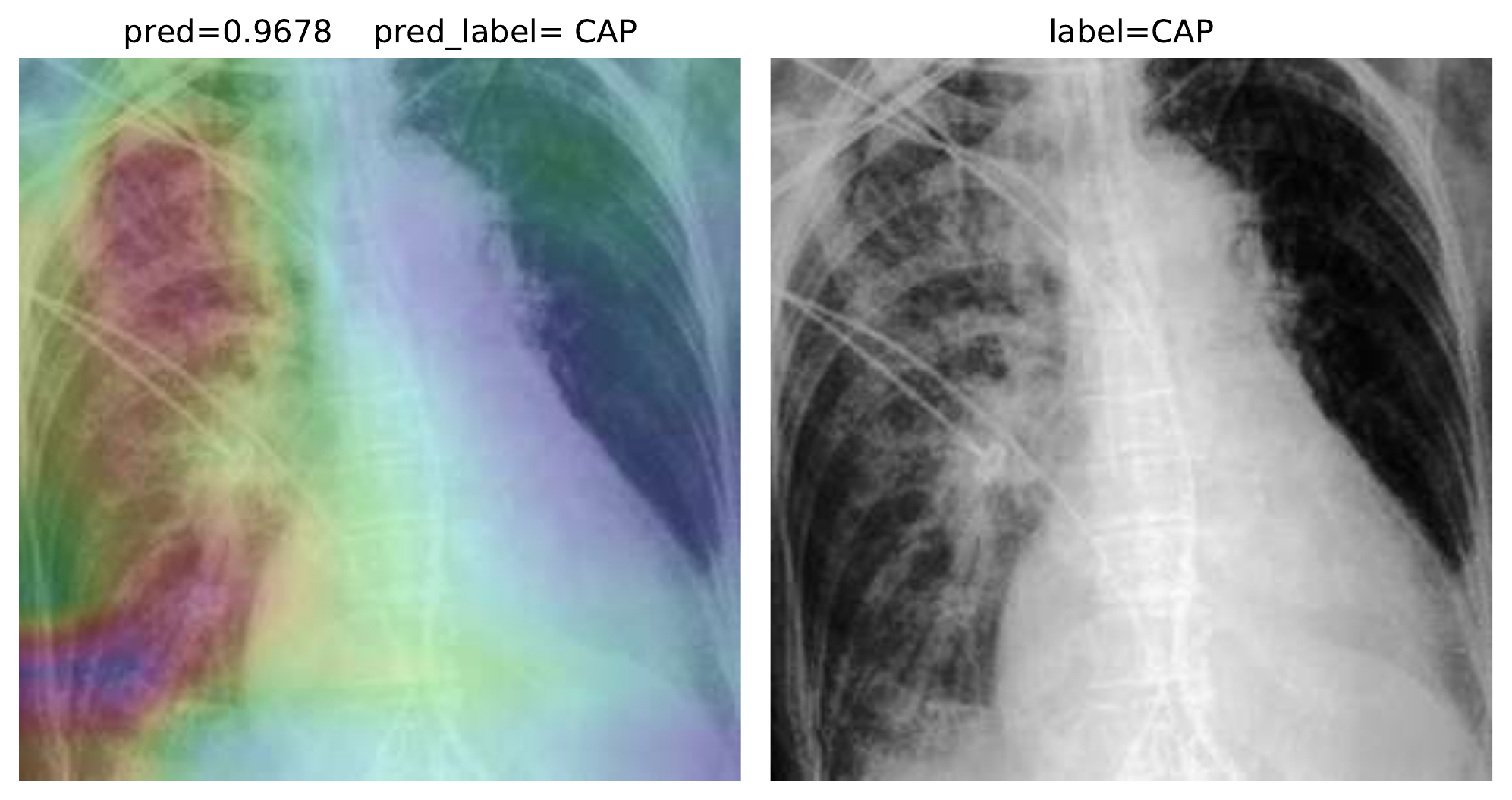}
        \label{fig:multiclass14}
    \end{subfigure}
    \hfill
    \begin{subfigure}{0.4\linewidth}
        \centering
        \includegraphics[width=\linewidth]{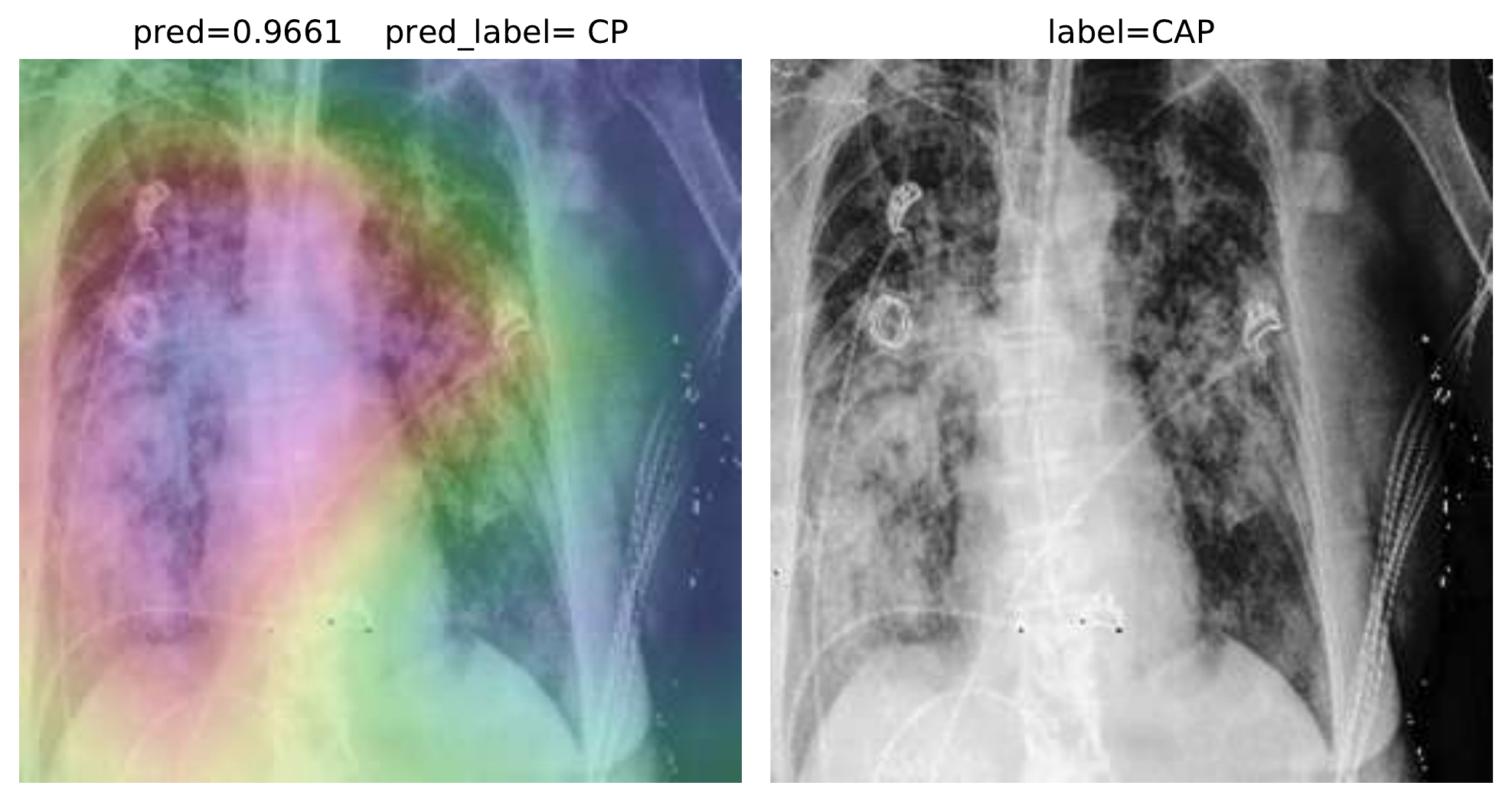}
        \label{fig:multiclass15}
    \end{subfigure}
    \hfill
    \begin{subfigure}{0.4\linewidth}
        \centering
        \includegraphics[width=\linewidth]{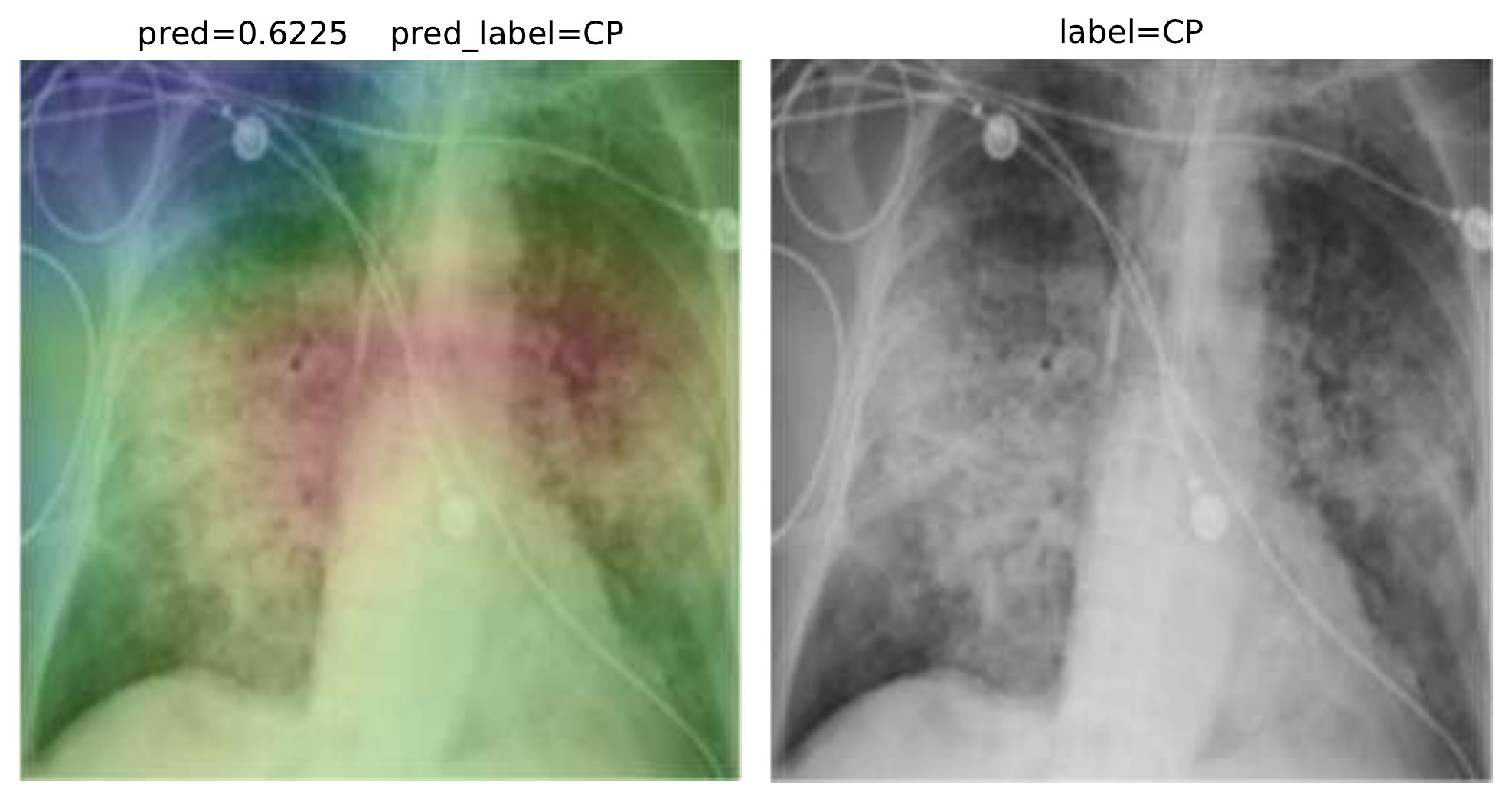}
        \label{fig:multiclass16}
    \end{subfigure}
\caption{Grad-CAMs from multiclass COVID-CXNet}
\label{fig:more_multi_cxnet}
\end{figure}

\end{document}